\documentclass[12pt,preprint]{aastex}

\usepackage{emulateapj5,apjfonts}
\usepackage{graphics}
\usepackage{amssymb}
\usepackage{onecolfloat}
\lefthead{Heitmann, Ricker, Warren, Habib}
\righthead{Robustness of Cosmological Simulations}

\begin{document}

\def\head{
  \vbox to 0pt{\vss
                    \hbox to 0pt{\hskip 440pt\rm LA-UR-04-5954\hss}
                   \vskip 25pt}

\title{Robustness of Cosmological Simulations I:
Large Scale Structure}
\author{Katrin~Heitmann\altaffilmark{1},
        Paul~M.~Ricker\altaffilmark{2,3},
        Michael~S.~Warren\altaffilmark{4}, and         
        Salman~Habib\altaffilmark{5}}

\affil{$^1$ ISR-1, ISR Division, The University of California, Los
Alamos National Laboratory, Los Alamos, NM 87545}
\affil{$^2$ Dept.\ of Astronomy, University of Illinois, Urbana, IL 61801}
\affil{$^3$ National Center for Supercomputing Applications, Urbana, IL 61801}
\affil{$^4$ T-6, Theoretical Division, The University of California, Los
Alamos National Laboratory, Los Alamos, NM 87545}
\affil{$^5$ T-8, Theoretical Division, The University of California, Los
Alamos National Laboratory, Los Alamos, NM 87545}

\date{today}

\begin{abstract}
  
The gravitationally-driven evolution of cold dark matter dominates the
formation of structure in the Universe over a wide range of length
scales. While the longest scales can be treated by perturbation
theory, a fully quantitative understanding of nonlinear effects
requires the application of large-scale particle simulation
methods. Additionally, precision predictions for next-generation
observations, such as weak gravitational lensing, can only be obtained
from numerical simulations. In this paper, we compare results from
several N-body codes using test problems and a diverse set of
diagnostics, focusing on a medium resolution regime appropriate for
studying many observationally relevant aspects of structure
formation. Our conclusions are that -- despite the use of different
algorithms and error-control methodologies -- overall, the codes yield
consistent results. The agreement over a wide range of scales for the
cosmological tests is test-dependent. In the best cases, it is at the
5\% level or better, however, for other cases it can be significantly
larger than 10\%. These include the halo mass function at low masses
and the mass power spectrum at small scales. While there exist
explanations for most of the discrepancies, our results point to the
need for significant improvement in N-body errors and their
understanding to match the precision of near-future observations. The
simulation results, including halo catalogs, and initial conditions
used, are publicly available.

\end{abstract}

\keywords{methods: N-body simulations ---
          cosmology: large-scale structure of the universe}
}

\twocolumn[\head]

\section{Introduction}
\subsection{Motivation}

There is strong evidence that structure formation in the Universe is
largely driven by gravitational instability. In the cosmological
`standard model' the evolution and structure of the mass distribution
is supposed to be dominated by `dark matter,' an as yet undetected
type of matter that interacts only gravitationally. In order to study
spatial and temporal aspects of the mass distribution today, it is
convenient to consider three length scales: very large length scales
where a linear treatment of the instability suffices, an intermediate
length scale where nonlinearities cannot be neglected yet the dynamics
of baryons are not of significant importance, and a high-resolution
regime where gas dynamics, star formation, etc. need to be taken into
account. While an analytic treatment is adequate at the largest
scales, because of the nonlinearity, both the intermediate and
small-scale regimes require a numerical approach.

In cold dark matter-dominated cosmologies, the transition from the
linear regime to the nonlinear regime at $z=0$ occurs at length scales
of order several $h^{-1}$ Mpc (or $k\sim 0.1~h$Mpc$^{-1}$). Depending
on the application, length scales larger than this can often be
satisfactorily treated by linear theory, augmented by perturbative
corrections as needed (Sahni \& Coles 1995). Galaxies exist within
dark matter halos, and typical halo size scales are $\sim 0.1-1$ Mpc.
Thus, in order to resolve the dynamics of halos, a spatial resolution
$\sim 10$ kpc appears to be necessary. (Here we will not be interested
in questions relating to the cores of halos and halo substructure,
where even higher resolution is needed.)

This is the first of two papers where we carry out comparison studies
of cosmological simulations focusing only on the dynamics cold dark
matter. Given the dominance of this component in the matter budget and
expected precision observations of the matter distribution from large
scale structure and lensing studies, however, we believe these studies
to be timely and important. Realistic simulations of clusters and
galaxy formation require high-resolution treatments of both dark
matter and gas dynamics including feedback from star formation, AGNs,
gas preheating, etc. There is not yet complete agreement on how all of
these effects should be incorporated; therefore, we believe that it is
too early to attempt a more complete study.

In this paper, we focus on medium resolution simulations, i.e., where
the spatial resolution is $\sim 10-100$ kpc. These simulations are
adequate for matter power spectrum computations, for weak lensing
studies, and for gathering cluster statistics. In the next paper, we
will focus on higher-resolution simulations which aim to resolve
individual galactic halos.

The main purpose of this work is to characterize the variation of
results from different cosmological simulation codes, all codes being
given exactly the same initial conditions, and all results being
analyzed in identical fashion. The point of the tests carried out in
this paper is to identify systematic differences between code results
and, to the extent possible, attempt to identify the reasons behind
gross discrepancies, should such arise.  The aim is not to tune the
codes to agree within some pre-defined error but rather to explore the
variability in the results when the codes are run using ``reasonable''
simulation parameters. In addition, it is also not our purpose to
establish which code is ``best'' for which application; while we have
often run codes with different force resolutions to illustrate general
points, we have made no effort to carry out a detailed benchmarking
exercise. Finally, the characterization of intrinsically statistical
issues such as limitations arising from finite sampling and studies of
the physical/observational relevance of the diagnostic tools employed
(e.g., halo finders) lie beyond the scope of this paper.

\subsection{N-body Simulation}

At the scales of interest to structure formation a Newtonian
approximation is sufficient to describe gravitational dynamics
(Peebles 1980). The task is therefore reduced to solving a
(collisionless) Vlasov-Poisson equation. This equation is a
six-dimensional partial differential equation, thus, on memory grounds
alone, a brute force approach cannot be applied. In addition, because
of its essential nonlinearity, the Vlasov-Poisson equation generates
ever-smaller spatial structures with time; it therefore becomes
necessary to utilize a robust scheme that avoids instabilities as
structure is created on sub-resolution scales. Having first made their
appearance in plasma physics, N-body codes form a now-standard
approach for dealing with this problem (Klypin \& Shandarin 1983;
Efstathiou et al. 1985; Sellwood 1987; Hockney \& Eastwood 1989;
Birdsall \& Langdon 1991; Bertschinger 1998; Couchman 1999).

In the N-body approach, the six-dimensional phase space distribution
is sampled by `tracer' particles and these particles are evolved by
computing the inter-particle gravitational forces. Since for $N$
particles this is a computationally intensive $O(N^2)$ problem,
approximation techniques are used to reduce the force computation to
$\sim O(N)$ or $\sim O(N\log N)$. These approximation techniques are
typically grid-based or employ multipole expansions, or are hybrids of
the two. Finally, the addition of gas dynamics is accomplished by
coupling the N-body solver to a hydro-code (other effects such as
star-formation, etc. can be treated by employing sub-grid techniques).

As stated above, the purpose of this paper is to investigate the
consistency of results from purely gravitational N-body codes in the
moderate resolution regime. (While performance issues are also
important, they will not be considered here.) In this regime, planned
observations are now reaching the point where it is necessary that
simulations be performed to low single-digit (percentage) accuracy
(Refregier et al. 2004). To establish an initial baseline, we
systematically compare results from the following codes: MC$^2$ ({\bf
  M}esh-based {\bf C}osmology {\bf C}ode)~(Habib et al.  2004), a
direct parallel particle-mesh (PM) solver; FLASH~(Fryxell et al.
2000\footnote{http://flash.uchicago.edu}), an adaptive-mesh-refinement
(AMR) grid code; HOT ({\bf H}ashed-{\bf O}ct {\bf T}ree)~(Warren \&
Salmon 1993), a tree-code; the tree code GADGET ({\bf GA}laxies with
{\bf D}ark matter and {\bf G}as int{\bf E}rac{\bf T}ions)~(Springel et
al. 2001\footnote{http://www.mpa-garching.mpg.de/gadget/right.html}); 
HYDRA, an adaptive-mesh P$^3$M-SPH code~(Couchman et al. 1995
\footnote{http://coho.mcmaster.ca/hydra/hydra\_consort.html}), and TPM,
a tree particle mesh code~(Xu 1995 and Bode et al.
2000\footnote{http://www.astro.princeton.edu/~bode/TPM/}). FLASH,
GADGET, HYDRA, and TPM are publicly available.

A variety of tests and diagnostic tools have been employed; in all
cases, the codes were run with identical initial conditions and the
results analyzed with identical diagnostic tools. We begin with the
Zel'dovich pancake collapse test to investigate issues of convergence
and collisionality~(Melott et al. 1997). We then turn to two
relatively realistic situations, the Santa Barbara cluster comparison
(Frenk et al.  1999) and `concordance' $\Lambda$CDM simulations.
Several diagnostics such as velocity statistics, halo catalogs, mass
functions, correlation functions and power spectra, etc. were used to
compare code results.  The initial conditions, outputs from all codes
at $z=0$, and halo catalogs used in this paper are publicly
available\footnote{http://t8web.lanl.gov/people/heitmann/arxiv/}.

The organization of the paper is as follows.  In Sec.~\ref{nbody}, we
give a short general introduction to N-body methods, followed by
summaries of the six different codes employed here. In
Sec.~\ref{sec:pan} we discuss results from the Zel'dovich pancake
test, focusing on questions regarding symmetry-breaking and
collisionality. In Secs.~\ref{sec:SB} and \ref{sec:SM}, we describe
results from the Santa Barbara cluster test and simulations based on
the parameters for the cosmological Standard Model as measured by WMAP
(Spergel et al. 2003) and other observations.  In Sec.~\ref{sec:disc}
we discuss our results for the different codes and summarize our
conclusions.

\section{The N-body Problem for Dark Matter}
\label{nbody}

The Vlasov-Poisson equation in an expanding Universe describes the
evolution of the six-dimensional, one-particle distribution function,
$f(\bf{x},\bf{p})$ (Vlasov equation):
\begin{equation}
{\partial f({\bf x},{\bf p}) \over\partial t}=
-\frac{{\bf p}}{ma^2}\cdot\nabla f({\bf x},{\bf p})+m\nabla\Phi({\bf
x})\cdot  
{\partial f({\bf x},{\bf p}) \over\partial {\bf p}},
\label{vlasov}
\end{equation}
where $\bf{x}$ is the comoving coordinate, ${\bf p}=ma^2\dot{\bf{x}}$,
$m$, the particle mass, and $\Phi({\bf x})$, the self-consistent
gravitational potential, determined by solving the Poisson equation,
\begin{equation}
\nabla^2\Phi({\bf x})=4\pi Ga^2[\rho({\bf x},t)-\rho_b(t)], \label{poisson}
\end{equation}
where $\rho_b$ is the background mass density and $G$ is the Newtonian
constant of gravitation. The Vlasov-Poisson system,
Eqns.~(\ref{vlasov}) -- (\ref{poisson}) constitutes a collisionless, 
mean-field approximation to the evolution of the full N-particle
distribution. As mentioned earlier, N-body codes attempt to solve
Eqns.~(\ref{vlasov}) -- (\ref{poisson}), by representing the
one-particle distribution function as
\begin{equation}
f(\bf{x},\bf{p})=\sum_{i=1}^N\delta(\bf{x}-\bf{x}_i)
                             \delta(\bf{p}-\bf{p}_i).
\label{nbrep}
\end{equation}
Substitution of Eqn.~(\ref{nbrep}) in the Vlasov-Poisson system of
equations yields the exact Newton's equations for a system of $N$
gravitating particles. It is important to keep in mind, however, that
we are not really interested in solving the exact N-body problem for a
finite number of particles, $N$. The actual problem of interest is the
exact N-body problem in the fluid limit, i.e., as
$N\rightarrow\infty$. For this reason, one important aspect of the
numerical fidelity of N-body codes lies in controlling errors due to
the discrete nature of the particle representation of
$f(\bf{x},\bf{p})$. Once the representation (\ref{nbrep}) is accepted,
errors also arise from time-stepping of the Newton's equations as well
as in solving the Poisson equation (\ref{poisson}). While heuristic
prescriptions are routinely followed and tested (see, e.g., Power et
al. 2003), a comprehensive theory of N-body errors does not yet exist.

Determining the forces for $N$ particles exactly requires $O(N^2)$
calculations. This is an unacceptable computational cost at the large
values of $N$ typical in modern N-body codes ($N\geq 10^7$),
especially as we are not interested in the exact solution in the first
place. Two popular approximate methods are used to get around this
problem -- the first introduces a spatial grid and the second employs
a multipole expansion. Hybrid algorithms that meld grid and
particle-force calculations are also common. The simplest $O(N\log N)$
grid code uses a particle-mesh (PM) methodology wherein the particle
positions are sampled to yield a density field $\rho(\bf{x})$ on a
regular grid. The Poisson equation is then solved by Fourier or other
(e.g., multi-grid) methods, and the force interpolated back on the
particles. Improvement on the grid resolution can be achieved by
adaptive mesh refinement (AMR) and by direct particle force
computation for particles within a few grid cells -- the
particle-particle particle mesh (P$^3$M) method. Tree codes are based
on the idea of approximating the gravitational potential of a set of
particles by a multipole expansion. All the particles are arranged in
a hierarchical group structure with the topology of a tree. Given a
point at which the potential needs to be evaluated and a particular
particle group, one decides if the point is sufficiently far away
(opening angle criterion) in order for a (suitably truncated)
multipole expansion to be used; if not, the subgroups of the parent
group are investigated. The process is continued until all groups are
fully searched down to the individual particle level.

Basic grid methods such as PM are fast, relatively simple to
implement, and memory-efficient. Consequently, they have good mass
resolution, but due to the limited dynamic range of the
three-dimensional spatial grid, only moderate force resolution. Tree
codes are gridless and have much better force resolution, but this
typically involves a performance and memory cost. Adaptive mesh and
hybrid algorithms such as tree-PM (TPM) constitute further steps in
attaining better compromises between mass and force resolution.  (Due
to discreteness/collisionality errors, excellent force resolution with
poor mass resolution is potentially just as problematic as the
reverse.)

\subsection{The Codes}

The codes utilized in this study have different algorithms, possess
different error modes and employ different error control
strategies. Thus, good convergence to a single solution is a strong
test of the validity of the N-body approach in modeling
dissipationless gravitational dynamics at the resolution scales probed
by the tests. We emphasize that the codes were run with no special
effort to optimize parameters separately for the individual test
problems; to conform to typical usage, only the recommended default
parameter values were used. Separate introductions to the codes are
given below.

\subsubsection{MC$^2$}

The multi-species MC$^2$ code suite~(Habib et al. 2004) includes a
parallel PM solver for application to large scale structure formation
problems in cosmology. In part, the code descended from parallel
space-charge solvers for studying high-current charged-particle beams
developed at Los Alamos National Laboratory under a DOE Grand
Challenge~(Ryne et al. 1998; Qiang et al. 2000). It exists in two
versions: a simpler version meant for fast prototyping and algorithm
development written in High Performance Fortran (HPF), and a
high-performance version based on Viktor Decyk's F90/MPI UPIC
framework~(Decyk \&
Norton\footnote{http://exodus.physics.ucla.edu/research/UPICFramework.pdf}).
This code has excellent mass resolution and has proven its efficiency
on multiple HPC platforms, with scaling being verified up to 2000
processors.

MC$^2$ solves the Vlasov-Poisson system of equations for an expanding
universe using standard mass deposition and force interpolation
methods allowing for periodic or open boundary conditions with second
and fourth-order (global) symplectic time-stepping and a Fast Fourier
Transform (FFT)-based Poisson solver. The results reported in this
paper were obtained using Cloud-In-Cell (CIC)
deposition/interpolation. The overall computational scheme has proven
to be remarkably accurate and efficient: relatively large time-steps
are possible with exceptional energy conservation being achieved.

\subsubsection{FLASH}

FLASH (Fryxell et al. 2000) originated as an AMR hydrodynamics code
designed to study X-ray bursts, novae, and Type~Ia supernovae as part
of the DOE ASCI Alliances Program.  Block-structured adaptive mesh
refinement is provided via the PARAMESH library~(MacNeice et al.
2000).  FLASH uses an oct-tree refinement scheme similar to Quirk
(1991) and de~Zeeuw \& Powell (1993). Each mesh block contains the
same number of zones ($16^3$ for the runs in this paper), and its
neighbors must be at the same level of refinement or one level higher
or lower (mesh consistency criterion). Adjacent refinement levels are
separated by a factor of two in spatial resolution. The refinement
criterion used is based upon logarithmic density thresholds. The mesh
is fully refined to the level at which the initial conditions contain
one particle per zone.  Blocks with maximum densities between 30 and
300 times the average are refined by one additional level. Those with
maxima between 300 and 3000 times the average are refined by two
additional levels, and each factor of 10 in maximum density above that
corresponds to an additional level of refinement. FLASH has been
extensively verified and validated, is scalable to thousands of
processors, and extensible to new problems~(Rosner et al. 2000; Calder
et al. 2000; Calder et al. 2002).

Numerous extensions to FLASH have been developed, including solvers
for thermal conduction, magnetohydrodynamics, radiative cooling,
self-gravity, and particle dynamics. In particular, FLASH now includes
a multigrid solver for self-gravity and an adaptive particle-mesh
solver for particle dynamics~(Ricker et al. 2004). Together with the
PPM hydrodynamics module, these provide the core of FLASH's
cosmological simulation capabilities. FLASH uses a variable-timestep
leapfrog integrator. In addition to other timestep limiters, the FLASH
particle module requires that particles travel no more than a fraction
of a zone during a timestep.

\subsubsection{HOT}

This parallel tree code~(Warren \& Salmon 1993) has been evolving for
over a decade on many platforms. The basic algorithm may be divided
into several stages (the method of error tolerance is described in
Salmon \& Warren 1994).  First, particles are domain decomposed into
spatial groups. Second, a distributed tree data structure is
constructed. In the main stage of the algorithm, this tree is
traversed independently in each processor, with requests for non-local
data being generated as needed. A \verb-Key- is assigned to each
particle, which is based on Morton ordering. This maps the points in
3-dimensional space to a 1-dimensional list, maintaining as much
spatial locality as possible.  The domain decomposition is obtained by
splitting this list into $N_p$ (number of processors) pieces. An
efficient mechanism for latency hiding in the tree traversal phase of
the algorithm is critical. To avoid stalls during non-local data
access, effectively explicit `context switching' is done using a
software queue to keep track of which computations have been put aside
waiting for messages to arrive. This code architecture allows HOT to
perform efficiently on parallel machines with fairly high
communication latencies (Warren et al. 2003). This code was among the
ones used for the original Santa Barbara Cluster Comparison Project
(Frenk et al. 1999) and also supports gas dynamics simulations via a
smoothed particle hydrodynamics (SPH) module (Fryer \& Warren 2002).
\newpage
\subsubsection{GADGET}

GADGET~(Springel et al. 2001) is a freely available N-body/hydro code
capable of operating both in serial and parallel modes. For the tests
performed in this paper the parallel version was used.  The
gravitational solver is based on a hierarchical tree-algorithm while
the gas dynamics (not used here) are evolved by SPH. GADGET allows for
open and periodic boundary conditions, the latter implemented via an
Ewald summation technique.  The code uses individual and adaptive time
steps for all particles.

\subsubsection{HYDRA}

HYDRA~(Couchman et al. 1995) is an adaptive P$^3$M (AP$^3$M) code with
additional SPH capability.  In this paper we use HYDRA only in the
collisionless mode by switching off the gas dynamics. The P$^3$M
method combines mesh force calculations with direct summation of
inter-particle forces on scales of two to three grid spacings. In
regions of strong clustering, the direct force calculations can become
significantly expensive. In AP$^3$M, this problem is tackled by
utilizing multiple levels of subgrids in these high density regions,
with direct force computations carried out on two to three spacings of
the higher-resolution meshes. Two different boundary conditions are
implemented in HYDRA, periodic and isolated.  The timestep algorithm
in the dark matter-only mode is equivalent to a leapfrog algorithm.

\subsubsection{TPM}

TPM~(Xu 1995 and Bode et al. 2000), a tree particle-mesh N-body
algorithm, is a hybrid code combining a PM and a tree algorithm. The
density field is broken down into many isolated high-density regions
which contain most of the mass in the simulation but only a small
fraction of the volume. In these regions the gravitational forces are
computed with the tree algorithm while for the bulk of the volume the
forces are calculated via a PM algorithm, the PM time steps being
large compared to the time-steps for the tree-algorithm. The time
integrator in TPM is a standard leap-frog scheme: the PM timesteps are
fixed whereas the timesteps for the tree-particles are variable.

\section{The Zel'dovich Pancake Test}
\label{sec:pan}
\subsection{Description of Test}

The cosmological pancake problem~(Zel'dovich 1970; Shandarin \&
Zeldovich 1989) provides a good simultaneous test of particle
dynamics, Poisson solver, and cosmological expansion. Analytic
solutions well into the nonlinear regime are available for both dark
matter and hydrodynamical codes~(Anninos \& Norman 1994), permitting
an assessment of code accuracy. After caustic formation, where the
analytic results fail, the problem continues to provide a stringent
test of the ability to track thin, poorly resolved features~(Melott
1983).  

As is conventional, we set the initial conditions for the pancake
problem in the linear regime. In a universe with $\Omega_0=1$ at
redshift $z$, a perturbation of wavenumber $k$ which collapses to a
caustic at redshift $z_c<z$ has comoving density and velocity given by 
\begin{eqnarray}
\label{Eqn:pancake soln 1}
\rho(x_e;z) & = & \bar{\rho}\left[1 -
{1+z_c\over1+z}\cos\left(kx_\ell\right)  
        \right]^{-1},\\
\label{Eqn:pancake soln 1.5}
v(x_e;z)    & = & -H_0(1+z)^{1/2}(1+z_c){\sin kx_\ell\over kx_\ell}\ ,
\end{eqnarray}
where $\bar{\rho}$ is the comoving mean density. Here $x_e$ is the
distance of  a point from  the pancake midplane, and $x_\ell$ is the
corresponding Lagrangian  coordinate,  found  by iteratively solving
\begin{equation}
\label{Eqn:perturb}
x_e = x_\ell - {1+z_c\over 1+z}{\sin kx_\ell\over k}\ .
\end{equation}
The Lagrangian coordinates for the dark matter particles $x_\ell$ are
assigned to lie on a uniform grid. The corresponding perturbed
coordinates $x_e$ are computed using
Eqn.~(\ref{Eqn:perturb}). Particle  velocities are assigned using
Eqn.~(\ref{Eqn:pancake soln 1.5}).    

As particles accelerate toward the midplane, their phase profile
develops a backwards ``S'' shape. At caustic formation the velocity
becomes multivalued at the midplane. The region containing multiple
streams grows in size as particles pass through the midplane. At the
edges of this region (the caustics, or the inflection points of the
``S''), the particle density is formally infinite, although the finite
force and mass resolution in the simulations keeps the height of these
peaks finite.  Some of the particles that have passed through the
midplane fall back and form another pair of caustics, twisting the
phase profile again.  Because each of these secondary caustics
contains five streams of particles rather than three, the second pair
of density peaks are higher than the first pair. In principle, this
caustic formation process repeats arbitrarily many times. In practice,
however, the finite number of particles and the finite force
resolution in simulations limit the number of caustics that can be
tracked.

In Melott et al.  (1997) the pancake simulation was used to test
different codes for unphysical collisionality and symmetry-breaking. 
It was found that tree and P$^3$M codes suffered from artifacts if the
value of the softening parameter was chosen too small, leading to
incorrect results for the pancake test.  We have carried out a set of
simulations to distinguish between lack of convergence due to
collisional effects and the failure of the test due to the inability
to strictly track the planar symmetry of the mass distribution.  Our
basic conclusions are that for the mass and force resolutions chosen
in the main cosmological simulations in this paper, collisional
effects are unlikely to be important.  While it is true that the
high-resolution codes fail at maintaining the symmetry of the pancake
test, the PM-code MC$^2$ does not.  Thus, one may verify whether this
failure is an issue in realistic simulations by comparing results for
the different codes, as we do later in the paper.  Our results are
consistent with the finding that, at least in the moderate resolution
regime, this failure in the pancake test does not translate to making
significant errors in more realistic simulations.

\subsubsection{Pancake Simulations}

Our main aim in performing the pancake simulations is to probe whether
the codes are working correctly in the dynamical range of interest.
For this purpose, small simulations are sufficient; what is important
is to match the mass resolution and the spatial resolution to the ones
used in the larger cosmological simulations.  Most of these larger
simulations used 256$^3$ particles, the box sizes varying between 64
Mpc, 90 Mpc, and 360 Mpc, with different cosmologies.  This leads to
particle masses of 1.08 $\cdot 10^9$ M$_\odot$, 1.9 $\cdot 10^9$
M$_\odot$, and 1.22 $\cdot 10^{11}$ M$_\odot$.  We perform the pancake
test in a $\sqrt{3} \cdot 10$~Mpc box, with 64$^3$ particles.  This
leads to a particle mass of 1.4 $\cdot 10^9$ M$_\odot$, which provides
a good representation for the cosmological test problems.  In order to
almost match the force resolution of the PM code with the other codes
we ran it using a 1024$^3$ grid in the cosmological simulations.  For
the 64~Mpc and 90~Mpc boxes, this is approximately equivalent to
running the pancake test with a 256$^3$ PM grid.

Based on these considerations, the following simulations were run: for
the cosmology we chose a SCDM model, i.e. $\Omega_m=1$, $H=50$
km/s/Mpc, $\Omega_b=\Omega_\Lambda=0$, in a $\sqrt{3} \cdot 10$~Mpc
box with 64$^3$ particles in all cases.  HOT was run with a (Plummer)
smoothing of 20 kpc while the smoothing for the HYDRA-simulation was
chosen to be 100 kpc. In AMR mode, the FLASH grid size was refined up
to three times, leading up to an effective 512$^3$ grid.  We varied
the mesh size for the MC$^2$ simulation between $64^3$ and 512$^3$ to
check convergence for increasing number of grid points. The TPM and
GADGET runs did not complete successfully for this test.

The initial conditions are set up such that the pancake normal is
aligned with the box diagonal, i.e., inclined at 54.7 degrees with
respect to the base plane. In order to check our results we performed
a high-resolution one-dimensional pancake simulation to serve as a
comparison template.  Due to the lack of an analytical solution, this
numerical comparison is important for verifying our results in the
nonlinear regime, after several caustics have formed.

\subsubsection{Results}

\begin{figure}
\includegraphics[width=70mm]{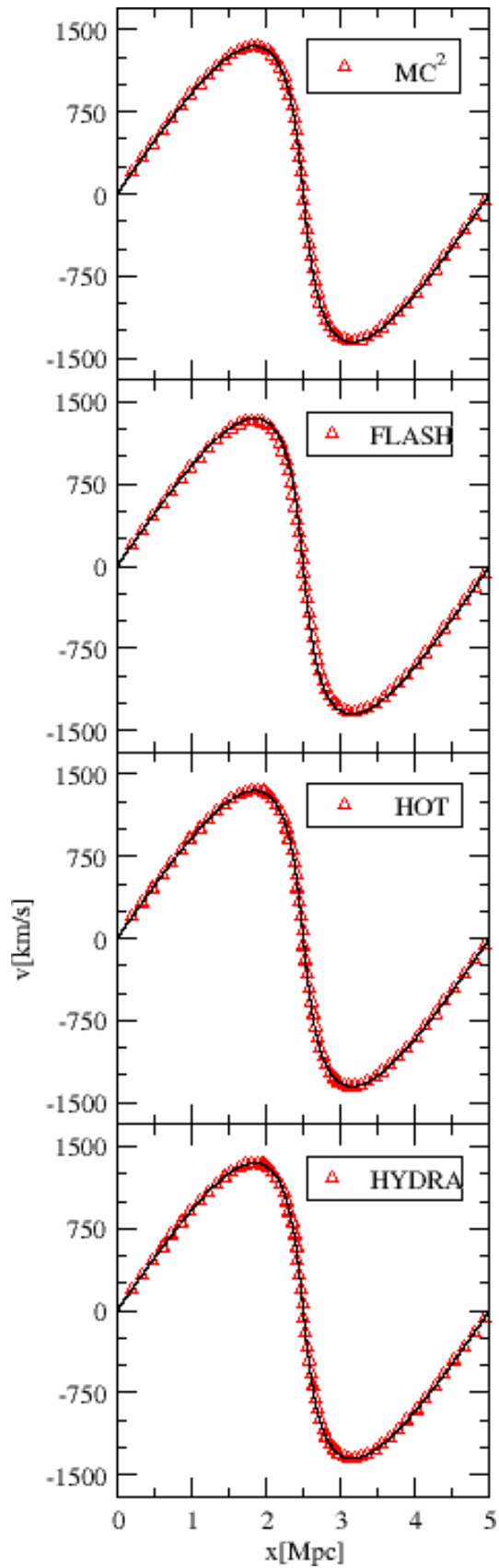}
\caption{Phase plane for the pancake test from four simulations
with $64^3$ particles at $z=7$, prior to caustic formation. As an
illustration, the FLASH result is taken from a low-resolution 64$^3$
mesh to demonstrate the resulting small inaccuracies near the curve
minimum and maximum. The solid lines are from a one-dimensional 
high-resolution simulation.}  
\label{plotone}
\end{figure}

\begin{figure}
\includegraphics[width=70mm]{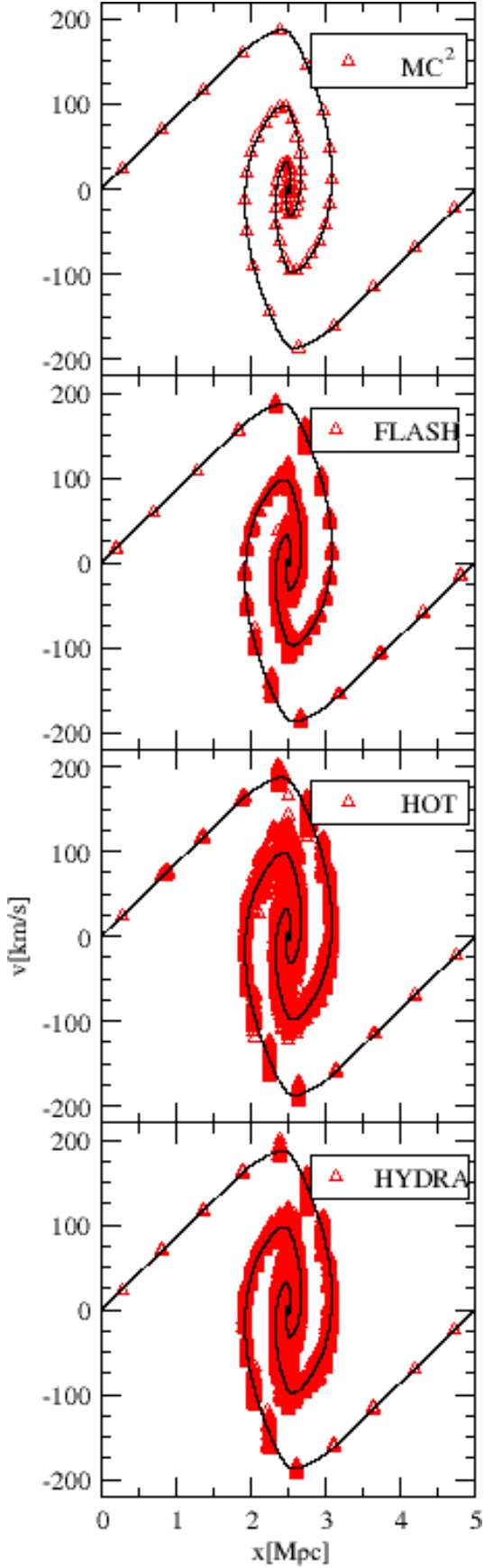}
\caption{Pancake test at $z=0$, $64^3$ particles, following
Fig.~\ref{plotone}. Very close to the center of the spiral, there is 
a seven-stream flow. Here, FLASH is run with an effective resolution 
equivalent to a $512^3$ mesh (For the equivalent resolution MC$^2$
results, see Fig.~\ref{plotthree}. For a discussion of all of the
results, see the text.}
\label{plottwo}
\end{figure}

\begin{figure}[t]
\includegraphics[width=85mm]{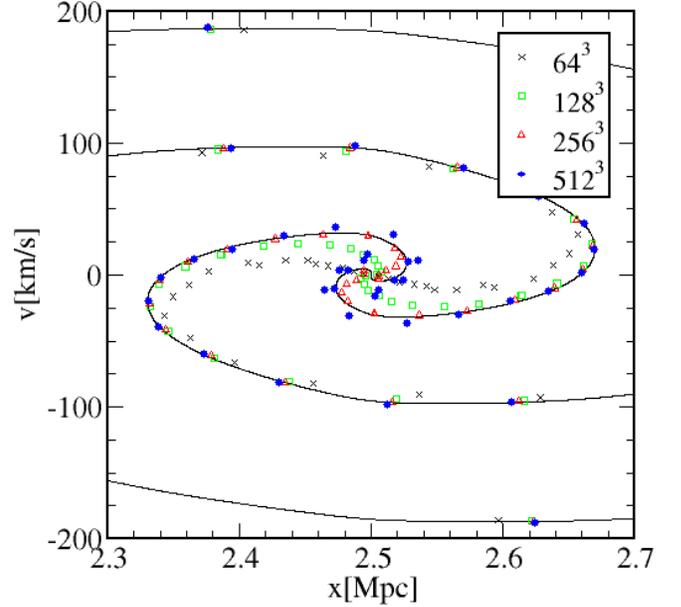}
\caption{Failure of convergence near the midplane for the pancake
test: MC$^2$ results, $64^3$ particles with four grid sizes at
$z=0$. Convergence fails at the final resolution reduction step (going
from a $256^3$ mesh to a $512^3$ mesh). See the text for a discussion
of these results.}  
\label{plotthree}
\end{figure}

The initial conditions were chosen so that the redshift at which the
first collapse occurs is $z_c= 5$.  We measured the positions and
velocities of the particles at $z=7$, i.e., before the collapse, and
also at $z=0$ after several caustics have formed.  The results are
displayed in phase-space plots, where positions and velocities are
projected onto the box diagonal.

At $z=7$, before caustic formation occurs, all codes perform very well
as can be seen in Fig.~\ref{plotone}. In addition, we performed a
convergence test with the PM-code MC$^2$. We ran a 64$^3$ particle
simulation on a mesh which was refined four times, from 64$^3$ to
512$^3$ mesh points (Fig.~\ref{plotone} shows the MC$^2$ result from
the $256^3$ mesh). At $z=7$, even the lowest resolution simulation
with a 64$^3$ mesh, e.g., the analogous result for FLASH in
Fig.~\ref{plotone}), traces the exact solution accurately. Expectedly,
increasing the number of mesh-points improves the result mainly near
the region of maximal density variation, i.e., at $x=2.5$~Mpc. No
failure of convergence was noted at the highest-resolution 512$^3$
simulation.

In contrast, by the time several caustics have formed, as shown in
Fig.~\ref{plottwo} for $z=0$, the code results display considerable
variation and lack of convergence. To understand this behavior, we
first turn to a convergence test with MC$^2$ shown in
Fig.~\ref{plotthree}. Here we have zoomed into the inner part of the
phase-space spiral by restricting the $x$-axis between 2.3 and
2.7~Mpc. While the 64$^3$ mesh simulation can track the first two
caustics, the inner part of the spiral is not resolved. A 128$^3$ mesh
provides some improvement, while a 256$^3$ mesh -- corresponding to a
resolution of 67 kpc -- leads to satisfactory agreement with the
essentially exact 1-D numerical solution (in Fig.~\ref{plottwo}, the
MC$^2$ results are shown for the $256^3$ mesh simulation).  This is
the level of MC$^2$ spatial resolution used for the cosmological
simulations.

As the mesh is refined further, collisional effects must enter at some
point. In fact, by refining the grid one more time and running the
64$^3$ particles on a 512$^3$ mesh (corresponding to a resolution of
34 kpc) the failure of convergence can be directly observed. While the
increase from a 64$^3$ mesh to a 256$^3$ mesh clearly led to a better
refinement of the inner spiral, i.e. the simulation was converging
towards the exact solution, a further increase of the spatial
resolution does {\em not} continue the improvement.  The particles on
the inner part of the spiral are moving away from the exact solution
instead of converging towards it.  Collisional effects destroy the
good convergence properties which were found for the smaller meshes.
Therefore, small-scale resolved structures at this high level of
spatial resolution can turn out to be incorrect. Convergence tests
performed with FLASH with uniform resolution give similar results.

\begin{table*}[t]
\begin{center}
\caption{\label{tabsbres} Softening lengths/mesh spacing for the Santa Barbara Runs} 
\begin{tabular}{lcccccc}
\hline\hline
\hfill& MC$^2$ & FLASH & HOT &GADGET & TPM & HYDRA \\
\hline
128$^3$ particles & 62.5 kpc  &250 kpc & 5 kpc (physical after z=9)&
10 kpc & 10 kpc & 10 kpc \\ 
256$^3$ particles & 62.5 kpc  &62.5 kpc & 5 kpc (physical after z=9)&
50 kpc & N/A & N/A \\ 
\hline\hline

\vspace{-1.5cm}

\tablecomments{Resolutions of the different codes for the different 
  Santa Barbara runs; the highest resolution runs were carried out with 
  HOT. If not so noted, the resolution is quoted in comoving coordinates.}
\end{tabular}
\end{center}
\end{table*}

It is clear, however, that the results in Fig.~\ref{plottwo} show
artifacts much worse than the mild lack of convergence observed in
Fig.~\ref{plotthree}. For example, the result from running FLASH with
refinement turned on, yielding an effective 512$^3$ mesh, shows severe
artifacts: the particle distribution is smeared out close to the
center of the spiral. Note that the smearing seen here cannot be
explained as a consequence of ``over-resolution'' as this effect is
not seen in the MC$^2$ results of Fig.~\ref{plotthree}, which have
the same effective resolution. A similar result was obtained with the
tree-code HOT, where the structure of the inner spiral is more or less
completely lost; the results from the AP$^3$M code HYDRA are not any
better. It is the failure of the AMR, tree, and AP$^3$M codes to
maintain the strict planar symmetry of the pancake collapse that is
apparently the root cause of the problem. To reiterate, the 256$^3$
and 512$^3$ PM runs roughly span the force resolutions used for the
other codes, and since only the 512$^3$ run shows a very mild failure
of convergence, force resolution alone cannot be the source of the
difficulty.

Our results provide a different and more optimistic interpretation of
the findings of Melott et al.  (1997). (See also Binney 2004.) While
high-resolution codes when run with small smoothing lengths (or
several refinement levels in the case of AMR) are not able to pass the
pancake test after the formation of several caustics, the main culprit
appears to be an inability to maintain the planar symmetry of the
problem and not direct collisionality (at least at the force
resolutions relevant for this paper), which would have been far more
serious. Whether the failure to treat planar collapse is a problem in
more realistic situations can be tested by comparing results from the
high-resolution codes against brute-force PM simulations. A battery of
such tests have been carried out in Secs.~\ref{sec:SB} and
\ref{sec:SM}. At the force resolutions investigated, these tests
failed to yield evidence for significant deviations.


\section{The Santa Barbara Cluster}
\label{sec:SB}

\subsection{Description of the Test}

Results from the Santa Barbara Cluster Comparison Project were
reported in 1999 in Frenk et al. (1999). The aim of this project was
to compare different techniques for simulating the formation of a
cluster of galaxies in a cold dark matter universe and to decide if
the results from different codes were consistent and reproducible.
For this purpose outputs from 12 different codes were examined,
representing numerical techniques ranging from SPH to grid methods
with fixed, deformable, and multilevel meshes. The starting point for
every code was the same set of initial conditions given either by a
set of initial positions or an initial density field.  Every simulator
was then allowed to evolve these initial conditions in a way best
suited for the individual code, i.e., implementations of smoothing
strategies, integration time steps, boundary conditions, etc. were not
specified but their choice left to the individual simulators. The
comparison was kept very general and practical; the main interest was
not to quantify the accuracy of individual codes but to investigate
the reliability of results from differing astrophysical simulations.
For this purpose, images of clusters at different epochs were
compared, as well as the mass, temperature, cluster luminosity, and
radial profiles of various dynamical and thermodynamic quantities.

In this paper we restrict ourselves to the dark matter component of
the simulation, run with periodic boundary conditions, and show
results only at $z=0$. In a departure from the original Santa Barbara
project, we enforce strict uniformity by using the same particle
initial condition for all the codes and using only one framework to
analyze the final outputs of the simulations, e.g., the strategy of
finding the center of the cluster is exactly the same for all
simulations. This procedure allows for a uniform comparison of the
different codes. In the following we briefly describe the simulations
performed and then analyze and compare the individual results.

\subsection{Simulations Performed}

The simulation of a single cluster in a small box is a time consuming
task for most cosmology codes.  The very high density region of the
cluster demands very small time steps to resolve short orbital
timescales. In order to avoid long run times, we therefore decided to
run the original Santa Barbara test with 256$^3$ particles on only a
subset of the tested codes: MC$^2$, HOT, FLASH (our three main codes),
and GADGET at a lower resolution than HOT. In addition we reduced the
initial conditions from 256$^3$ particles to 128$^3$ particles by
averaging over every eight particles and ran all six codes with these
initial conditions.  In this second set of simulations,
lower-resolution FLASH results are used to convey some measure of
sensitivity to resolution limits.  The cosmology employed in the Santa
Barbara test is SCDM, i.e., $\Omega_m=1.0$ and the Hubble constant
$H=50$~km/s/Mpc.  The box size is 64~Mpc.  The properties of the
cluster were all measured at $z=0$.  Table~\ref{tabsbres} summarizes
the resolutions at which the different simulations were performed.

\subsection{Results}

\begin{table*}
\begin{center}
\caption{\label{tab1} Cluster properties measured at $z=0$}
\begin{tabular}{lccccccc}
\hline\hline
\hfill& MC$^2$ & FLASH & HOT & GADGET & HYDRA & TPM &Average\\
\hline
\hfill&\hfill&\hfill&\hfill&\hfill&\hfill&\hfill&\hfill\\
\underline{$r_{200}$ [Mpc]}&\hfill&\hfill&\hfill&\hfill&\hfill&\hfill&\hfill\\
\hfill&\hfill&\hfill&\hfill&\hfill&\hfill&\hfill&\hfill\\
128$^3$& 2.753  & 2.753 & 2.753 & 2.753 & 2.753 & 2.753 &2.753\\
Residual& --& --& --& --& --& -- &\hfill\\
256$^3$&2.753&2.753&2.753&2.753&N/A&N/A&2.753\\
Rel. Residual& --& --& --& --& N/A &N/A&\hfill\\
\hfill&\hfill&\hfill&\hfill&\hfill&\hfill&\hfill&\hfill\\
\hline
\hfill&\hfill&\hfill&\hfill&\hfill&\hfill&\hfill&\hfill\\
\underline{$m_{\rm tot}$ [$10^{15}$ M$_\odot$]}&\hfill&\hfill&\hfill&\hfill&\hfill&\hfill
&\hfill\\
\hfill&\hfill&\hfill&\hfill&\hfill&\hfill&\hfill&\hfill\\
128$^3$& 1.215 & 1.196 &  1.220 &1.214 &1.218 & 1.212 & 1.213\\
Rel. Residual & 0.002  & -0.01 &  0.006 & 0.001 & 0.005 & 0.0004 &\hfill\\
256$^3$ & 1.208 & 1.208 & 1.215 & 1.246 & N/A & N/A & 1.220 \\
Rel. Residual &  -0.01 & -0.01 & -0.004  & 0.02 & N/A & N/A &\hfill\\
\hfill&\hfill&\hfill&\hfill&\hfill&\hfill&\hfill&\hfill\\
\hline
\hfill&\hfill&\hfill&\hfill&\hfill&\hfill&\hfill&\hfill\\
\underline{$\sigma_{\rm DM}$} [km/s]&\hfill&\hfill&\hfill&\hfill&\hfill&\hfill&\hfill\\
\hfill&\hfill&\hfill&\hfill&\hfill&\hfill&\hfill&\hfill\\
128$^3$& 989.72& 900.93 & 988.15 &1009.66&1011.77 & 1007.03 &984.54\\
Rel. Residual & 0.005 & -0.08 &  0.004 &0.026 & 0.028 & 0.023 &\hfill\\
256$^3$& 998.09 & 1016.79 &  1005.62 &1012.05 & N/A & N/A &1008.14 \\
Rel. Residual & -0.01  & 0.009 &  -0.003 &0.004 & N/A & N/A &\hfill \\ 
\hfill&\hfill&\hfill&\hfill&\hfill&\hfill&\hfill&\hfill\\
\hline\hline
\end{tabular}
\end{center}
\end{table*}

As in the original Santa Barbara test paper (``SB paper'' in the
following) we study some specific properties of the cluster at $z=0$,
the velocity and density profiles, and projected images of the cluster
itself.

\begin{figure*}
\begin{center}
\parbox{15.2cm}{

\parbox[t]{15.2cm}{
\parbox[t]{15.2cm}
{\begin{center}
\includegraphics[width=152mm]{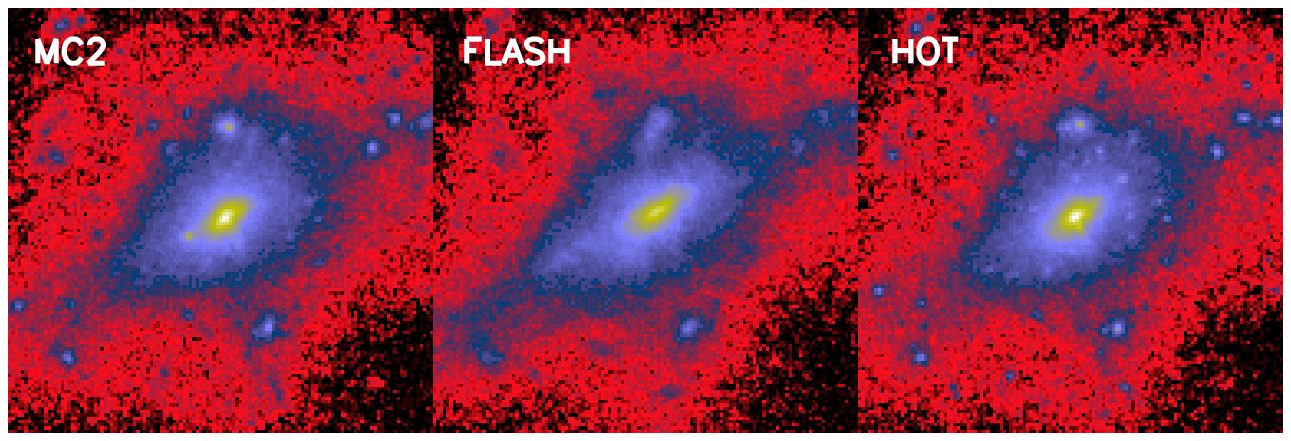}
\end{center}}}

\vspace{-0.8cm}

\parbox[t]{15.2cm}{
\parbox[t]{15.2cm}
{\begin{center}
\includegraphics[width=152mm]{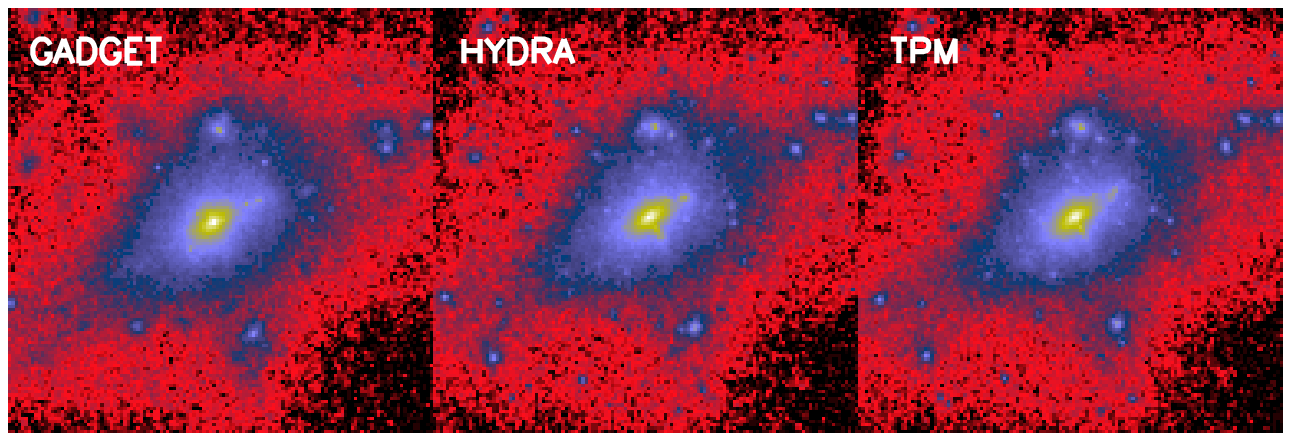}
\end{center}}}

\vspace{-0.7cm}

\parbox[t]{15.2cm}{
\parbox[t]{15.2cm}
{\begin{center}
\includegraphics[width=152mm]{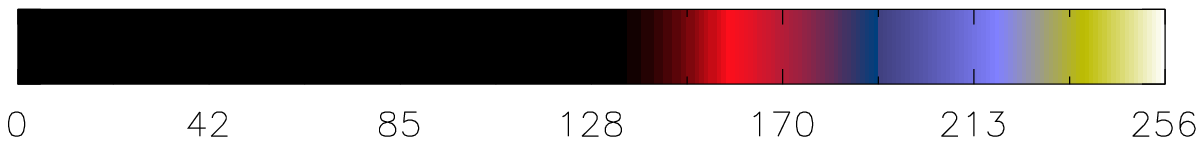}
\end{center}}}}
\caption{Santa Barbara cluster simulation: Projected dark matter
  density in logarithmic units at $z=0$, 128$^3$ particles projected
  on a 1024$^2$ mesh, no smoothing. The dynamic range of the density
  variation in this figure is roughly 5 orders of magnitude. The lower
  force resolution used here for the FLASH simulation is apparent in
  the figure.}
\label{plotfour}
\end{center}
\end{figure*}

\begin{figure*}
\begin{center}
\parbox{20cm}{
\parbox[t]{20cm}{
\parbox[t]{18.2cm}
{\begin{center}
\includegraphics[width=182mm]{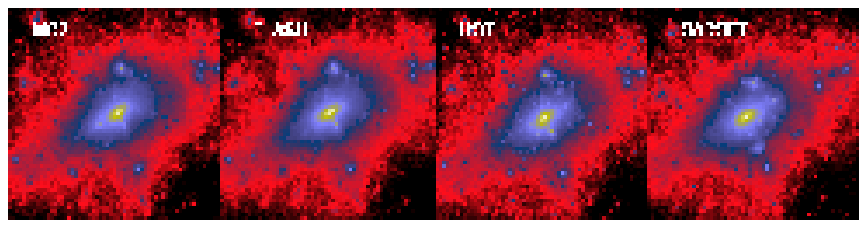}
\end{center}}}

\vspace{-0.7cm}

\parbox[t]{20cm}{
\parbox[t]{18.2cm}
{\begin{center}
\includegraphics[width=182mm]{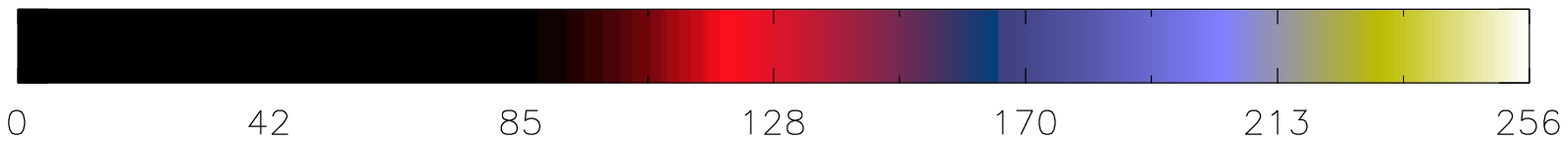}
\end{center}}}}
\caption{Projected dark matter density following
Fig.~\ref{plotfour}, but for $256^3$ particles. The FLASH force
resolution is better by a factor of 4 than for the $128^3$ particle
run while the GADGET force resolution is 5 times worse. The force
resolution for MC$^2$ and HOT are the same as in the $128^3$ runs 
(see also Table \ref{tabsbres}). Note the increased definition of
substructure independent of the force resolution.}
\label{plotfive}
\end{center}
\end{figure*}

We start our discussion with a useful but qualitative comparison,
using images of the two-dimensional projected density of the
cluster. All results are shown at $z=0$.  Figs.~\ref{plotfour} and
\ref{plotfive} display the cluster in an 8~Mpc box from the 128$^3$
and the 256$^3$ simulations, respectively. In order to provide the
best direct comparison we have not filtered the images. (A Gaussian
smoothing of 250 kpc as used in the original paper washes out far too
many details of the cluster which are real.) The density is projected
onto a 1024$^2$ $xy$-plane grid and displayed in logarithmic
units. The pixel size is 62.5 kpc which is substantially larger than
the formal spatial resolution in the higher-resolution runs.

The overall appearance of the cluster -- orientation, shape, size,
highest density region in the center -- agrees very well for all
simulations.  Comparing the 128$^3$ simulation from MC$^2$, HOT,
HYDRA, and TPM shows that many small scale features are reproduced
with high fidelity.  All four simulations show three small high
density regions in the NE as well as three high density regions in the
SW, and a high density region in the upper left corner of the
plot. The FLASH image shows most of these features as well, but due to
the lower resolution employed some of the very small features are
missing -- especially on the edge of the cluster.  The GADGET results
show some small deviations compared to the other simulations: The
three high density regions in the NE visible in the other results
reduce to two in the GADGET image, and the lower left corner doesn't
show as much structure.  Overall one gets the impression that the
GADGET image is not at $z=0$, but slightly shifted. This may be due to
a non-optimal setting of the maximum timestep in the GADGET
configuration file ($0.01$). While this value is appropriate at low
redshifts, in the linear regime GADGET will take timesteps equal in
size to the MaxSizeTimestep parameter.  We have not investigated
whether making this parameter smaller would resolve this discrepancy.
Sub-structure is clearly present in the HYDRA, HOT, and TPM
images. However, there is a lack of one-to-one agreement at the level
of these individual features.

The situation for the images from the 256$^3$ simulations shown in
Fig.~\ref{plotfive} is very similar.  Due to better mass resolution
the cluster appears to be smoother than in the 128$^3$ simulation.
The agreement of the MC$^2$, FLASH, and HOT results is excellent, as
almost all obvious features (except the sub-structure) are identical.
The HOT image shows more substructure than the MC$^2$ and FLASH images
which is to be expected due to the higher force resolution of the
tree-code compared to the mesh-codes.  As in the 128$^3$-particle
simulation, the GADGET image seems to be taken at a slightly different
redshift than the other ones.  In summary the result of this
qualitative comparison is very good -- much better than in the
original paper.

We continue with a more quantitative comparison of cluster properties.
In Table~\ref{tab1} results for the radius $r_{200}$ of the cluster
measured in Mpc, the total mass measured in $10^{15}$ M$_\odot$, and
the rms velocity dispersion of the cluster in km/s are listed. We also
give the average measurement for each quantity and the deviation from
the average for the simulations.

Independent of the number of particles (128$^3$ or 256$^3$) all codes
lead to exactly the same result for the radius of the cluster
$r_{200}=2.753$ Mpc, which is slightly larger than the result from the
original test, $r_{200}=2.70$ Mpc with an rms scatter of
0.04~Mpc.

The average mass for the cluster is $1.21\cdot 10^{15}$ M$_\odot$ for
the 128$^3$ particle simulation and $1.22\cdot 10^{15}$ M$_\odot$ for
the 256$^3$ simulation. This is a little higher than the average mass
found in the original test; the highest mass reported from two runs in
the SB paper was $1.21\cdot 10^{15}$ M$_\odot$.  The deviation from
the average for all runs is of the order of 1\%, for some codes even
sub-1\%. Here we quote the relative residual calculated as $(\langle
x\rangle-x)/\langle x\rangle$, $\langle x\rangle$ being the average of
all results. In the original paper the agreement was reported to be
sub-10\%, an order of magnitude worse than our result. For the radius,
the results from all the runs agree to three significant figures so no
residuals are shown.

The average of the velocity dispersion $\sigma_{\rm DM}$ (following
the convention of the original paper, the one-dimensional velocity
dispersion $\sigma_{\rm DM}=\sigma/\sqrt{3}$ with $\sigma$ being the
three-dimensional velocity dispersion) is around 1000~km/s, slightly
higher than in the SB paper. The agreement is remarkably good for both
128$^3$ and 256$^3$ particle simulations with a deviation of only
around 1\%. Due to the resolution limits, the 128$^3$-FLASH simulation
produces a value 8\% lower than the average.  In the original paper
the differences in the individual code results were again much less
satisfactory than in our comparison, with deviations of up to 10\%. We
discuss the reasons for the improved results later below.

Finally, we compare three different cluster profiles: the dark matter
density, velocity dispersion, and the radial velocity as functions of
radius. These profiles were obtained by first determining the center
of the cluster, and then averaging the particle information in 15
spherical shells evenly spaced in log radius. While we have shown all
results for all codes down to the smallest radius, the results from
MC$^2$, FLASH, and GADGET for the 256$^3$ particle runs can be only
trusted down to $r=0.06$~Mpc because of the resolution employed. For
the 128$^3$ particle runs the force resolution of HOT, GADGET, HYDRA,
and TPM is sufficient to generate correct results over the entire
measured $r$-range while for MC$^2$ resolution artifacts are expected
to be seen at $r<0.06$~Mpc and for FLASH at $r<0.2$~Mpc. Following the
original SB paper, the main plots are in the lower panel with the
average over the code results shown with a solid curve (we have
removed the FLASH results from the average in the $128^3$ particle
simulations). In the upper panel the residuals are displayed,
calculated as $\ln x - \ln\langle x\rangle $ for the dark matter
density and velocity dispersion profiles and $x -\langle x\rangle $
for radial velocity profiles.

\begin{figure}[t]
\includegraphics[width=80mm]{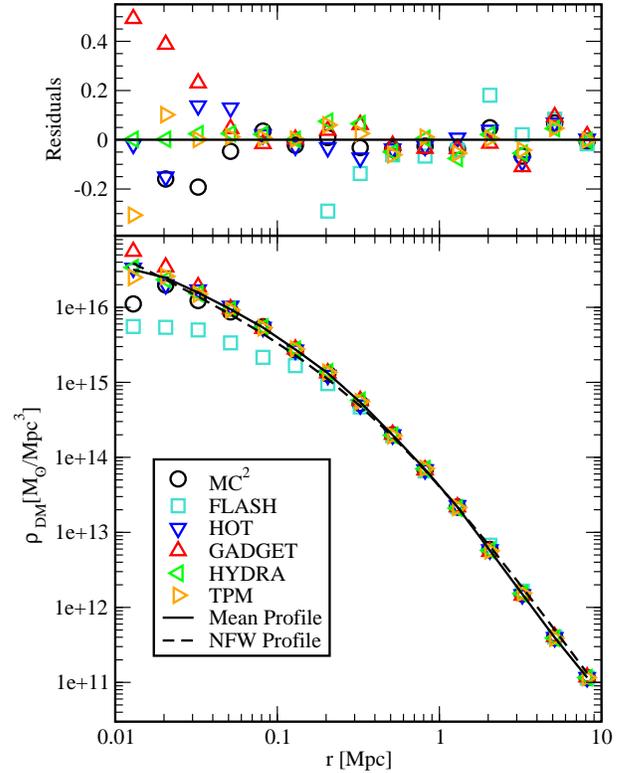}
\caption{Dark matter density profile for the Santa Barbara cluster, 
$128^3$ particles. The mean profile (solid curve) is a simple average 
over all the simulations, with the exception of the FLASH results
which were run at low resolution.}   
\label{plotsix}
\end{figure}

The dark matter density profile for the 128$^3$ particle runs is shown
in Fig.~\ref{plotsix}. Down to a radius of $r=0.25$~Mpc the
agreement of all codes is very good, at the sub-10\% level. For
smaller radii the FLASH profile falls off faster due to resolution
limitations.  The residuals for FLASH in the region $r<0.1$~Mpc are
not shown.  The agreement of the remaining five codes is still very
good (better than 10\%) until $r=0.06$~Mpc is reached, the resolution
limit of the MC$^2$ simulation. At the smallest radii $r<0.05$~Mpc
HOT, HYDRA, and TPM are still in very good agreement while the GADGET
profile is somewhat steeper. For comparison we plot a profile
suggested by Navarro, Frenk, \& White (1995) (``NFW profile'' in the
following) with the concentration parameter $c=7.12$.  We calculated
the optimal concentration parameter based on the average mass and
$r_{200}$ obtained from our simulations using a publicly available
code from Navarro
\footnote{http://pinot.phys.uvic.ca/~jfn/mywebpage/home.html}. The
value we found is smaller than in the original SB paper where $c=7.5$
was chosen. The lower value here results from the higher cluster mass
as compared to the original SB paper. The profile is an excellent
match to the simulation results.

\begin{figure}[hb]
\includegraphics[width=80mm]{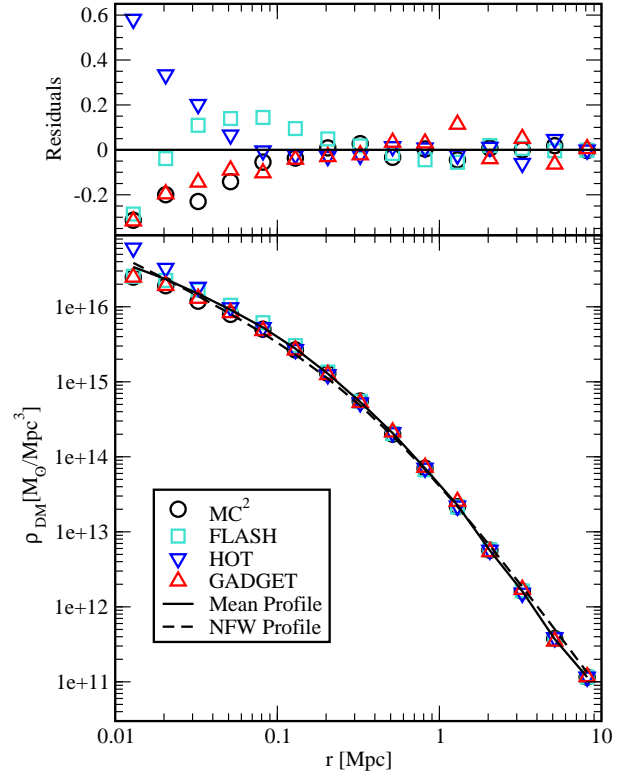}
\caption{Dark matter density profile for the Santa Barbara cluster,
256$^3$ particle simulations. Unlike Fig.~\ref{plotsix}, here MC$^2$
and FLASH are run with the same effective force resolution.}
\label{plotseven}
\end{figure}

In Fig.~\ref{plotseven} we show the result for the four 256$^3$
particle runs. Down to $r=0.06$~Mpc, the resolution limit for MC$^2$,
FLASH, and GADGET in this simulation, the agreement of all runs is
sub-8\%, again excellent. (The agreement in the original paper was
20\% or better.) In this simulation, only HOT was run with a very
small smoothing length, thus at small radii it is the only code with
enough resolution to provide converged results. The other three codes
were run with roughly equivalent force resolution and they do provide
consistent results at small radii, the fall-off of the halo density
for $r<0.1$ Mpc resulting from the force smoothing. The corresponding
NFW profile is shown for comparison.

\begin{figure}
\includegraphics[width=80mm]{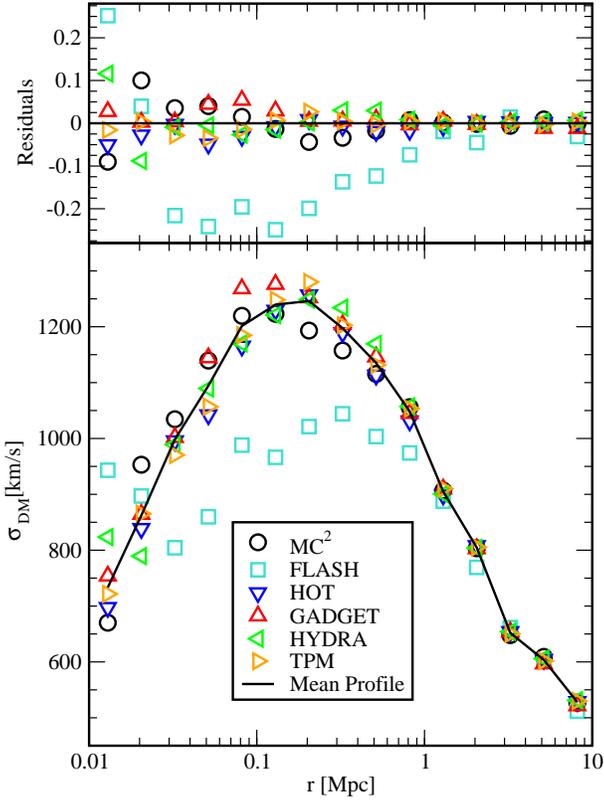}
\caption{Dark matter velocity dispersion profile for the Santa Barbara
  cluster, $128^3$ particles. The low-resolution FLASH results are not
  included in the mean profile.}
\label{ploteight}
\end{figure}

\begin{figure}
\includegraphics[width=80mm]{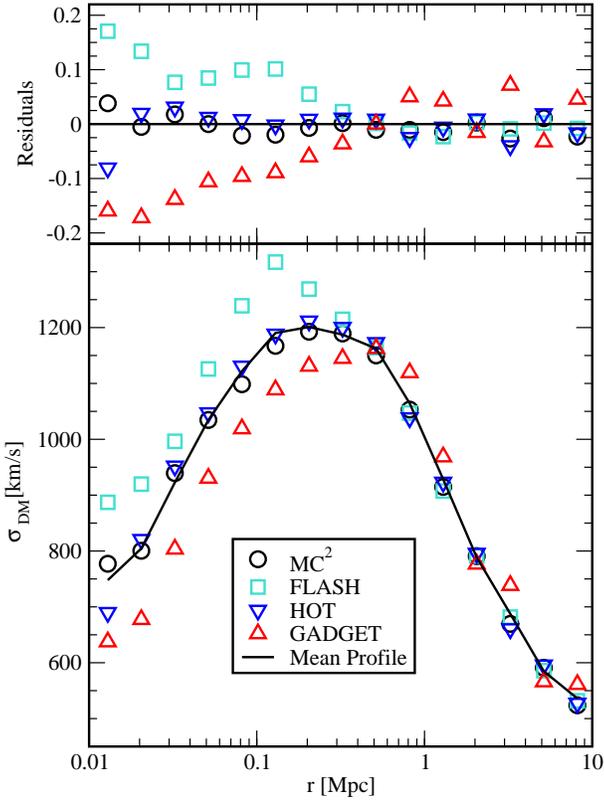}
\caption{Dark matter velocity dispersion profile for the Santa Barbara
cluster, 256$^3$ particle simulations.} 
\label{plotnine}
\end{figure}

\begin{figure}
\includegraphics[width=80mm]{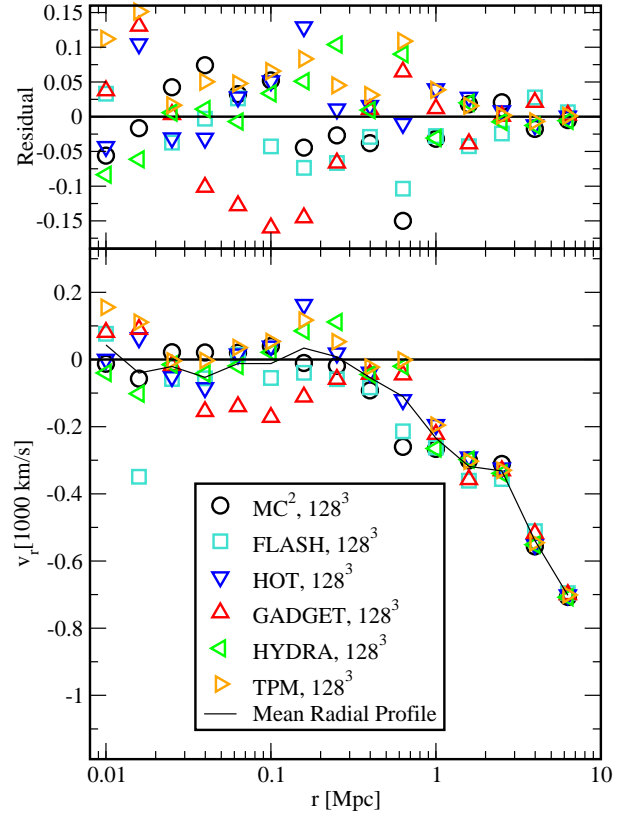}
\caption{Dark matter radial velocity profile for the Santa Barbara 
  cluster, $128^3$ particles. The low-resolution FLASH results are not
  included in the mean profile.}
\label{plotten}
\end{figure}

\begin{figure}
\includegraphics[width=80mm]{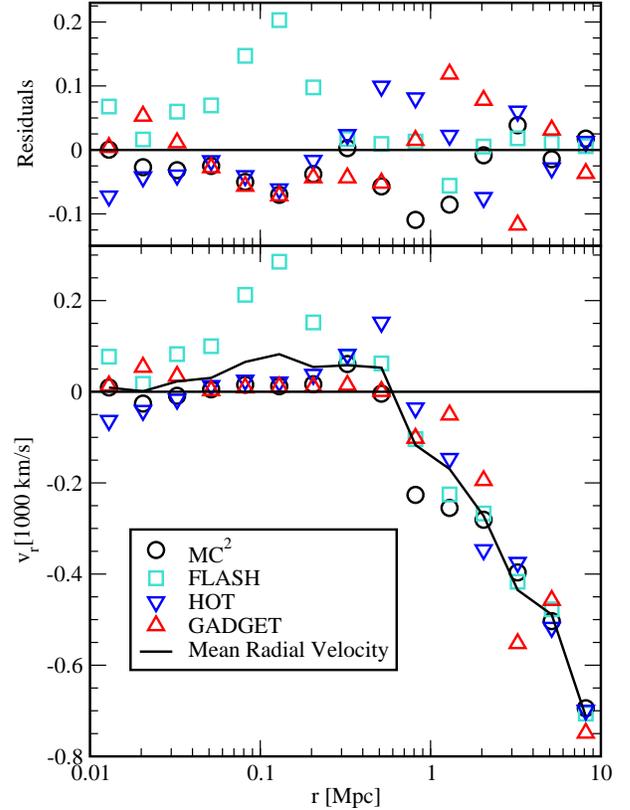}
\caption{Dark matter radial velocity profile for the Santa Barbara 
cluster, $256^3$ particle simulations.} 
\label{ploteleven}
\end{figure}

In Figs.~\ref{ploteight} and \ref{plotnine} the dark matter
velocity dispersion profiles are shown for the 128$^3$ particle and
256$^3$ particle runs.  In both cases the velocity dispersion peak is
higher than in the original SB paper, 1250~km/s for the 128$^3$ runs
and 1200~km/s for the 256$^3$ runs, while in the SB comparison the
average maximum was around 1100 km/s. This could be due to the higher
force resolutions in the present runs, a conclusion consistent with
the higher $\sigma_{DM}$ peak values obtained in the original paper by
the higher-resolution codes. This argument can also be checked by the
low-resolution FLASH-128$^3$ run: The maximum here is around
1000~km/s. The general agreement in both the 128$^3$ and 256$^3$
particle cases is very good for the outer shells, better than 5\%.  In
the inner shells the agreement is better than 10\%, with only the
low-resolution FLASH-128$^3$ run disagreeing up to 20\%. Following the
results for the density profile, the overall agreement between the
different runs is better by a factor of two than in the original SB
paper.  We do not observe a large scatter of the results in the last
bin as reported in the SB paper. There it was argued that the scatter
is due to noise arising from subclustering but we see no evidence to
support this.

Finally, we compare the radial velocity profiles. The results are
shown in Figs.~\ref{plotten} and \ref{ploteleven}. Net infall is seen
at roughly the same radius as in the original paper, between 600~kpc
and 700~kpc for both cases. The scatter in the 128$^3$ particle
simulations is around 150~km/s (note that the residual here is quoted
without the logarithm), while the scatter in the 256$^3$ simulations
is slightly smaller with 100~km/s. The original SB result has a
scatter of 200~km/s. Overall the agreement of the different codes is
very satisfactory, however the 256$^3$ particle FLASH simulation shows
a relatively high radial velocity in the inner region.

Our results for the Santa Barbara cluster comparison for the six
tested codes are very positive. The agreement of all measured
quantities across the codes is much better than in the original
test. This could be due to several reasons: (i) Every code was run
with exactly the same initial particle positions and velocities,
resulting in a more uniform comparison (using only the initial density
field can lead to slightly different initial conditions since every
simulator would have to implement the Zel'dovich step to obtain
positions and velocities). (ii) We used the same analysis code for all
simulations, thereby eliminating any uncertainties introduced by
variations in diagnostics routines, e.g., estimating the center of the
cluster, etc.  (iii) Most of the simulations were run with very
similar resolution, varying from 5~kpc to 64~kpc for the high
resolution runs, while the resolutions in the original test varied
between 5~kpc and 960~kpc. (iv) The present test used dark-matter only
simulations.


\section{Tests with the Cosmological Concordance Model}
\label{sec:SM}

\begin{table*}
\begin{center}
\caption{\label{tab2}Softening lengths/mesh spacings for the 
$\Lambda$CDM runs}  
\begin{tabular}{lcccccc}
\hline\hline
\hfill& MC$^2$ & FLASH & HOT &GADGET & HYDRA & TPM \\
\hline
LCDMs, 256$^3$ particles & 90 kpc  & 90 kpc &  10 kpc (physical after z=9)&
20 kpc & 40 kpc & 10 kpc \\ 
LCDMb, 256$^3$ particles & 350 kpc & 350kpc & 20 kpc (physical after z=9)&
20 kpc & 40 kpc & 20 kpc \\ 
\hline\hline
\end{tabular}
\end{center}
\end{table*}

\subsection{Test Description}

In this Section, we study results from two realistic N-body
cosmological simulations of a $\Lambda$CDM universe, a ``small'' 90
Mpc box and a ``large'' 360 Mpc box using $256^3$ particles. These
boxes straddle a representative range of force and mass resolutions
for state-of-the-art large scale structure simulations designed to
study power spectra, halo mass functions, weak lensing, etc. The 90
Mpc box is somewhat small as a representative simulation due to lack
of large-scale power. Nevertheless, we chose this volume for code
comparison because it allows a relatively high resolution for MC$^2$,
which we are using as a collisionless standard.

The initial linear power spectrum was generated using the fitting
formula provided by Klypin \& Holtzman (1997) for the transfer
function. This formula is a slight variation of the common BBKS fit
(Bardeen et al. 1986). It includes effects from baryon suppression but
no baryonic oscillations. We use the standard Zel'dovich approximation
(Zel'dovich 1970) to provide the initial particle displacement off a
uniform grid and to assign initial particle velocities. The starting
redshift for both box-sizes is $z_{in}=50$. In each case, we choose
the following cosmology: $\Omega_m=0.314$ (where $\Omega_m$ includes
cold dark matter and baryons), $\Omega_b=0.044$,
$\Omega_\Lambda=0.686$, $H_0=71$~km/s/Mpc, $\sigma_8=0.84$, and
$n=0.99$. These values are in concordance with the measurements of
cosmological parameters by WMAP (Spergel et al. 2003). Both
simulations used 256$^3$ simulation particles, implying particle
masses of 1.918$\cdot 10^9$~M$_\odot$ for the small box and
1.227$\cdot 10^{11}~$M$_\odot$ for the large box.

The analysis of the runs is split into two parts. First, the particle
positions and velocities and their two-point functions are studied
directly. We measure velocity distributions, the mass power spectrum,
and correlation functions, and compare position slices from the
simulation box directly. Second, halo catalogs are generated and
investigated. Here we compare velocity statistics, correlation
functions, mass functions, and for some specific mass bins, the
positions, masses, and velocities of selected individual halos.

The MC$^2$-simulations are carried out on a 1024$^3$ mesh for both
physical box sizes.  A detailed convergence study with MC$^2$ will be
presented elsewhere~(Habib et al. 2004). For the test simulations,
FLASH's base grid is chosen to be a 256$^3$~mesh, with two levels of
mesh refinement leading to an effective resolution equivalent to a
1024$^3$ mesh in high density regions. The softening length for the
other four codes are chosen in such a way that sufficient resolution
is guaranteed and reasonable performance on a Beowulf cluster is
achieved. The individual softening lengths used ranged between 10 to
40~kpc. These choices of softening length are consistent with the mass
resolution set by the number of particles and this is borne out by the
results of Sec. \ref{hmf}. In Table~\ref{tab2} the force resolutions
from the different codes and runs are summarized.  LCDMs refers to the
90~Mpc~box simulation, LCDMb to the 360~Mpc~box simulation.

\subsection{Results}

We use the following color coding for the different codes in all
comparison plots: MC$^2$ results are shown with black lines and
circles, FLASH with turquoise lines and squares, HOT with blue lines
and down-triangles, GADGET with red lines and up-triangles, HYDRA with
green lines and left-triangles, and TPM with orange lines and
right-triangles. As an arbitrary convention, all residuals shown are
calculated with respect to the results from GADGET, rather than from
an average over the results from all the codes. This is done mainly in
order to avoid contaminations of averages from lower resolution
results or from individual simulations with some other systematic
source of error.

\subsubsection{Positions and Velocities of the particles}
\begin{figure*}
\begin{center}
\parbox{30cm}{
\parbox[t]{30cm}{
\parbox[t]{9.0cm}
{\begin{center}
\hspace{0cm}\includegraphics[width=73mm]{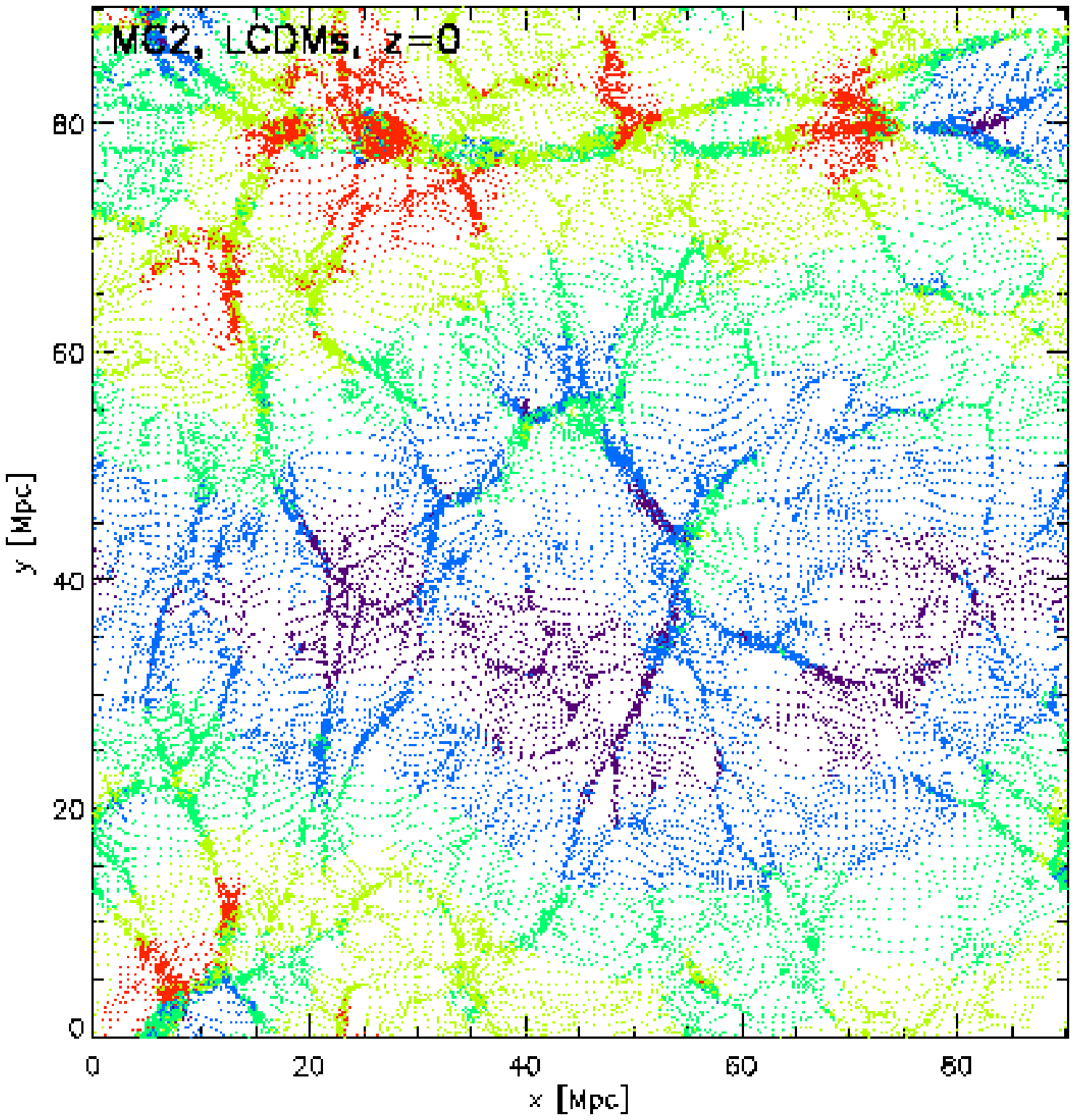}
\end{center}}
\parbox[t]{9.0cm}
{\begin{center}
\hspace{0cm}\includegraphics[width=73mm]{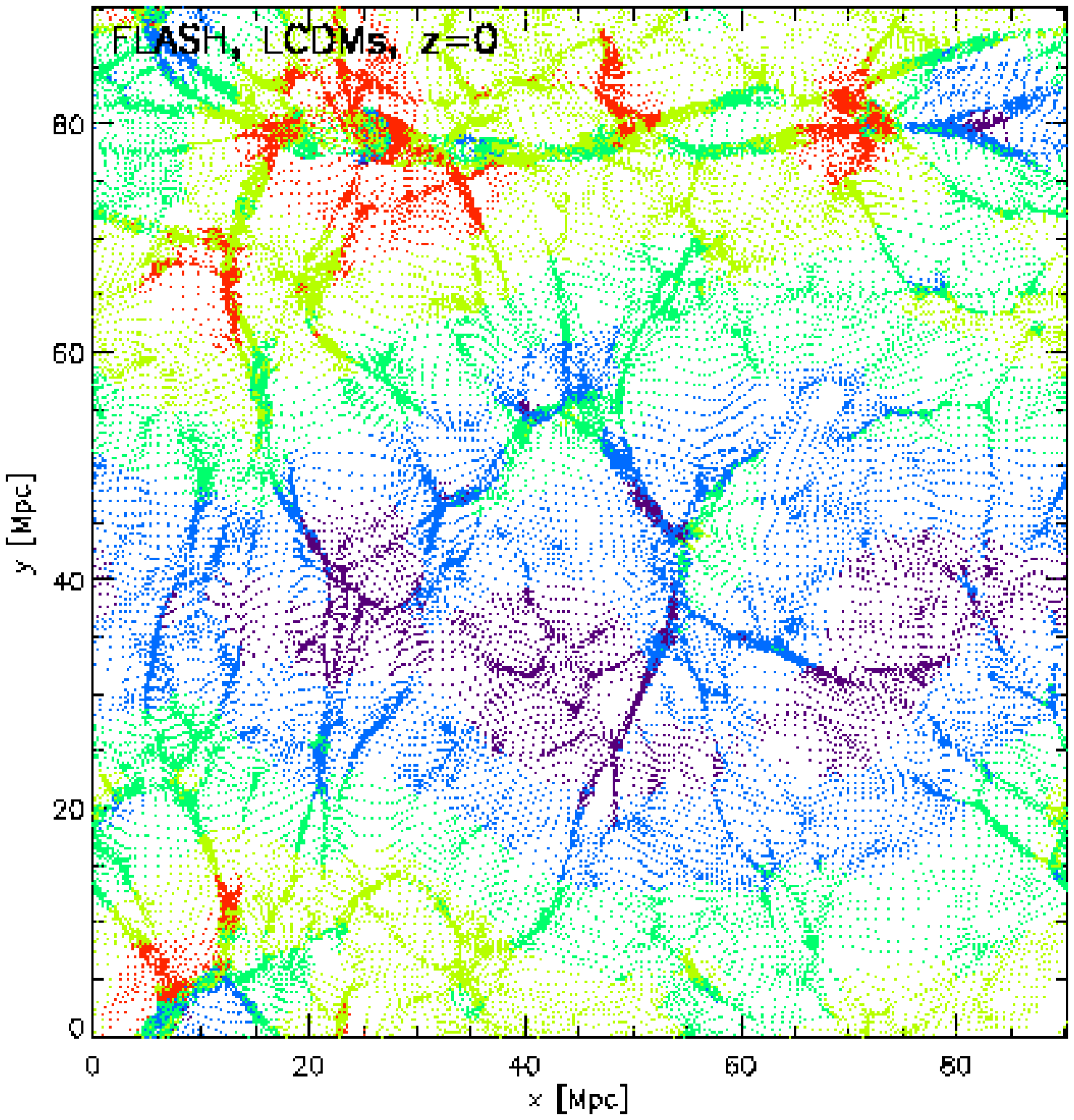}
\end{center}}}

\vspace{-0.7cm}

\parbox[t]{30cm}{
\parbox[t]{9.0cm}
{\begin{center}
\hspace{0cm}\includegraphics[width=73mm]{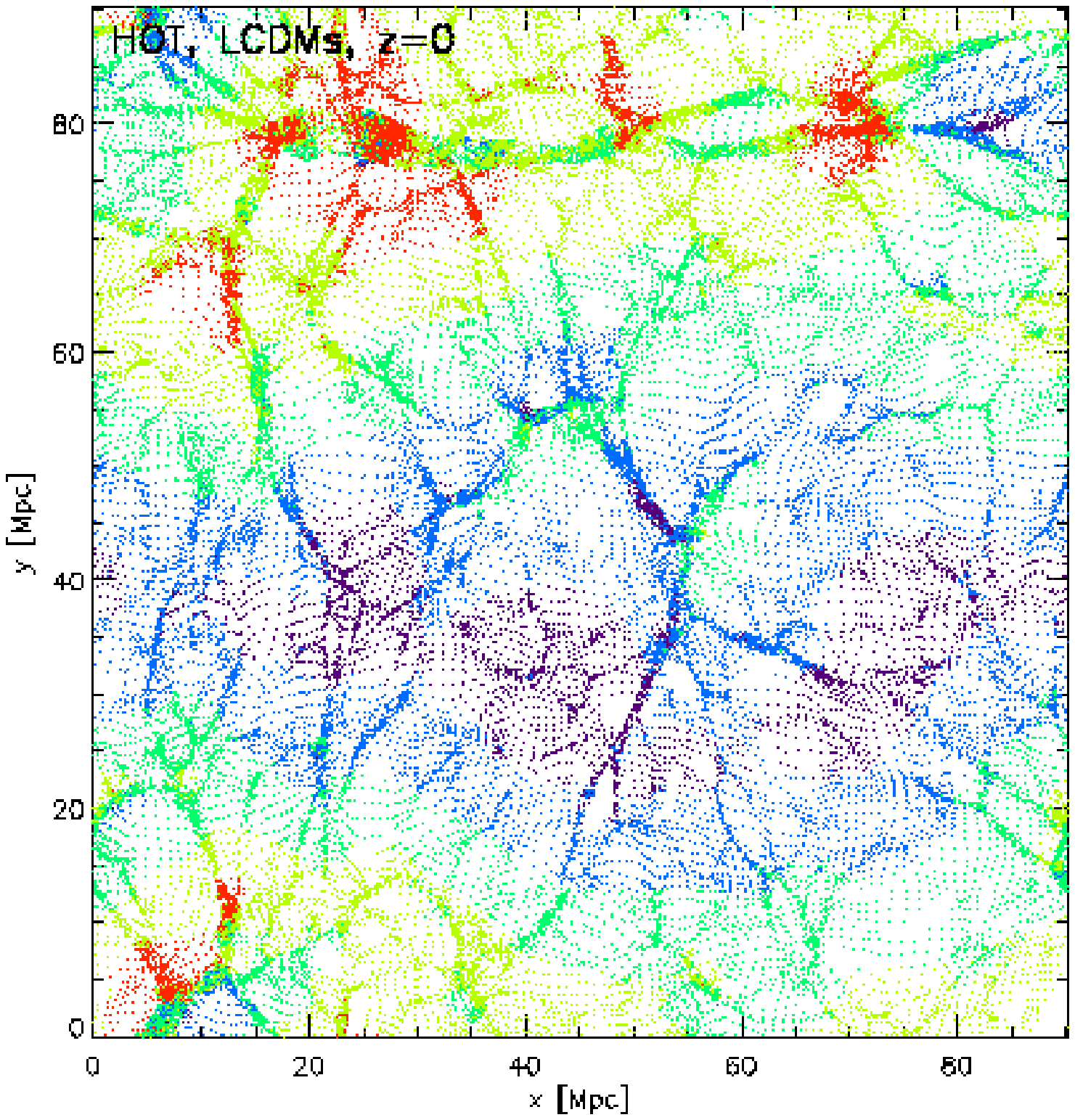}
\end{center}}
\parbox[t]{9.0cm}
{\begin{center}
\hspace{0cm}\includegraphics[width=73mm]{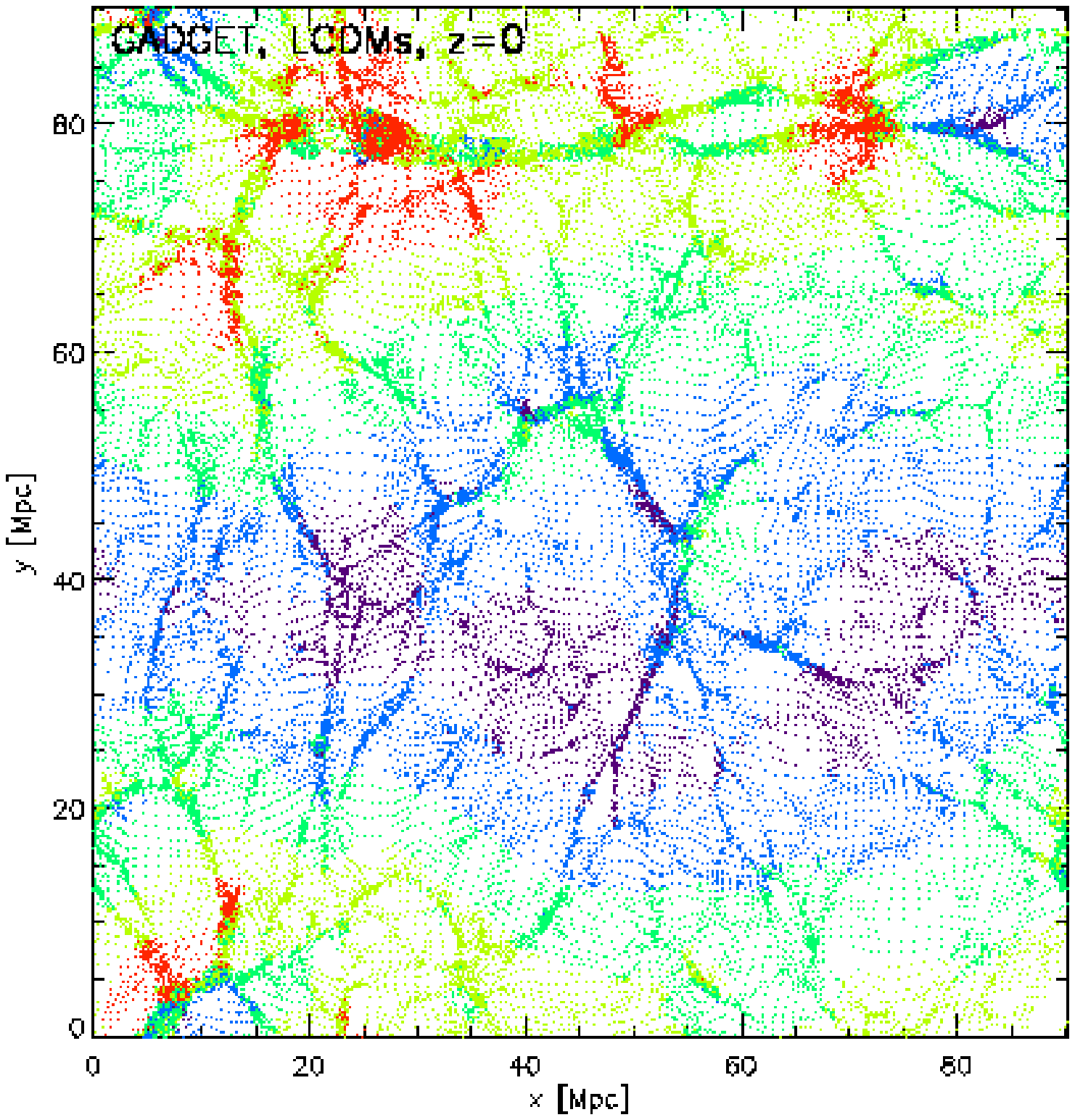}
\end{center}}}

\vspace{-0.7cm}

\parbox[t]{30cm}{
\parbox[t]{9.0cm}
{\begin{center}
\hspace{0cm}\includegraphics[width=73mm]{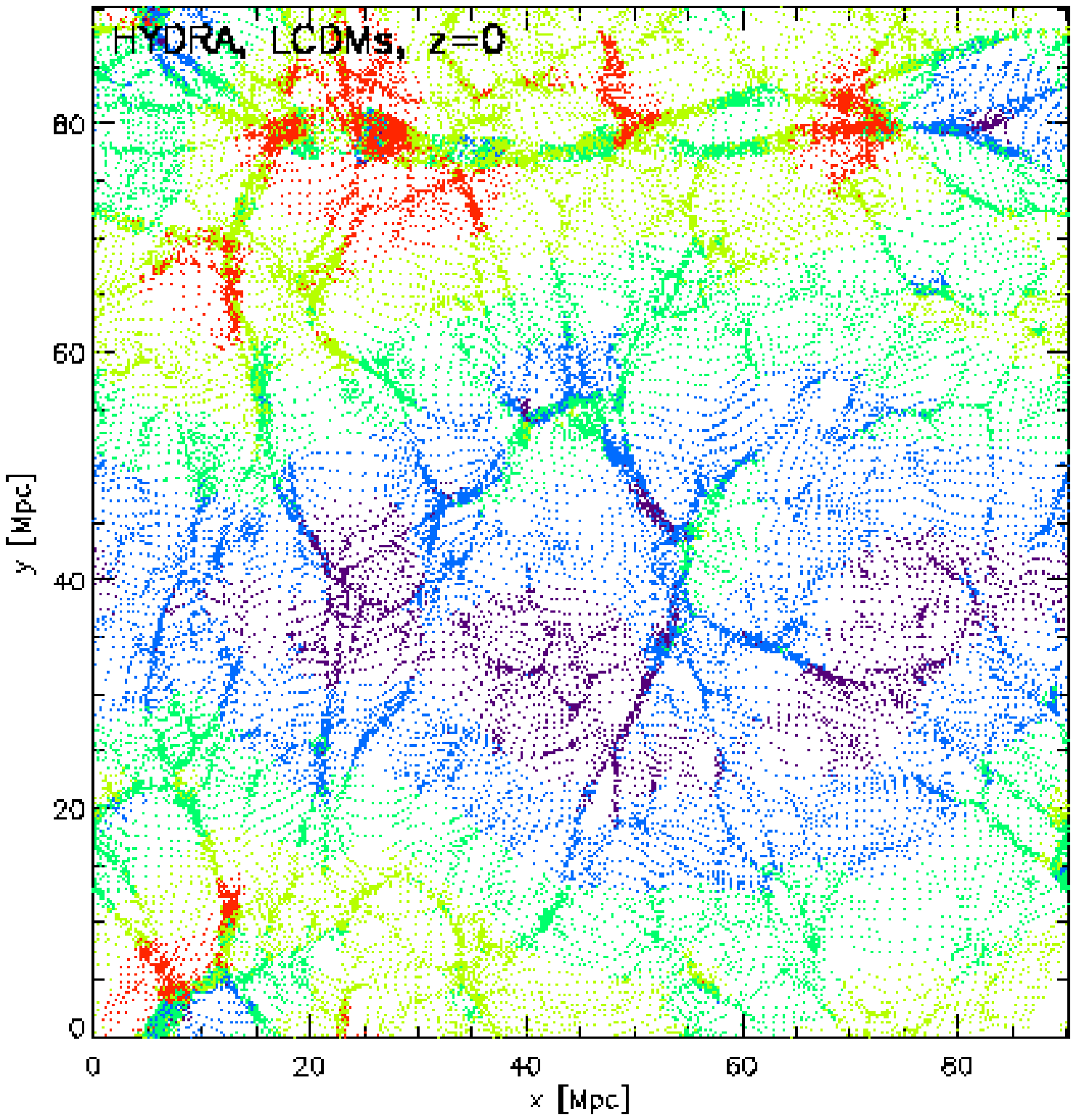}
\end{center}}
\parbox[t]{9.0cm}
{\begin{center}
\hspace{0cm}\includegraphics[width=73mm]{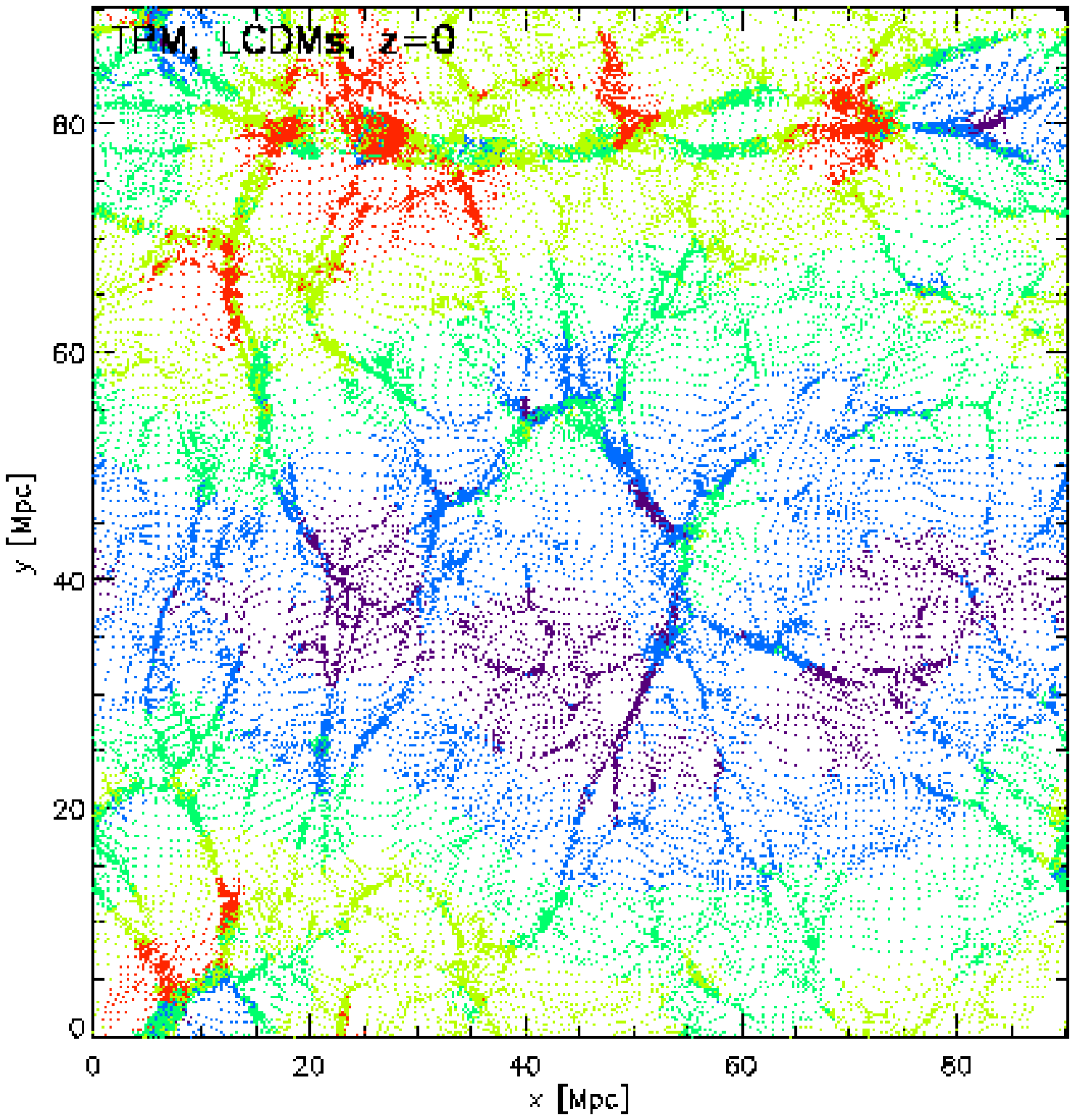}
\end{center}}}}
\caption{Comparison of the particle positions and velocities 
  in a 1~Mpc thick slice of the 90~Mpc box. The colors show four
  (uniform) velocity bins between 0 and 440 km/s. Dark blue
  corresponds to a velocity between 0 and 110 km/s, light blue to a
  velocity between 110 km/s and 220 km/s, dark green to a velocity
  between 220 km/s and 330 km/s, and red corresponds to the
  highest velocities between 330 km/s and 440 km/s.}  
\label{plottwelve}
\end{center}
\end{figure*}

\begin{figure*}
\begin{center}
\parbox{30cm}{
\parbox[t]{30cm}{
\parbox[t]{9cm}
{\begin{center}
\includegraphics[width=73mm]{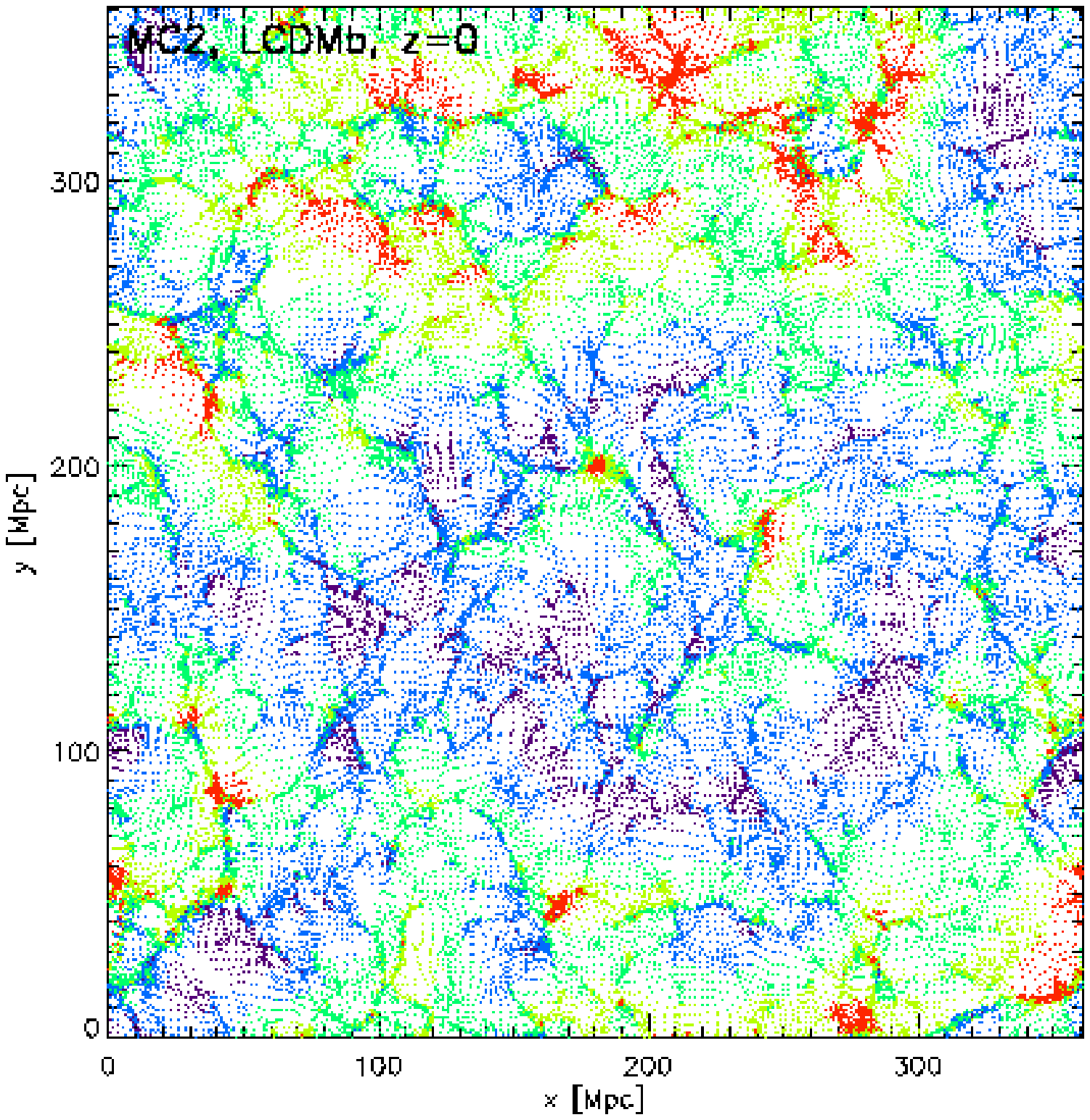}
\end{center}}
\parbox[t]{9cm}
{\begin{center}
\includegraphics[width=73mm]{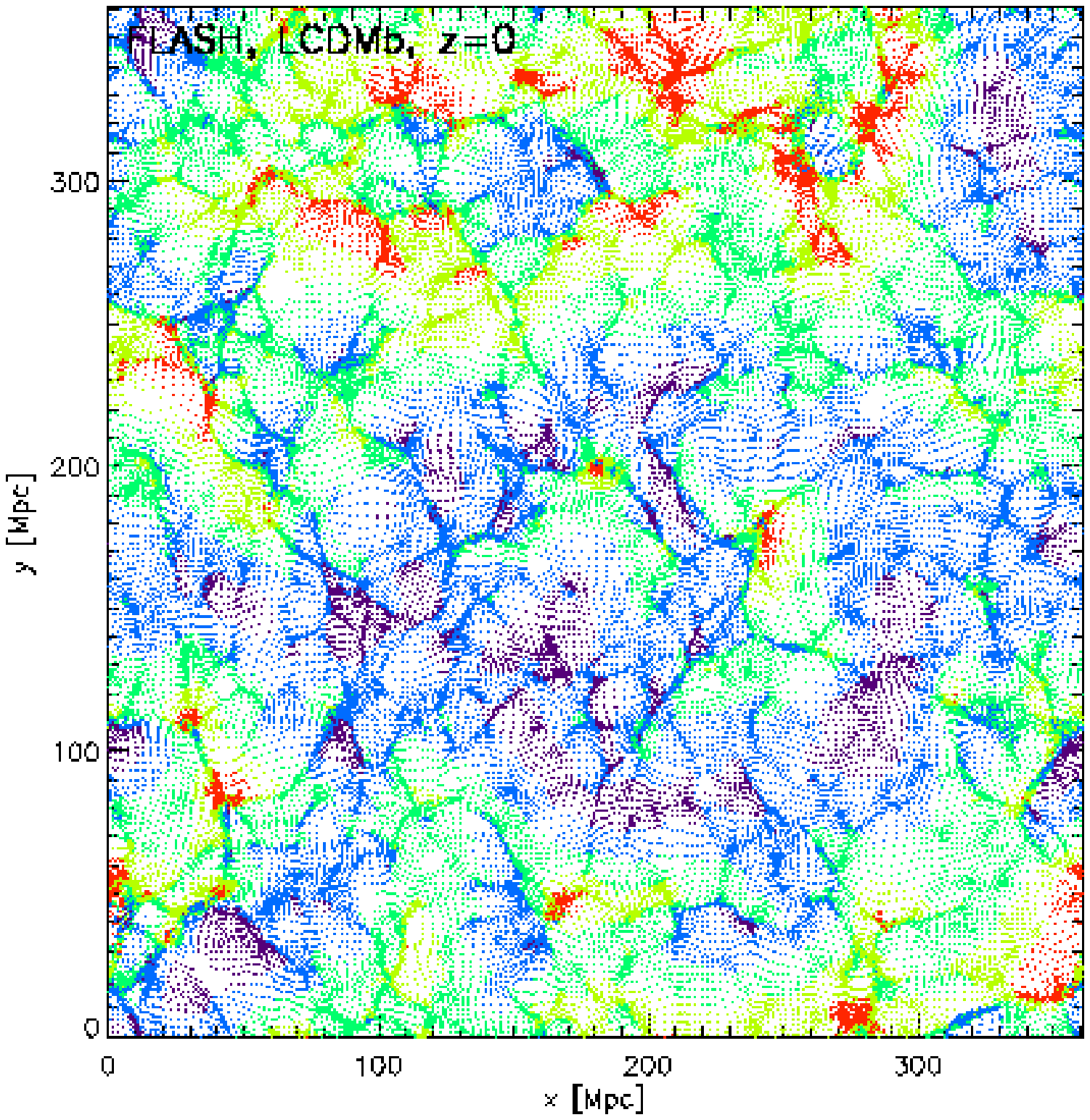}
\end{center}}}

\vspace{-0.7cm}

\parbox[t]{30cm}{
\parbox[t]{9cm}
{\begin{center}
\includegraphics[width=73mm]{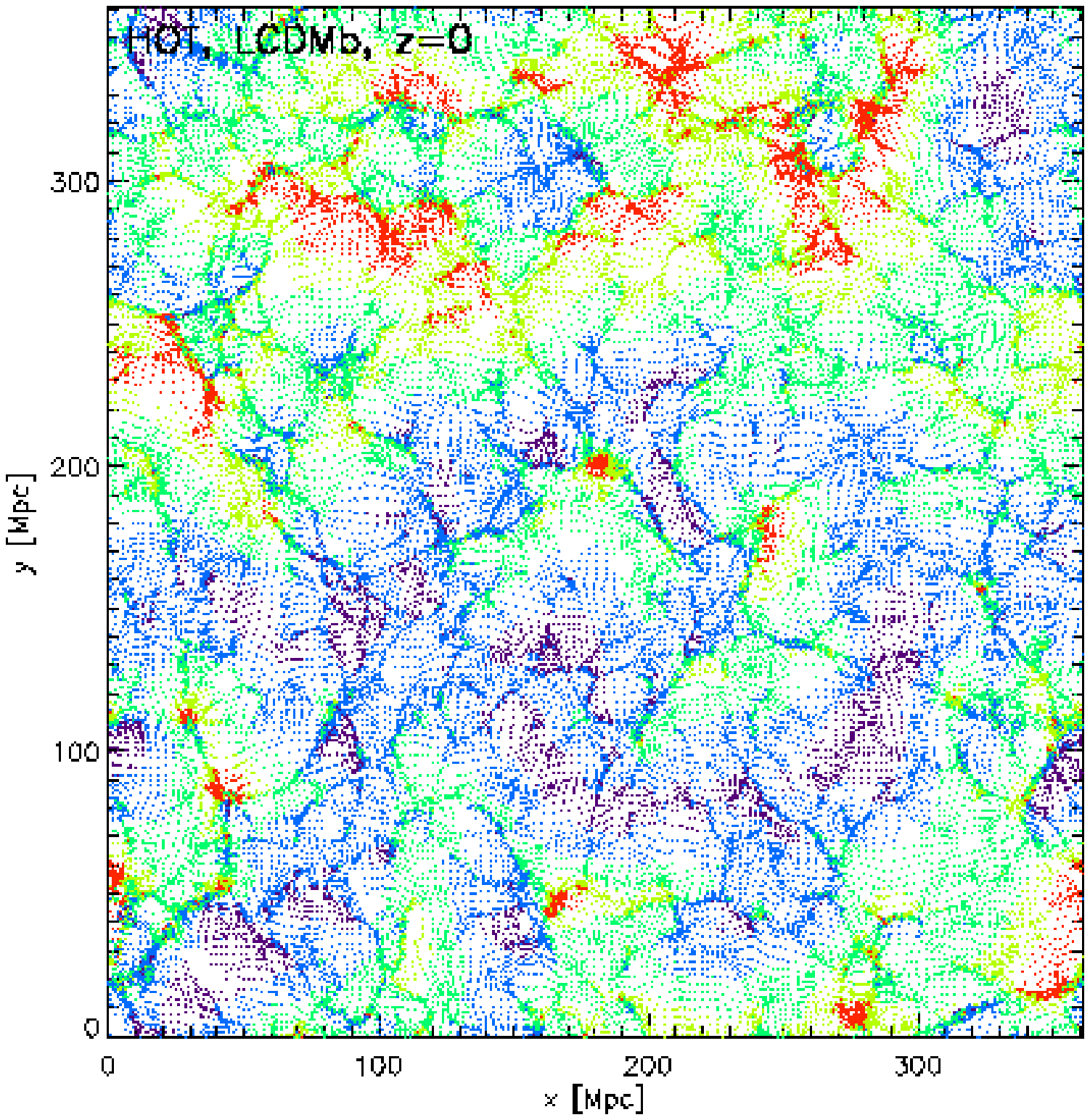}
\end{center}}
\parbox[t]{9cm}
{\begin{center}
\includegraphics[width=73mm]{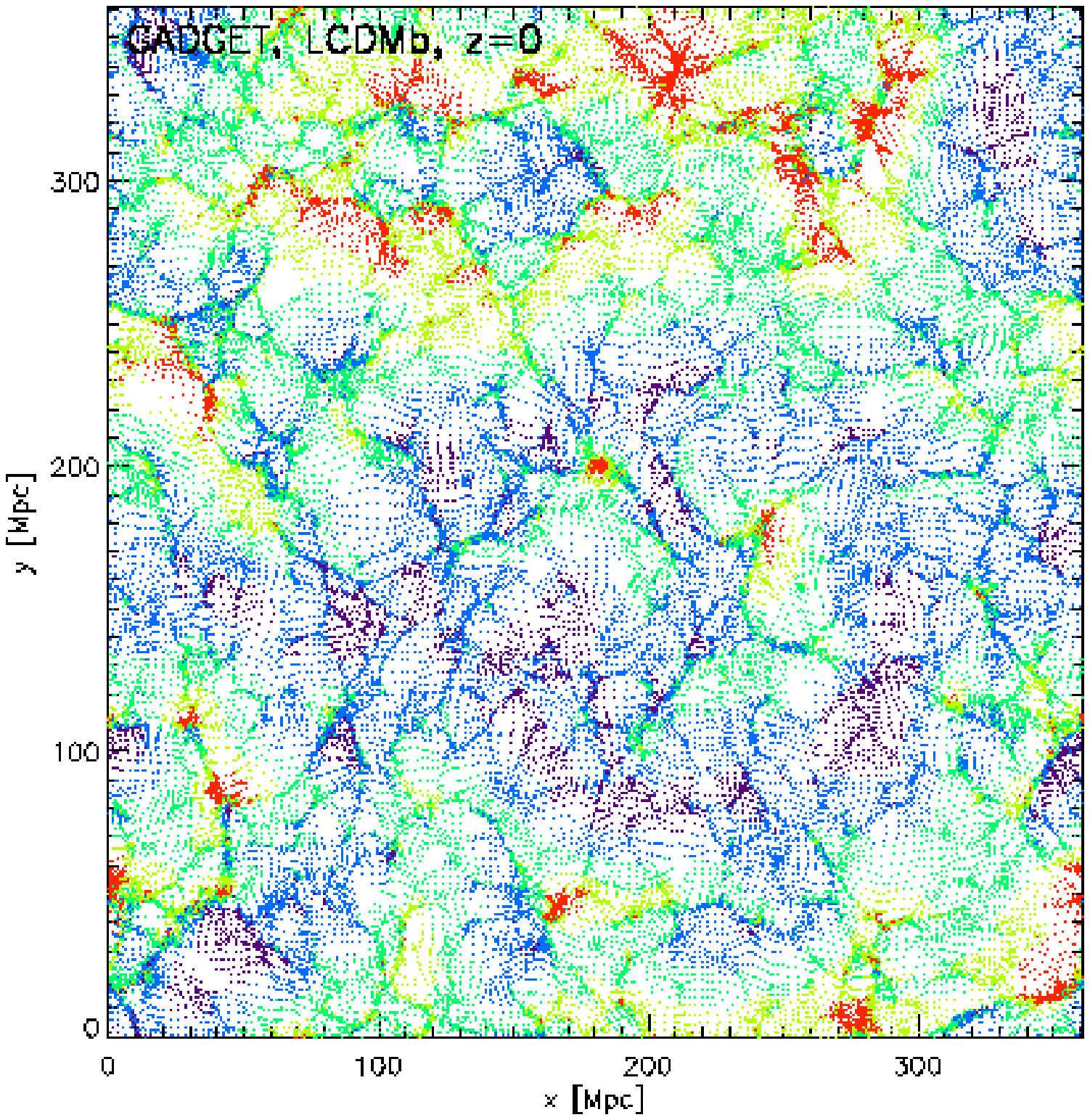}
\end{center}}}

\vspace{-0.7cm}

\parbox[t]{30cm}{
\parbox[t]{9cm}
{\begin{center}
\includegraphics[width=73mm]{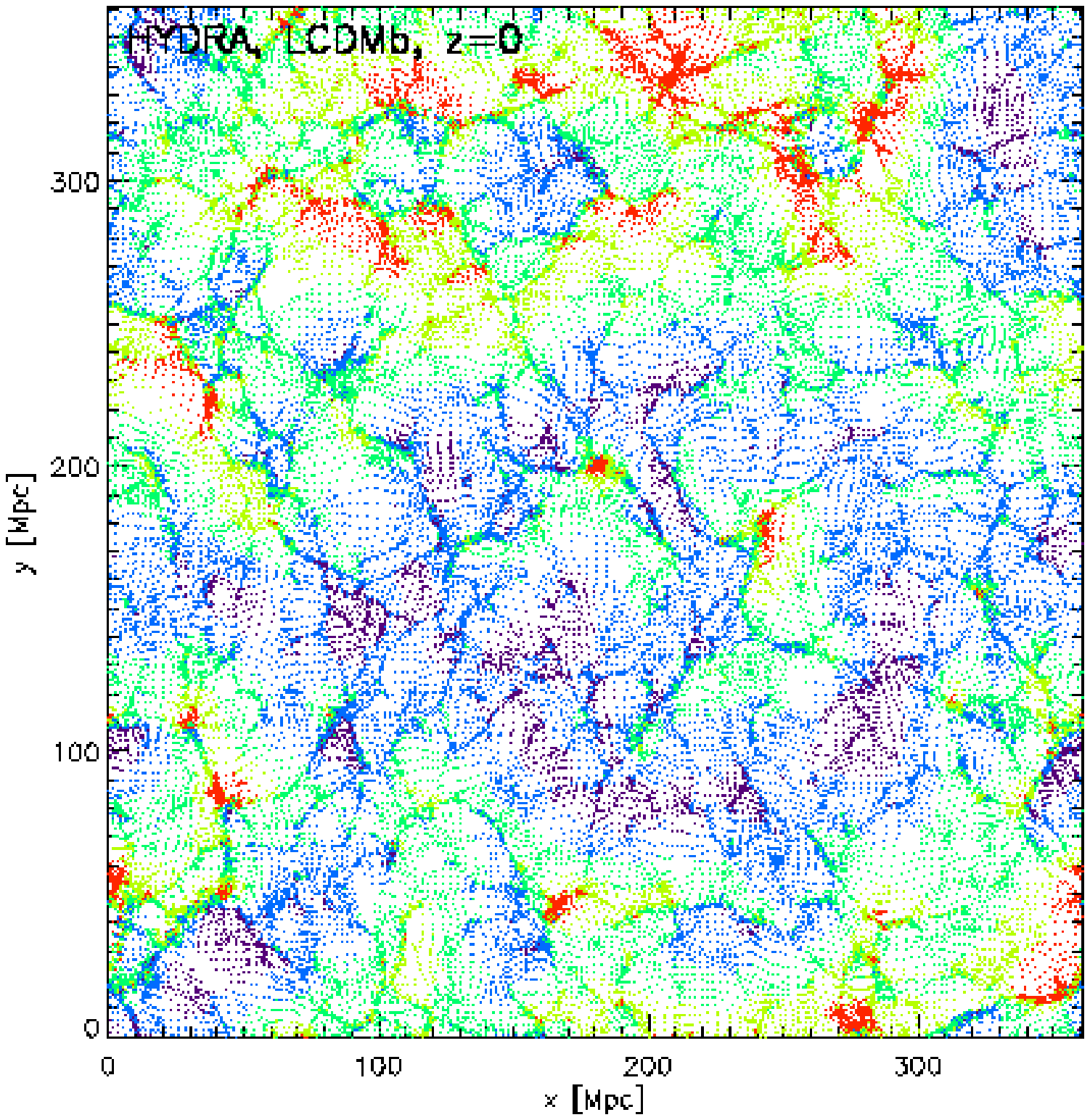}
\end{center}}
\parbox[t]{9cm}
{\begin{center}
\includegraphics[width=73mm]{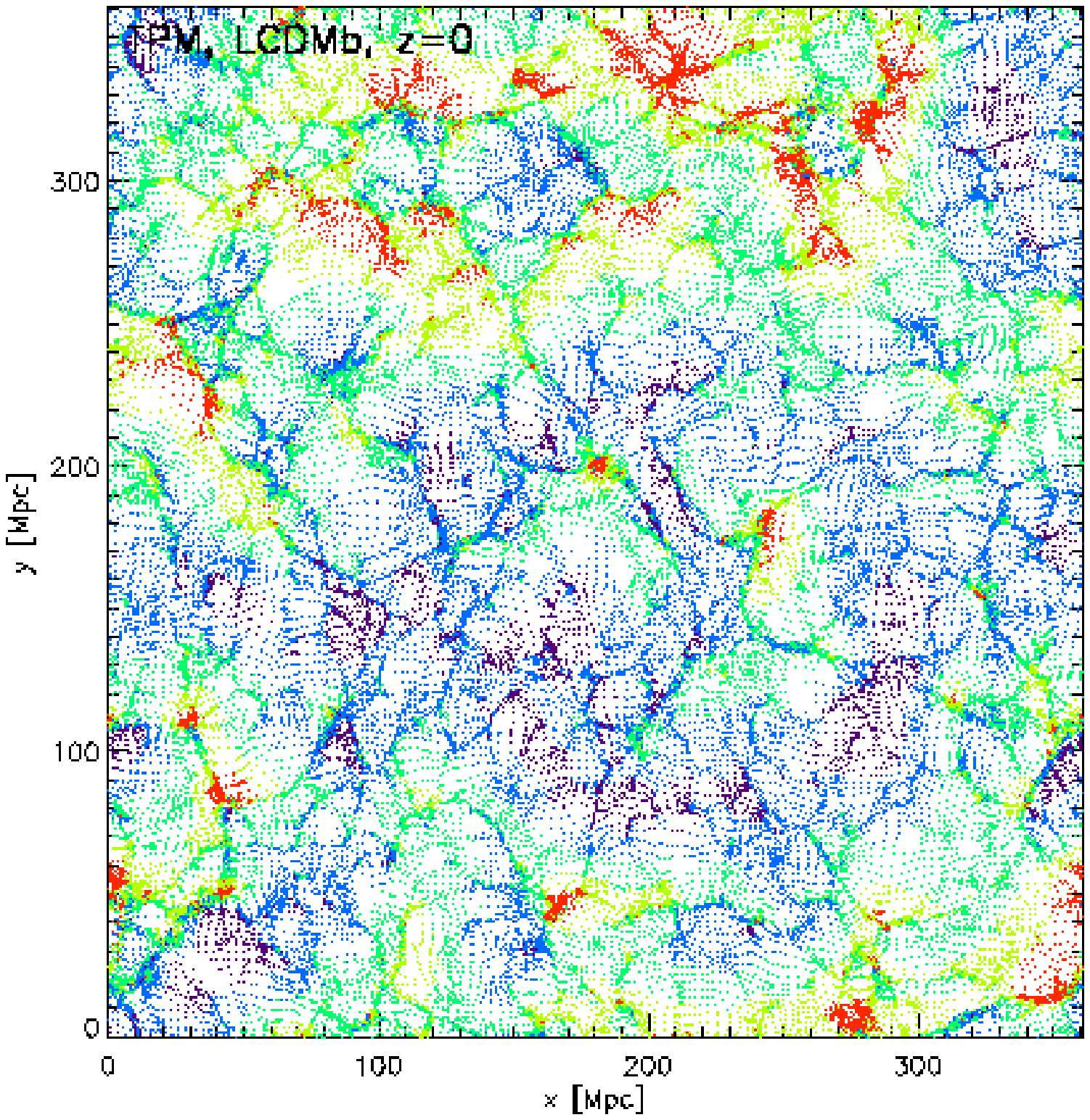}
\end{center}}}}
\caption{{Comparison of the particle positions and velocities 
  in a 3~Mpc thick slice of the 360 Mpc box. As in the small box, the
  velocities are divided into 4 equal bins of size 175~km/s with a
  similar color-coding, the maximum speed being 700~km/s.}}
\label{plotthirteen}
\end{center}
\end{figure*}

In order to first compare the codes qualitatively and picture the
`cosmic web' it is helpful to look at the particles directly.  In
Figs.~\ref{plottwelve} and \ref{plotthirteen} we show the results for the six
codes.  For the small box (Fig.~\ref{plottwelve}) we cut out a 1~Mpc
thick slice in the $xy$-plane through the center of the box.  The
velocities of the particles are divided into 4 equal bins between 0
and 440~km/s.  Each particle is individually colored according to the
bin it belongs to. The agreement of the 6 codes is very impressive,
with a visual impression of almost particle by particle agreement (if
one were to zoom in, this would of course not hold in the small-scale
virialized regions).  Even though the grid codes MC$^2$ and FLASH were
run at lower resolution than the other four codes, in this qualitative
comparison they are almost indistinguishable from the higher
resolution runs. The results for the big box are shown in
Fig.~\ref{plotthirteen}.  Here we have cut out a 3~Mpc thick slice in the
$xy$-plane through the center of the box. Again, even though the
resolutions of the different simulations are different, the plots are
very similar.

\subsubsection{Velocity Statistics of the Particles}

We now turn to a more detailed comparison of the particle
velocities. We measure the distribution of the particle speeds
$v=\sqrt{v_x^2+v_y^2+v_z^2}$ in bins of 25~km/s for the small box and
of 30~km/s for the large box. In order to compare the distributions
from the different codes we only include velocities between 1~km/s and
2500~km/s in the small box and between 1~km/s and 3000~km/s for the
big box. 

\begin{figure}[b]
\includegraphics[width=80mm]{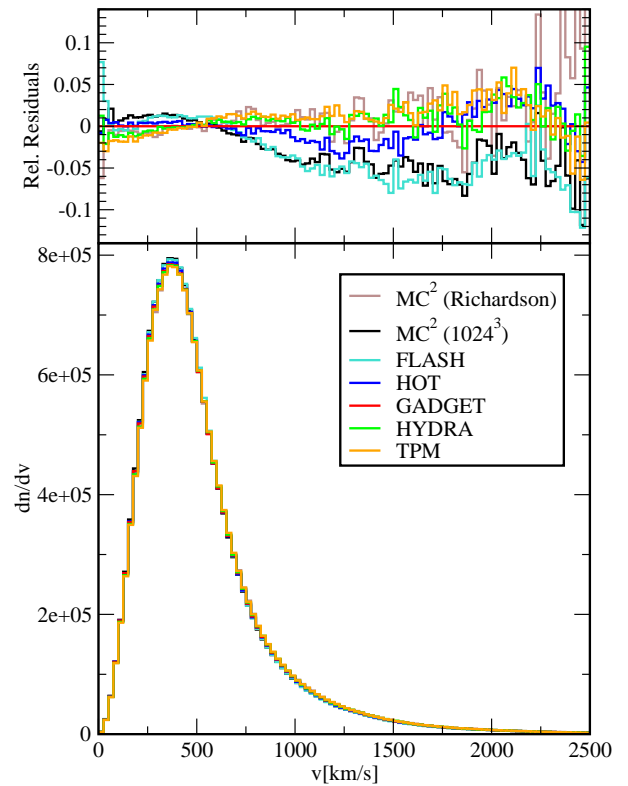}
\caption{Particle velocity distribution 
  for all six codes at $z=0$, 90 Mpc~box.  The upper panel
  shows the relative residuals in comparison with
  GADGET.}
\label{plotfourteen}
\end{figure}

Let us first consider the results for the 90~Mpc box.  In
Fig.~\ref{plotfourteen} the comparison of the velocity distribution
for all 6 codes is shown. HOT, GADGET, HYDRA, and TPM are almost
indistinguishable. In the upper panel of the plot the relative
residual in comparison to GADGET is shown.  The relative deviation of
HOT, HYDRA and TPM from GADGET is below 3\% up to 1500 km/s, while
MC$^2$ (on a 1024$^3$ mesh) and FLASH deviate in this region by less
than 5\%. By combining a 512$^3$ mesh simulation and the 1024$^3$ mesh
simulation from MC$^2$, it is possible to predict the result for the
continuum from the PM simulations using Richardson extrapolation with
linear convergence (the order of convergence was found numerically
using a range of grid sizes). This reduces the systematic deviation
from the higher resolution codes very significantly ($\sim 5\%$ up to
2000 km/s), as evidenced by the top panel of
Fig.~\ref{plotfourteen}. The main significance of this result lies in
the fact that MC$^2$ is not collisional up to the resolution employed
in the cosmological tests, thus the Richardson extrapolation to a
higher effective resolution does not bring in any artificial heating
from collisions. The much improved agreement with the higher
resolution codes is strong evidence that, over the velocity range
studied, collisional effects are subdominant in those codes as well.
[Details of the extrapolation methodology and how it may be used to
improve resolution for mesh codes will be provided in a forthcoming
paper (Habib et al. 2004).] The deviation of FLASH from GADGET is very
similar to the deviation of the 1024$^3$ simulation from MC$^2$,
indicating that most relevant regions were refined in the FLASH
simulation to the highest level. Overall, this first quantitative
comparison may be considered very satisfactory.

\begin{figure}[b]
\includegraphics[width=80mm]{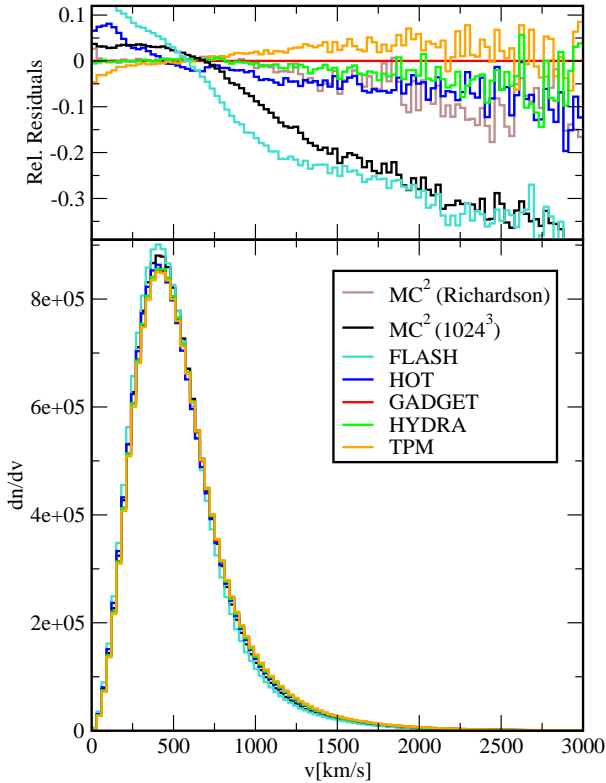}
\caption{Particle velocity distribution following Fig.~\ref{plotfourteen}
  for the 360~Mpc box. Results from the grid codes deviate at large
  velocities as their resolution is not adequate to resolve high
  density regions. Note the small systematic deviations at high
  velocities for the codes run a high resolution.}
\label{plotfifteen}
\end{figure}

Next, we analyze the corresponding results for the 360 Mpc box. The
comparison of all six codes is shown in
Fig.~\ref{plotfifteen}. Overall, the results are not as good as for
the smaller box. Nevertheless, HYDRA and TPM agree with GADGET at
$\sim 10\%$ over the whole range and $\sim 5\%$ up to
1500~km/s. Relative to these codes, the velocities generated by HOT
are slightly lower overall, however, the agreement is still around 6\%
over most of the range. The raw MC$^2$ and FLASH results deviate from
the higher resolution runs up to 40\% for very large velocities as the
grid resolution is not adequate to resolve high density regions
properly. Once again, Richardson extrapolation using a 512$^3$ and a
1024$^3$~mesh succeeds in converting the MC$^2$ results to those very
similar to the higher-resolution simulations (less than 5\% difference
from GADGET out to 1500~km/s). Note that at large velocities,
systematic deviation trends across the codes are clearly visible,
especially in the large box.

\begin{figure}[b]
  \includegraphics[width=90mm]{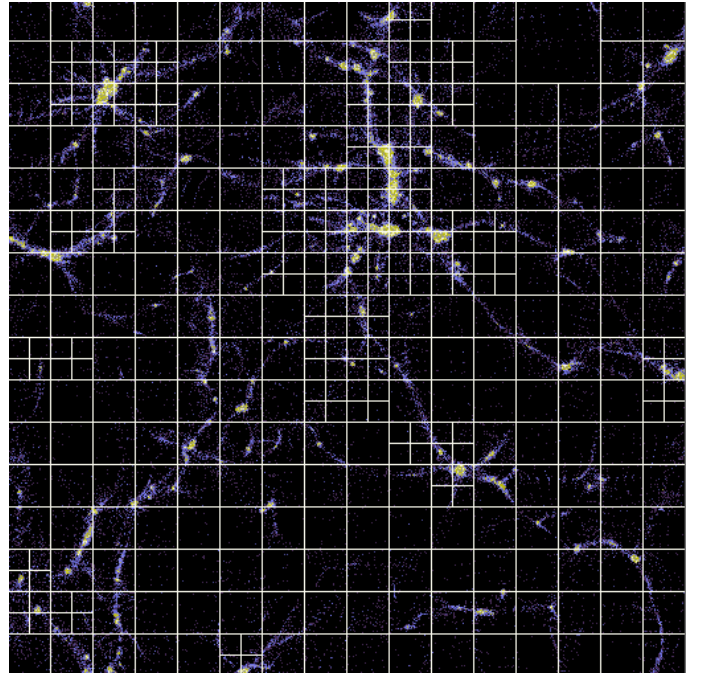}
\caption{Refinement levels from a FLASH-AMR simulation superposed on a
partial density slice from the center of the 360~Mpc simulation box. The 
actual highest resolution grid scale is 16 times smaller than the 
smallest box shown in the figure.} 
\label{plotsixteen}
\end{figure}

\begin{figure}[t]
\includegraphics[width=80mm]{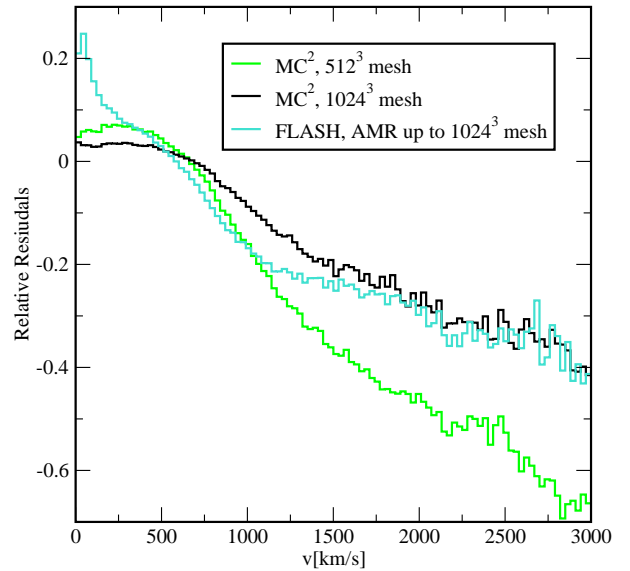}
\caption{Comparison of the relative residuals for particle velocities
from the MC$^2$ 512$^3$ grid and 1024$^3$ grid runs with the FLASH
results, 360 Mpc~box. At higher velocities, the AMR results match
better with the 1024$^3$ grid results, while at lower velocities, they
agree with the 512$^3$ grid results. See the text for a discussion.} 
\label{plotseventeen}
\end{figure}

Around the peak of the distribution, MC$^2$ performs slightly better
than FLASH. This is easy to understand: As velocities are highest in
the high density regions (e.g., the clumped regions in
Fig.~\ref{plotthirteen}), the effective resolution in these regions
controls the accuracy with which the tail of the velocity distribution
is determined. Fig.~\ref{plotsixteen} shows a log-density slice from the
simulation with the AMR-grid superimposed onto it. The high density
regions are appropriately refined by FLASH (smaller
submeshes). However, the lower density regions are treated with poorer
resolution, therefore, we expect good agreement with a 1024$^3$ mesh
simulation for the higher velocities, and for the lower velocities,
good agreement with the corresponding resolution PM simulation.  In
Fig.~\ref{plotseventeen} we show the relative residuals with respect to
GADGET of the 512$^3$~mesh and 1024$^3$ mesh MC$^2$ simulations, and
the AMR-FLASH simulation.  In the high velocity tail the higher
resolution simulation from MC$^2$ and FLASH agree very well, while for
the lower velocity region the MC$^2$ 512$^3$ mesh simulation agrees
better with FLASH (at late times, near z=0, the main box is run by
FLASH at a resolution of 512$^3$).

Finally, we compare results for $v_{12}(r)$, the (pair-averaged)
relative pairwise streaming velocity considered as a function of the
separation $r$ between two mass tracers. If the tracers are dark
matter particles then the pair conservation equation [following the
standard BBGKY approach (Davis \& Peebles 1977; Peebles 1980)],
connects the $v_{12}(r)$ field to the two-point correlation function of
the particle distribution $\xi(r)$. Measurements of $v_{12}(r)$ from
peculiar velocity surveys provide useful constraints on $\Omega_m$ and
$\sigma_8$ (Feldman et al. 2003). While theoretical predictions for
these measurements are best carried out with the corresponding
velocity statistics for halos, computing $v_{12}(r)$ for particles and
comparing this with the same quantity calculated for halos is used to
determine halo velocity bias. The pairwise peculiar velocity
dispersion $\sigma_{12}(r)=\langle[{\bf v}({\bf r})-\langle{\bf
v}({\bf r})\rangle]^2/3\rangle^{1/2}$ is also an important statistical
quantity, connected to $\Omega_m$ and the two-point correlation
function via the Cosmic Virial Theorem (Peebles 1976a and 1976b).

\begin{figure}
\includegraphics[width=80mm]{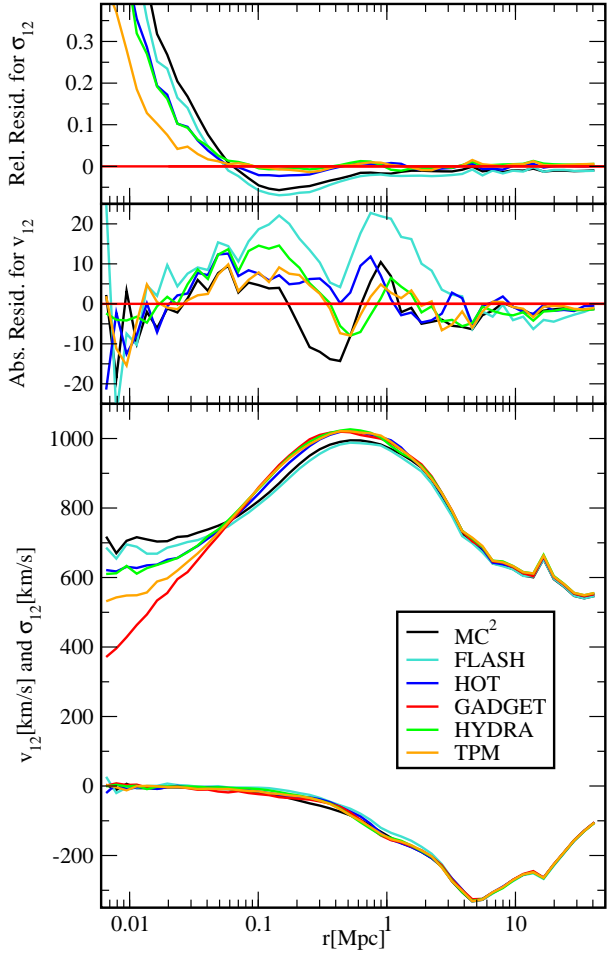}
\caption{Comparison of the relative particle pairwise streaming velocity 
  $v_{12}(r)$ (lower set of curves) and the pairwise peculiar velocity
  dispersion $\sigma_{12}(r)$ (upper set) for all six codes at $z=0$,
  90 Mpc box. At small distances, the GADGET results for
  $\sigma_{12}(r)$ deviate from those of the other codes. For a
  discussion of this problem, see Sec. \ref{subsec:partcorr}.}
\label{ploteighteen}
\end{figure}

\begin{figure}
\includegraphics[width=80mm]{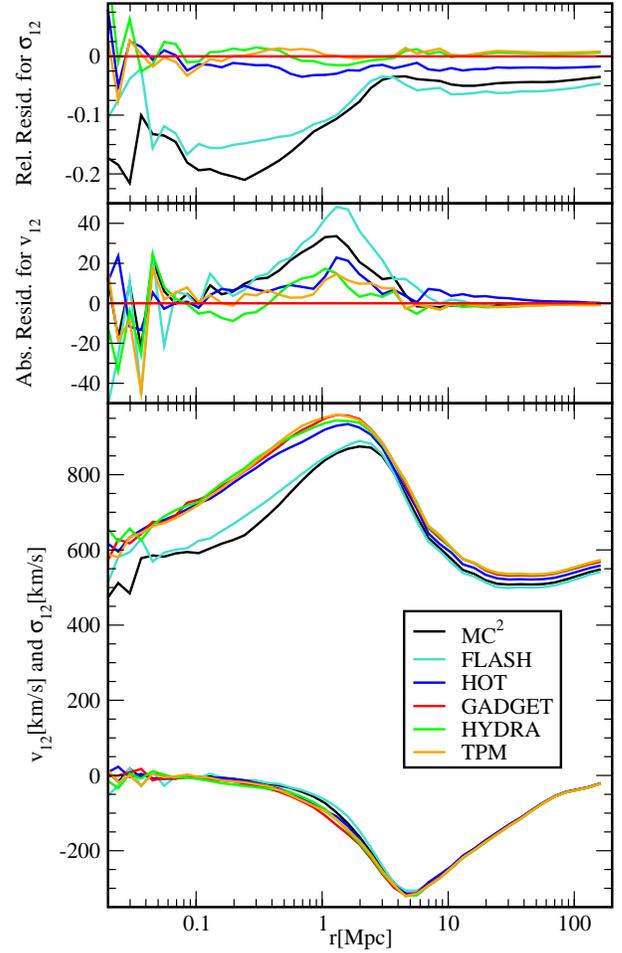}
\caption{Comparison of the relative particle pairwise streaming velocity 
  $v_{12}(r)$ (lower set of curves) and the pairwise peculiar velocity
  dispersion $\sigma_{12}(r)$ (upper set) for all six codes at $z=0$,
  360 Mpc box. The lower resolution of the grid codes is more apparent
  in the results for $\sigma_{12}(r)$.}
\label{plotnineteen}
\end{figure}

Results for the two quantities are displayed in
Figs.~\ref{ploteighteen} and \ref{plotnineteen} for the small and
large boxes respectively. Absolute residuals with respect to GADGET
are provided for the $v_{12}(r)$ results (since $v_{12}$ can vanish),
while relative residuals are given for $\sigma_{12}(r)$. The $v_{12}$
results for the small box show a small bump in the GADGET results with
respect to the other codes at $r\simeq 0.5$~Mpc, otherwise the results
are better than 10\%, and, in the actual range of physical interest,
$r> 1$~Mpc, (except for the FLASH results), the agreement is better
than 5\%. (See Sec. \ref{subsec:partcorr} for a discussion of issues
with the time-stepping in GADGET.) The results for $\sigma_{12}(r)$
from the high resolution codes are very good, with agreement at the
few \% level achieved, except on small scales where resolution
limitations are manifest. The results from the lower-resolution mesh
codes, as expected, show a small 5\% suppression in $\sigma_{12}(r)$
in the region of the peak.  (Here we have not attempted extrapolation
to improve the results.)  Results from the large box for $v_{12}$
again show a systematic difference between the other high-resolution
codes and GADGET at roughly 10\% in the region around 1~Mpc. The
results for $\sigma_{12}(r)$ for the high-resolution codes are very
consistent, with better than 5\% agreement across the range. The mesh
codes (without extrapolation) are at a particular disadvantage in this
test since the peak in $\sigma_{12}(r)$ occurs relatively close to
their resolution limit.

\clearpage

\subsubsection{Power Spectrum of the Particle Density Field}
\label{pps}

The next quantitative comparison performed for the particles is the
measurement of the power spectrum $P(k)$. The power spectrum is
obtained in the following way: First the density is calculated with a
CIC (Cloud-In-Cell) scheme on a 1024$^3$ spatial grid.  Then a
1024$^3$ FFT is employed to obtain the density in $k$-space and
squared to yield the power spectrum. (In order to compensate for the
CIC softening we deconvolve the $k$-space density with the CIC window
function.)  Finally, the three-dimensional power spectrum is binned to
get the one-dimensional power spectrum. We did not attempt to
compensate for particle shot noise or use a bigger FFT as the aim here
is to look for systematic differences between codes rather than
actually obtain a more accurate $P(k)$. The power spectrum will be
studied in more detail in later work.

\begin{figure}
\includegraphics[width=80mm]{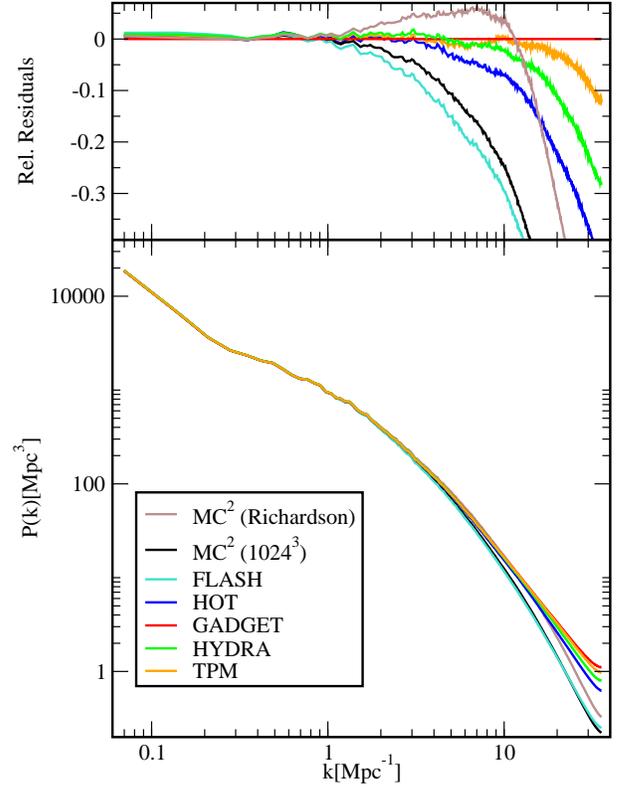}
\caption{Comparison of the mass density power spectra for all six
  codes at $z=0$, 90~Mpc box. The Richardson-extrapolated result from
  MC$^2$ is also shown. The early fall-off of the two grid codes is
  expected and consistent with the resolution employed. The high
  result from GADGET is discussed in Sec. \ref{subsec:partcorr}.}
\label{plottwenty}
\end{figure}

We begin again with the small box, the results being shown in
Fig.~\ref{plottwenty}. The lower panel shows the results from the six
codes and in addition a power spectrum generated by Richardson
extrapolation from a 512$^3$ and a 1024$^3$ mesh MC$^2$ simulation.
The upper panel displays the relative residuals with respect to
GADGET.  The agreement between the HOT, GADGET, HYDRA, and TPM is very
good, better than 10\% up to $k=10$~Mpc$^{-1}$. For higher $k$ the HOT
and HYDRA simulations have slightly less power than the GADGET and TPM
simulations. The results from the two mesh codes, MC$^2$ and FLASH,
are in good agreement with the other runs up to a few Mpc$^{-1}$. For
higher $k$, the limited resolution of these two runs leads to larger
discrepancies, as expected. In these runs, the slightly higher
resolution of MC$^2$ compared to FLASH is again visible. The
Richardson extrapolation result from MC$^2$ differs from the GADGET
simulation by roughly 5\% up to $k$=10~Mpc$^{-1}$. For larger $k$ the
difference becomes much larger. This is as expected since the
extrapolation procedure cannot generate power which was not present in
the underlying simulations; it can only improve the accuracy of the
results on scales where convergence holds. Due to this property, it
functions as an excellent test of code convergence.

\begin{figure}
  \includegraphics[width=80mm]{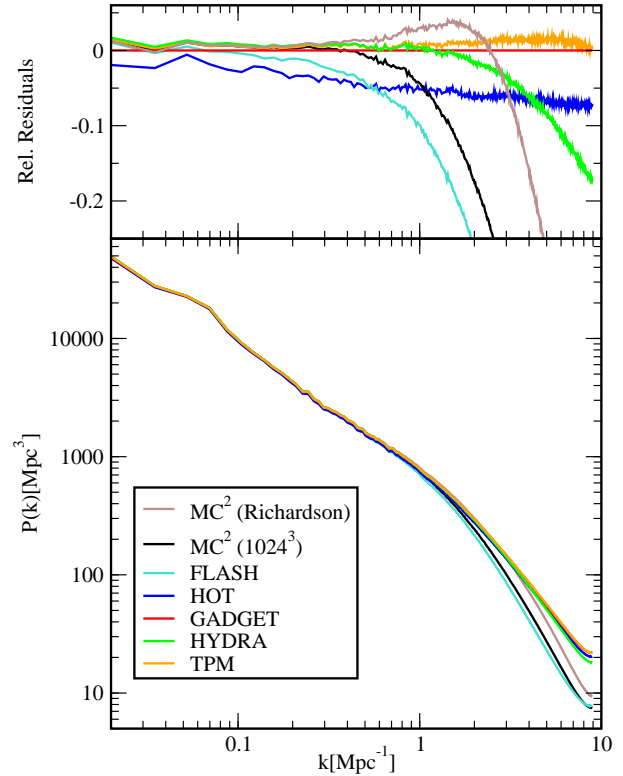}
\caption{Mass density power spectrum results for the 360~Mpc box, 
  following Fig.~\ref{plottwenty}.} 
\label{plottwentyone}
\end{figure}

Next we compare the results for the large box, as shown in
Fig.~\ref{plottwentyone}.  The overall result is very similar to the small
box. Over roughly two orders of magnitude in $k$ the high resolution
runs from HOT, GADGET, HYDRA, and TPM agree better than 5\%.  The TPM
simulation agrees with GADGET over the entire range to better than
2\%. The HYDRA run shows again a slight fall-off of the power with $k$
before the other high resolution codes do. The results from HOT appear
to be systematically low in the normalization. The MC$^2$ result shows
good agreement (sub-5\%) with GADGET up to $k$=1~Mpc$^{-1}$, while the
FLASH power spectrum falls off faster already at $k$=0.4~Mpc$^{-1}$.
This is consistent with the fact that in the large box less of the
volume was refined ($14.3\%$) than in the small box ($40.8\%$). As
seen for the small box, Richardson extrapolation again leads to a
large improvement: the difference with GADGET becomes sub-5\% up to
$k$=4~Mpc$^{-1}$.

As with the results obtained for the particle velocity statistics, the
results for the power spectrum are also very satisfactory. From the
point of view of force resolution, all six codes agree over the
$k$-range for which consistent results are to be expected: The four
high resolution codes agree at the sub-5\% level in the main regions
of interest, while the the results for the two mesh codes are also
consistent with the expected resolution and convergence properties of
these codes.


\subsubsection{Particle Correlation Function}
\label{subsec:partcorr}

After comparing the power spectrum we now investigate its real-space
counterpart, the correlation function generated from the particle
positions. The $O(N^2)$ calculation is straightforward, though
tedious. In principle the correlation function contains the same
information as the power spectrum, however, since they were obtained
in different ways, comparison against the correlation function
provides a simple cross-check. The correlation functions were
generated by randomly choosing 65,536 particles from each simulation,
and calculating the number of particles contained within
logarithmically spaced radial bins using a direct $N^2$ search.

\begin{figure}[b]
  \includegraphics[width=80mm]{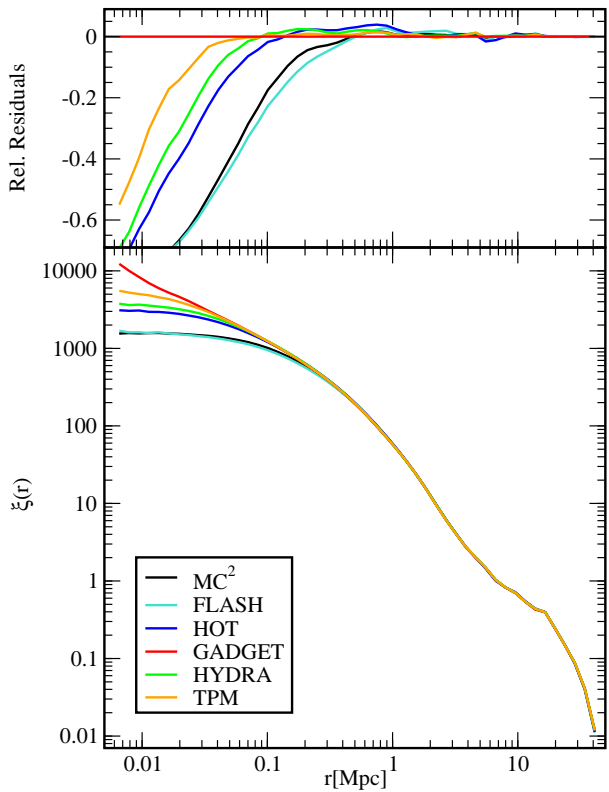}
\caption{Comparison of the particle two-point correlation functions
  from the six codes at $z=0$ for the 90~Mpc box. See the text for a
  discussion of the rise in the GADGET results at small distances. The
  results for the grid codes are consistent with the employed
  resolution.}
\label{plottwentytwo}
\end{figure}

The results for the small box are shown in Fig.~\ref{plottwentytwo}.
As always, the lower panel shows the results from the six different
codes while the upper panel displays the relative residuals with
respect to the GADGET simulation.  The agreement of all codes is very
good for large $r$, between 200~kpc and 40~Mpc, the deviation from
GADGET for all other codes being less than 5\%. For $r$ smaller than
200~kpc the limited resolution of the MC$^2$ and FLASH simulations
becomes apparent, the correlation function falling off much faster
than for the other simulations. At $r\approx$~20~kpc the correlation
function from GADGET rises much more steeply than from the other
simulations. This is an artifact due to time integration errors: The
design of GADGET's multistep integrator is prone to the build-up of
secular integration errors which can manifest themselves in effective
energy losses of some particles, which in turn boosts the phase-space
density in halo centers. While this can be avoided with finer (and
more expensive) time-stepping, this is not an optimum solution. GADGET
has been updated 2 years ago to avoid this problem, but this version
is not public yet. \footnote{We thank Volker Springel for providing
  this information.} This explains the large relative residuals with
respect to GADGET at small scales. Up to this point HOT, HYDRA, and
TPM agree better than 5\% with GADGET.

\begin{figure}[t]
\includegraphics[width=80mm]{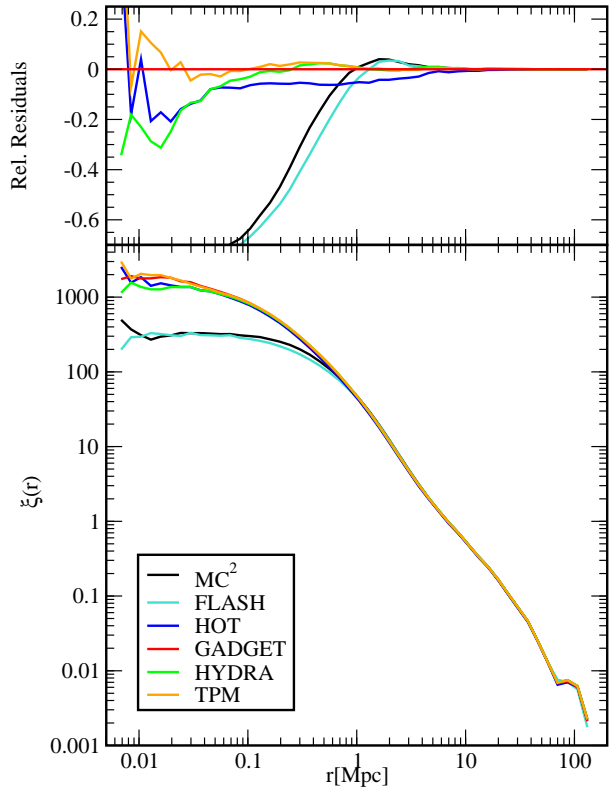}
\caption{Particle two-point correlation function for the 360~Mpc 
  box, following Fig.~\ref{plottwentytwo}.} 
\label{plottwentythree}
\end{figure}

The results for the large box are shown in Fig.~\ref{plottwentythree}
and are similar to those from the small box. For large $r$ the
agreement is once again excellent, between 0.8~Mpc and 160~Mpc all
codes agree to better than 5\%. At small $r$, FLASH and MC$^2$ suffer
from degraded resolution, and the disagreement between the other codes
becomes larger. The correlation function from HOT is systematically
lower in the region 0.1~Mpc$<r<$5~Mpc, while HYDRA's correlation
function, due to the slightly lower resolution employed than HOT,
GADGET, and TPM, falls off a little earlier than GADGET and TPM. For
very small $r$ the discrepancy between the codes is at the 10\% level,
but here the codes are being pushed too close to their resolution
limits.

Overall, as expected, the results for the correlation functions are
very similar to those for the power spectrum. The resolution limits of
the grid codes MC$^2$ and FLASH are clearly exposed at the correct
scales, with the four high resolution codes agreeing well with each
other until scales close to the resolution limits are reached.

\subsubsection{Halo Mass Function}
\label{hmf}

We now turn from comparing particle position and velocity statistics
to comparing results extracted from halo catalogs. In many ways, it is
the halo information that is of more direct relevance to comparison
with cosmological observations, so these tests are of considerable
importance. The halo catalogs were generated using a
Friends-Of-Friends (FOF) algorithm (Davis et al. 1985). This algorithm
defines groups by linking together all pairs of particles with
separation less than $b{\bar n}^{-1/3}$, where $\bar n$ is the mean
particle density. This leads to groups bounded approximately by a
surface of constant local density, $\rho/\bar \rho \approx 3/(2\pi
b^3)$. We have chosen a linking length of $b=0.2$ corresponding to a
density threshold $\rho/\bar\rho\approx 60$. As discussed in Lacey \&
Cole 1994, for a spherical halo with a density profile $\rho
(r)\propto r^{-2}$, this local density threshold corresponds to a mean
overdensity $\langle \rho \rangle/\bar \rho \approx 180$. 

In order to confirm the results obtained with this halo finder we also
analyzed two of the small box simulations (GADGET and MC$^2$) with a
different FOF halo finder\footnote{We thank Stefan Gottl\"ober for
generating the alternate set of halo catalogs.}  with a linking length
of $b=0.17$ corresponding to a density threshold $\rho/\bar\rho
\approx 97$. (For a discussion on the dependence of the mass function
on the chosen linking length for a FOF halo finder, see, e.g., Lacey
\& Cole 1994.) We checked that the two FOF halo finders gave the same
results for $b=0.17$.

\begin{figure}[t]
\includegraphics[width=83mm]{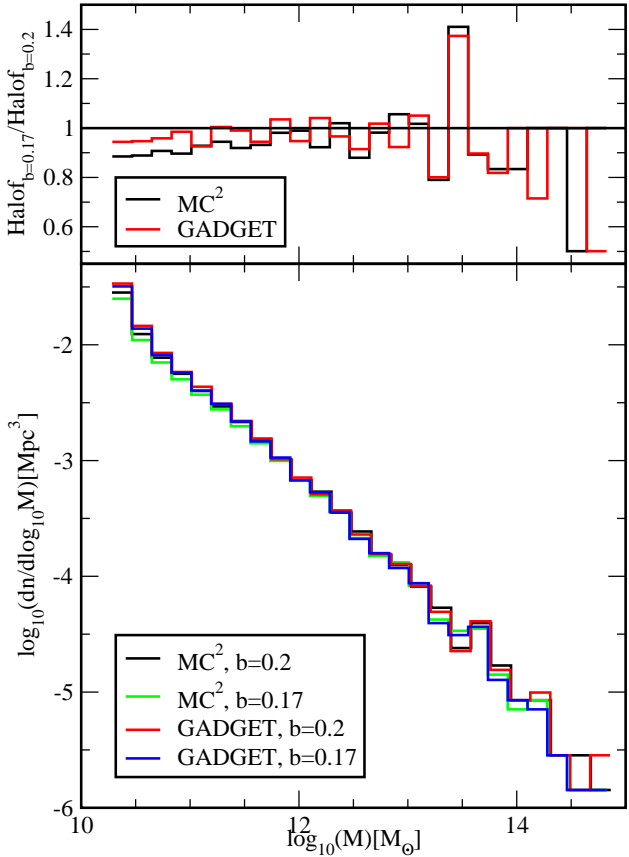}
\caption{Comparison of results from two FOF halo finders with linking
lengths of $b=0.2$ and $b=0.17$ run on the MC$^2$ and GADGET
simulations at $z=0$ for the 90~Mpc box. The top panel shows the ratio
of the $b=0.17$ mass function and the $b=0.2$ mass function for the
two codes.}  
\label{plottwentyfour}
\end{figure}

The comparison between the halo mass functions for the two choices of
linking length is shown in Fig.~\ref{plottwentyfour}. In the lower
panel the mass functions found with the two halo finders for GADGET
and MC$^2$ are shown. In the upper panel we show the ratio of the two
mass function results for MC$^2$ and the ratio of the two mass
functions found for GADGET. The ratio is the $b=0.17$ mass function
divided by the $b=0.2$ mass function.  The smaller linking length
should find fewer halos in the small mass region region and tend to
``break up'' heavier halos in the high mass region. Consequently the
ratio should show a dip at both extremes (since there are far fewer
high mass halos, overall the smaller linking length will lead to fewer
halos). Our results are very consistent with this expectation and also
with the fact that more MC$^2$ halos are ``lost'' by the $b=0.17$
finder at small masses -- due to the lower resolution, these halos are
more diffuse as compared to the GADGET halos.

\begin{figure}[b]
\includegraphics[width=83mm]{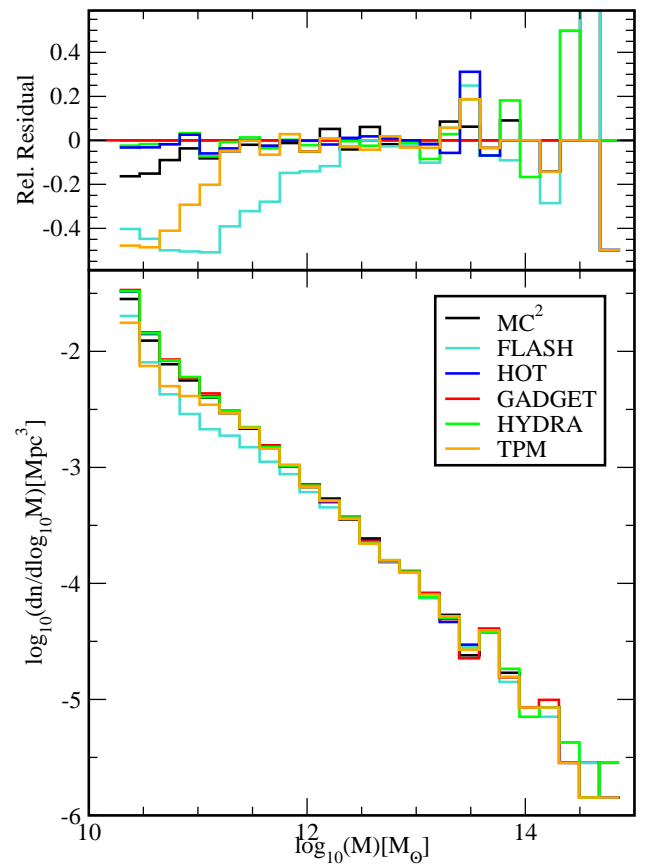}
\caption{Comparison of the halo mass function for all six codes at 
  $z=0$, 90~Mpc box. See the text for a discussion of code 
  systematics.}
\label{plottwentyfive}
\end{figure}

\begin{table*}
\begin{center}
\caption{\label{halonum} Numbers of halos found in the different simulations}
\begin{tabular}{lcccccc}
\hline\hline
\hfill& MC$^2$ & FLASH & HOT &GADGET & TPM & HYDRA \\
\hline
LCDMs &  49087 halos  &  32494 halos & 54417 halos  &55854 halos &   
34367 halos & 54840 halos \\
LCDMb & 65795 halos  &  33969 halos& 73357 halos  & 72980 halos &  
55967 halos & 74264 halos \\
\hline\hline
\end{tabular}
\end{center}
\end{table*}

Discounting the high mass end, there is still a systematic variation
of up to 10\% in the two halo mass functions.  This result should be
kept in mind when comparing the mass function to semi-analytic fits:
even the same type of halo finder with slightly different parameter
choices can easily affect the mass function at the several \%
level. Furthermore, changing the mass resolution can also lead to
systematic effects on halo masses using the same FOF algorithm. These
aspects will be considered in detail elsewhere (Warren et al. 2005).

Next we analyze the halo catalogs generated with the same FOF halo
finder for all six codes. We start by comparing the numbers of halos
found in the different simulations. The results are listed in
Table~\ref{halonum}. The smallest halos considered contain 10
particles, leading to a minimum halo mass for the small box of
$2\cdot10^{10}$~M$_\odot$. In this box, the most massive halos have
masses of the order of $7\cdot 10^{14}$ M$_\odot$. Allowing 10
particles to constitute a halo is a somewhat aggressive definition,
and we are not arguing that this is appropriate for cosmological
analysis. However, as a way of comparing codes, the very small halos
are the ones most sensitive to force resolution issues and therefore
tracking them provides a useful diagnostic of resolution-related
deficiencies as well as providing clues to other possible problems.

Largely due to the generous definition of what constitutes a halo, the
variation in the total number of halos is rather large: For the small
box, HOT, GADGET and HYDRA found the largest number of halos, around
55000, while FLASH and TPM found almost 40\% fewer halos.  The number
of halos from the MC$^2$ simulation is in between these two extrema, a
little closer to the higher numbers.  For the large box the agreement
of the number of identified halos from HOT, GADGET, and HYDRA, around
73000, is much better, but again the TPM simulation leads to almost
20000 fewer halos, and from the FLASH simulation, only 33000 halos
were identified.  The result from MC$^2$ is slightly lower than the
results from HOT, GADGET, and HYDRA but much higher than the halos
identified from TPM and FLASH.

We have generated mass functions from the different halo catalogs
using a FOF algorithm with linking length $b=0.2$.  The comparison of
the mass functions is shown in Fig.~\ref{plottwentyfive}. As mentioned
above, far fewer halos were found in the FLASH and TPM simulations,
which becomes very apparent in this figure: At the low mass end of the
mass function the results from these two codes are discrepant up to
40\%. In contrast, MC$^2$, HOT, GADGET, and HYDRA agree very well, to
better than 5\%.

\begin{figure}[b]
\includegraphics[width=83mm]{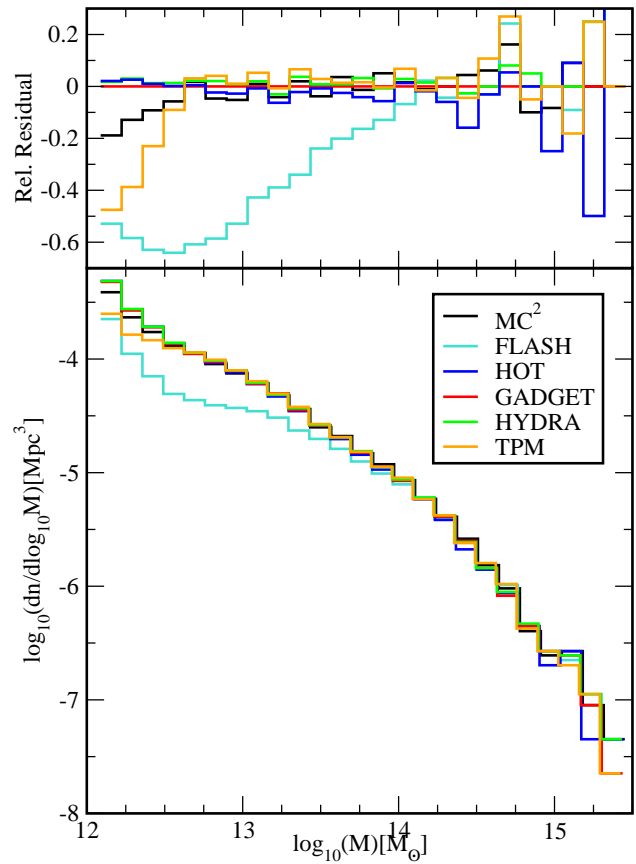}
\caption{Comparison of the halo mass function for all six codes at 
  $z=0$, 360~Mpc box. For discussion of code systematics, see text.} 
\label{plottwentysix}
\end{figure}

In case of the grid codes, the fall-off at the low-end of the mass
function can be easily connected to the resolution limit. To see this,
given a particular halo mass, one computes a corresponding scale
radius, say, e.g., $r_{180}$ and compares this with the grid
resolution, $\Delta$. The scale radius $r_{180}$ is defined to be such
that the mean matter density within it is 180 times the mean
background density (numerically, its value is roughly that of the
virial radius). Clearly, if $\Delta > r_{180}$, halos in the
particular mass region will be suppressed, as the finite force
resolution will prevent the formation of compact objects on this
scale. We have verified that the threshold for the fall-off in the
mass function for MC$^2$ is set by the approximate condition
$\Delta\sim r_{180}$ in both the small and large boxes. One may wonder
whether the lack of smaller halos has any effect on the mass function
at masses higher than the resolution threshold set by $\Delta\sim
r_{180}$. As shown by our results from both the mass function, and the
comparison of individual halo masses below, this is not the case. The
particles that would have ended up in smaller halos are not lost, and
end up streaming into the larger halos so that there is no systematic
bias in the mass function above the resolution threshold.

The result found for FLASH is also easy to understand. The base grid
in the FLASH simulation was chosen to be $256^3$ which was then
refined up to 1024$^3$, as explained above. At the beginning of the
simulation this base resolution is not sufficient to form small
halos. As the simulation proceeds forward in time, the resolution
increases only where the density crosses the AMR threshold, elsewhere
the small halos cannot be recovered. Therefore AMR codes will lead to
very good results for the large halos and their properties, since for
these the mesh will be appropriately refined. But without the
implementation of AMR-specific initial conditions which help to solve
the problem of starting with a pre-refined base grid, the mass
function will be suppressed for small masses. This work is now in
progress (Lukic et al. 2005). The condition $\Delta\simeq r_{180}$
imposed using the base $256^3$ grid correctly predicts the fall-off
point in the mass function for the FLASH simulations. [A similar
result for the mass function was found by O'Shea et al. (2003), in
which GADGET was compared to ENZO, an AMR-code developed by Bryan \&
Norman (1997,1999): The mass function from the ENZO simulation was
lower for small masses than the one obtained from GADGET.]

We are unable to explain the fall-off in the mass function for the TPM
simulations (seen in both the small and large boxes) in terms of a
simple force resolution argument. It is possible that this effect is
due to the grid-tree hand-off, but at present this remains a
speculation.

The results for the large box show trends similar to the small box
results. The overall agreement of the six codes shown in
Fig.~\ref{plottwentysix} is better than for the small box. The mass
function from the TPM simulation is again lower than the others by
roughly 40\% for the small mass halos. The FLASH result for the small
halos is up to 60\% lower than that from GADGET. As in the small box,
the condition $\Delta\simeq r_{180}$ determines the mass below which
halos are lost.  The deviation from GADGET for the other codes is less
than 2\% over a large range of the mass function, a pleasing result.

\subsection{Halo Power Spectrum}

\begin{figure}
\includegraphics[width=80mm]{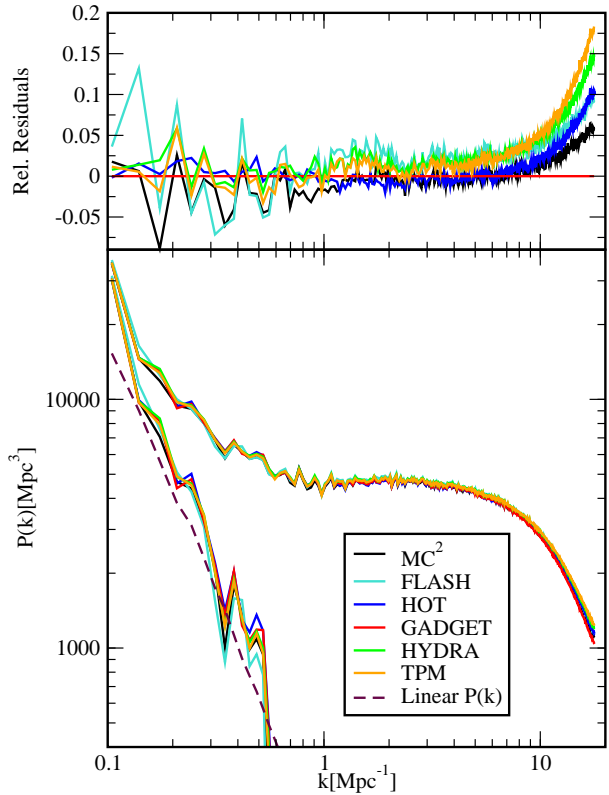}
\caption{Comparison of the halo power spectrum for all six codes at 
  $z=0$, 90~Mpc box. Only halos with more than 100 particles were
  used. The upper set of curves in the lower panel has 
  not been noise-subtracted while the lower set of curves was
  corrected for shot-noise. The dashed line is the linear power
  spectrum from the corresponding initial condition realization.}
\label{plottwentyseven}
\end{figure}

\begin{figure}[t]
\includegraphics[width=80mm]{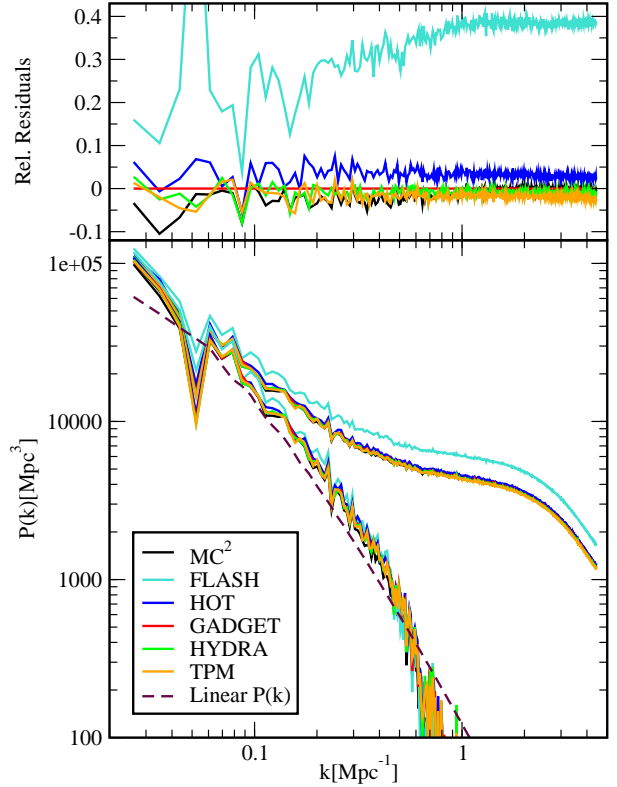}
\caption{Comparison of the halo power spectrum for all six codes at 
  $z=0$, 360~Mpc box, following Fig.~\ref{plottwentyseven}. The upward
  bias of the FLASH result is discussed in the text.}
\label{plottwentyeight}
\end{figure}

The distribution of dark matter halos is biased with respect to the
overall distribution of the dark matter (Mo \& White 1996). One way
of estimating this bias is to compare the halo power spectrum with the
(linear or nonlinear) power spectrum of the particle density
field. Another method is to compare the respective correlation
functions (halo-halo vs. particle-particle) (Colin et al. 1999;
Kravtsov \& Klypin 1999). We discuss both the halo power spectrum and
the correlation function below.

In order to minimize problems with selection effects due to the
differing code force resolutions and possible systematic uncertainties
with halo finders, for the spatial analysis of halos we selected only
those halos with more than a 100 particles. In the small box, this
implies a halo mass threshold of $2\cdot 10^{11}$~M$_\odot$ which, as
comparison with Fig.~\ref{plottwentyfive} shows, succeeds in avoiding
the low halo mass region where MC$^2$ and TPM are somewhat deficient.
For the larger box, the corresponding halo mass threshold is $1.2\cdot
10^{13}$~M$_\odot$ which also avoids the low-mass region
(Fig.~\ref{plottwentysix}). For both boxes the FLASH results, because
of the AMR issues discussed above, continue to have fewer halos until
a higher threshold is reached. In the small box there are roughly 6000
halos for each code, except for FLASH which had less than 5000.

The halo power spectrum was computed in essentially the same way as
the particle power spectrum in Sec.~\ref{pps} by first generating a
CIC density field and then using a $512^3$ FFT. Since the number of
halos is much smaller than the number of particles, here shot noise
subtraction is essential to obtain physically sensible results at
small scales. We show both the noise-subtracted and raw halo power
spectra in Figs.~\ref{plottwentyseven} and \ref{plottwentyeight} for
the small and large boxes respectively. However, for code comparison
purposes the direct halo power spectra are more relevant as noise
subtraction induces large changes in the power spectra. Therefore, the
relative residuals are computed for these quantities.

The results for the small box show a roughly $5\%$ deviation till
$k=10$~Mpc$^{-1}$. Note that this agreement is much better than the
case for the particle power spectrum where the large $k$ regime is
dominated by small-scale particle motions. The case for the large box
is very similar except for the FLASH results. The reason why the FLASH
results are higher is because, (i) as earlier noted, there are fewer
halos overall and (ii) the small mass halos are disproportionately
absent with a resulting increase in bias.

\subsubsection{Halo Correlation Function}

The halo correlation functions were computed in the same way as the
particle correlation function in Section~\ref{subsec:partcorr}. Unlike
the statistics of the raw particle distribution, these correlation
functions provide a more direct measure of the galaxy distribution and
are an essential input in the halo model framework for studying the
clustering of galaxies (for a review, see Cooray \& Sheth
2002). 

\begin{figure}[t]
\includegraphics[width=80mm]{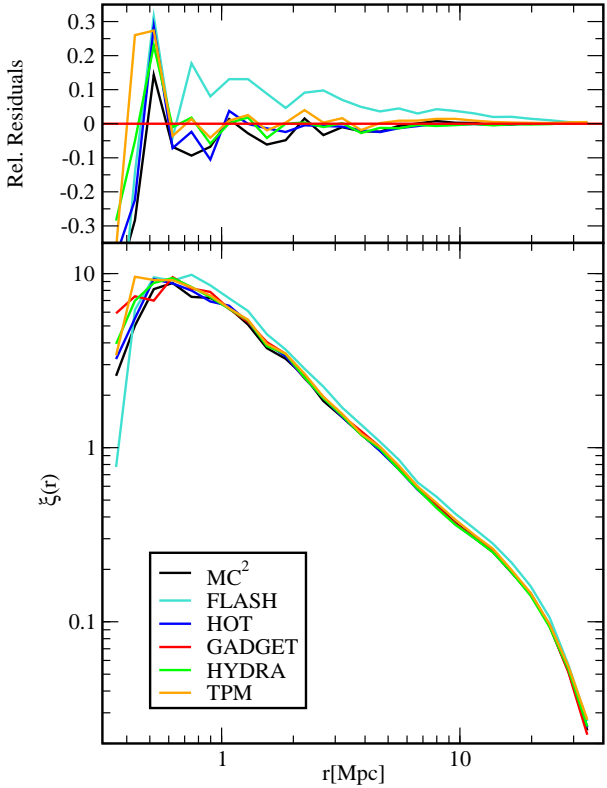}
\caption{Halo two-point correlation function for the 90~Mpc box. 
  Only halos with more than 100 particles are considered, corresponding 
  to a threshold halo mass of $2\cdot 10^{11}$~M$_\odot$.}
\label{plottwentynine}
\end{figure}

\begin{figure}[hb]
\includegraphics[width=80mm]{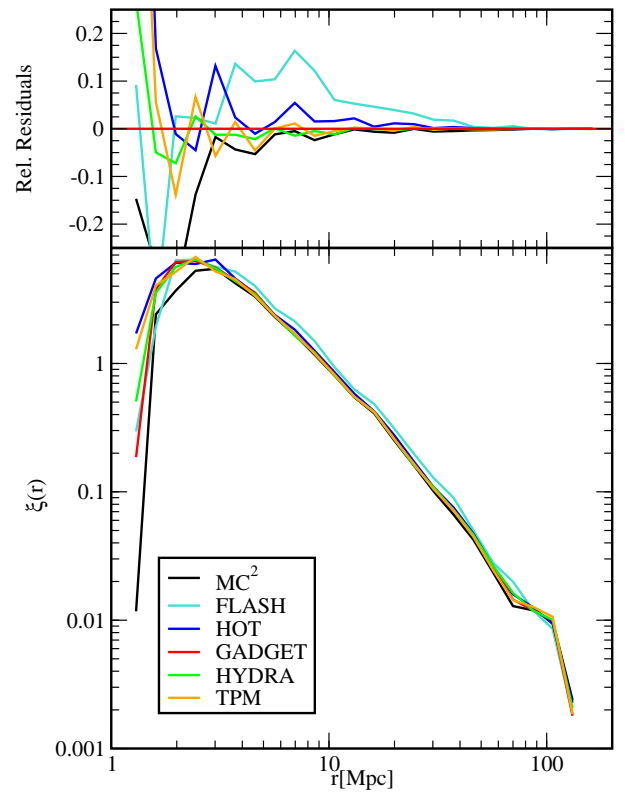}
\caption{Halo two-point correlation function following 
  Fig.~\ref{plottwentynine} for the 360~Mpc box. In this case the lower halo 
  mass cutoff is $1.2\cdot 10^{13}$~M$_\odot$.} 
\label{plotthirty}
\end{figure}

The results for the small box are shown in Fig.~\ref{plottwentynine}.  For
$r\ge 0.6$~Mpc the deviations of the codes from the GADGET result are
smaller than 10\%; for $r\ge 1$~Mpc the agreement is better than
5\%. The FLASH result here is slightly worse because of the systematic
undercounting of halos above the mass threshold of $2\cdot
10^{11}$~M$_\odot$. The results for the large box, shown in
Fig.~\ref{plotthirty}, are very similar. For $r\ge 3$~Mpc the agreement
of the codes with GADGET is again better than 5\%, with the exception
of FLASH which disagrees over most of the range at about 10\% (for the
same reason as earlier). The dips in $\xi(r)$ at the largest values of
$r$ in both boxes is a finite volume artifact, while the suppression
at the smallest values of $r$ is largely due to the length scale
corresponding to the halo mass cutoffs in the two cases, approximately
$0.1 - 0.2$ Mpc for the small box and $0.5 - 1$ Mpc for the large
box. One expects the two-point function results to lose the
suppression on length scales at roughly twice these values, and the
simulation results are consistent with this expectation.

\subsubsection{Halo Velocity Statistics}

The single-point halo velocity statistics for the individual halos
were computed using the more aggressive definition of 10 particles per
halo, while the calculations of $v_{12}(r)$ and $\sigma_{12}(r)$ used
halos with at least 100 particles as described above. To facilitate
direct comparison, the individual velocity distributions were
normalized by dividing by the total number of halos for each
simulation.

\begin{figure}
\includegraphics[width=80mm]{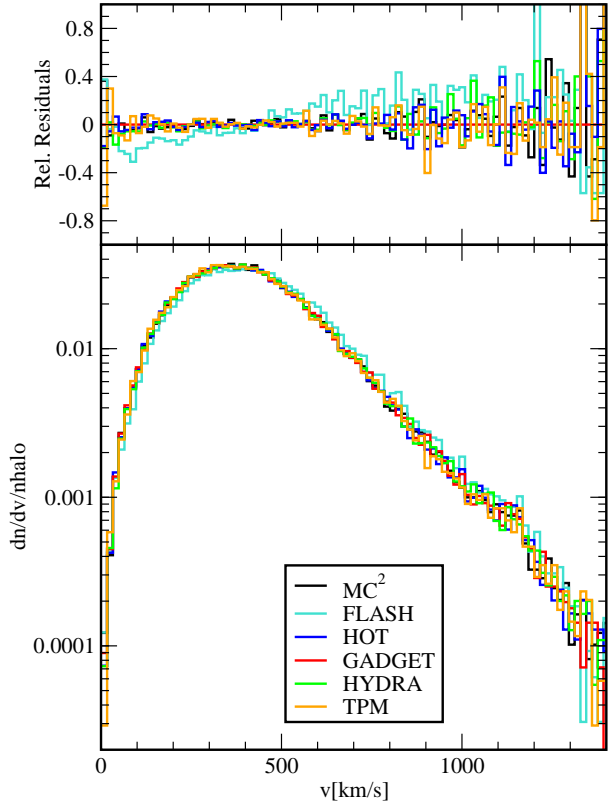}
\caption{Comparison of the halo velocity distribution for
all six codes at $z=0$, 90~Mpc box. This comparison uses all halos,
i.e., halos with a minimum number of 10 particles.}
\label{plotthirtyone}
\end{figure}

\begin{figure}
\includegraphics[width=80mm]{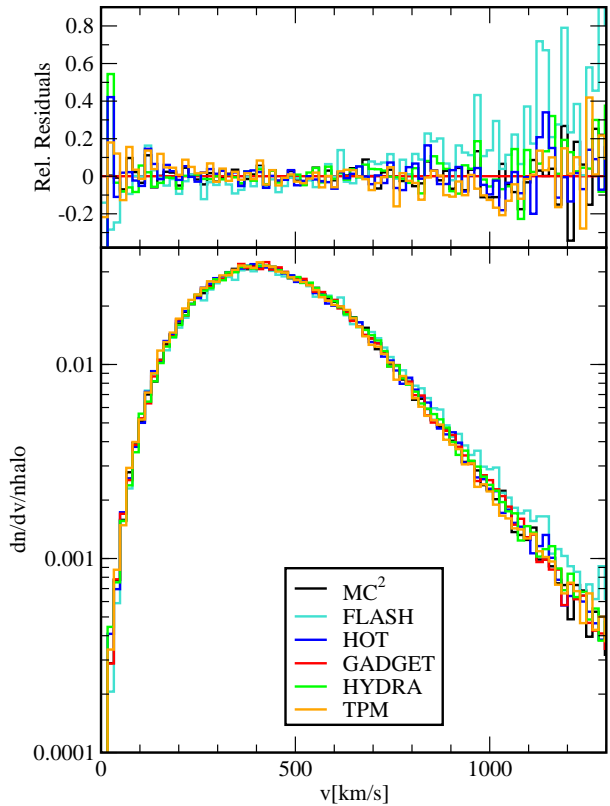}
\caption{Comparison of the halo velocity distribution for
all six codes at $z=0$, 360~Mpc box, following
Fig.~\ref{plotthirtyone}.} 
\label{plotthirtytwo}
\end{figure}

In Fig.~\ref{plotthirtyone} we present the halo velocity distribution
for the small box. Aside from the skewed distribution from the FLASH
results due to the missing lower-mass halos, all the other codes are
in reasonably good agreement. We have defined the velocity of a halo
as the average velocity taken over all the constituent particles;
consequently, there is an inherent statistical scatter in the
comparison of the results. Light halos will have a large velocity
scatter, and as light halos contribute over the entire range of
velocities, this will lead to scatter in the comparative results.
(Note that in this analysis we are accepting halos with only 10
particles.) The larger the halo mass, the less the scatter.
Furthermore, heavier halos tend to have lower velocities and this
trend is easy to spot in the better agreement below and around the
peak of the distribution in Fig.~\ref{plotthirtyone}. Up to 750~km/s
the agreement is at the $\pm 10$\% level. The results from the large
box are shown in Fig.~~\ref{plotthirtytwo}.

Another perspective, less sensitive to halo finder vagaries (White
2002), will appear in the later analysis of direct comparisons of
individual halos, given below. For halos with more than 30 particles
which are simultaneously identified in the results from all the codes,
the velocity scatter in the small box is approximately $\sim 10$ km/s
or less, with a corresponding value of $\sim 20$ km/s for the larger
box. As halos get heavier, there are often individual cases with
relatively significant mass variation due to the FOF halo finder, and,
as expected for these halos, the velocity can then show less
agreement. Nevertheless, for the halos that are simultaneously found
by the FOF finder in all the codes, the velocity agreement is
impressive, better than a few percent.

\begin{figure}[h]
\includegraphics[width=80mm]{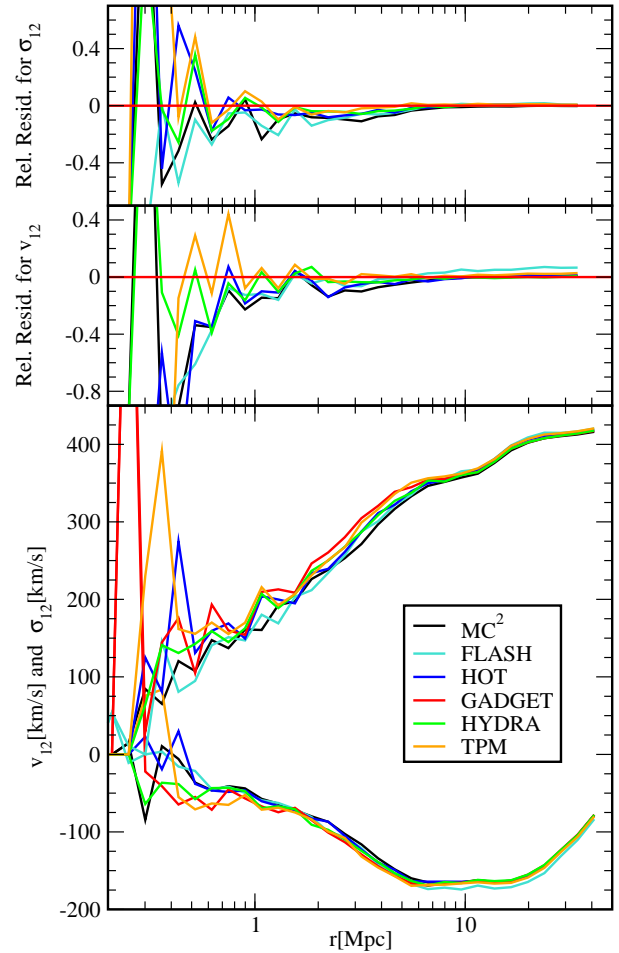}
\caption{Relative mean pairwise velocity $v_{12}(r)$ (lower set of
  curves) and the velocity dispersion $\sigma_{12}(r)$ (upper set)
  from halos with more than 100 particles (M$_{halo}>2\cdot
  10^{11}$~M$_\odot$) for the 90 Mpc box.}   
\label{plotthirtythree}
\end{figure}

\begin{figure}[t]
\includegraphics[width=80mm]{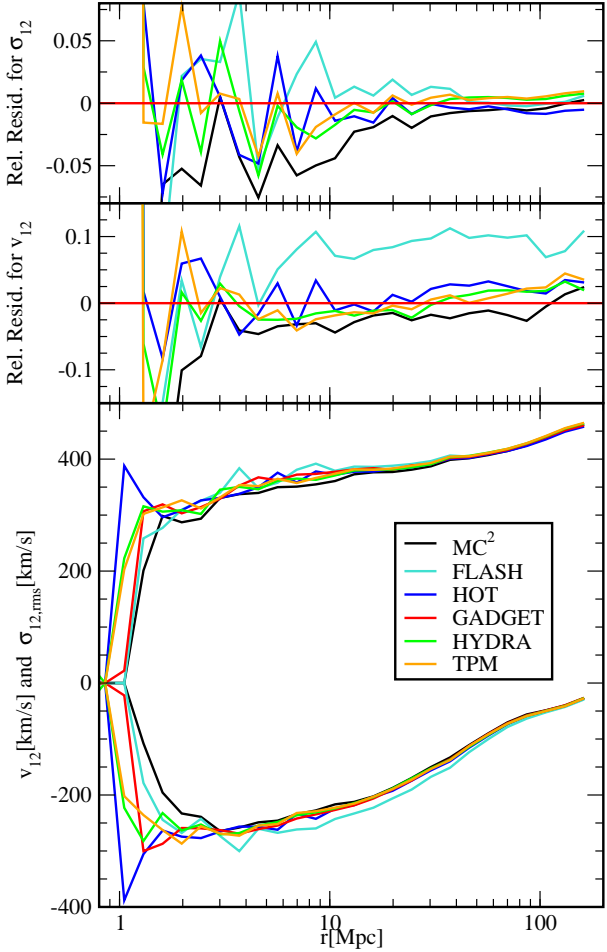}
\caption{Results from the 360~Mpc box following
Fig.~\ref{plotthirtythree}. The lower halo mass cutoff here is $1.2\cdot
10^{13}$~M$_\odot$.} 
\label{plotthirtyfour}
\end{figure}

The relative pairwise velocity for the halos and the corresponding
dispersion is shown for the small and large boxes in
Figs.~\ref{plotthirtythree} and \ref{plotthirtyfour}. These results should be
compared with the corresponding data from the individual particle
analysis (Figs.~\ref{ploteighteen} and ~\ref{plotnineteen}). The
points to note are that, as is well-known, $v_{12}(r)$ is relatively
insensitive as to whether particles or halos are considered, whereas
this is not true of $\sigma_{12}(r)$, which is significantly lower for
the halos.

Because of the halo sizes corresponding to the lower mass cutoffs, the
smallest pair-distance scale at which the small box results should
begin to make sense is $r\sim 0.5$ Mpc, with a corresponding
condition, $r\sim 2$ Mpc for the larger box. As an inspection of
Figs. ~\ref{plotthirtythree} and \ref{plotthirtyfour} demonstrates, it is only
after these conditions are satisfied that the code results converge,
as to be expected. The results from the large box are substantially
better than the results from the small box. This is because the halos
here have much larger masses (by a factor of 60) and are taking part
in a smoother regime of the flow. Beyond $r\sim 2$ Mpc, the codes give
consistent results at the 5\% level for $\sigma_{12}(r)$, and, except
for FLASH, also for $v_{12}(r)$. For the lower mass halos in the
smaller box, the situation is not as good. At the lower end of the
acceptable $r$ range, the error for both quantities is $\sim 20$\%,
improving to better than 5\% at a separation of a few Mpc. From the
point of view of comparison to observations this is still quite
acceptable, as present observational results have far worse errors and
statistics limitations (Feldman et al. 2003).

\vspace{1.75cm}

\newpage

\subsubsection{Direct Comparison of Positions, Masses, and Velocities 
of Individual Halos}

\begin{figure}[b]
\includegraphics[width=80mm]{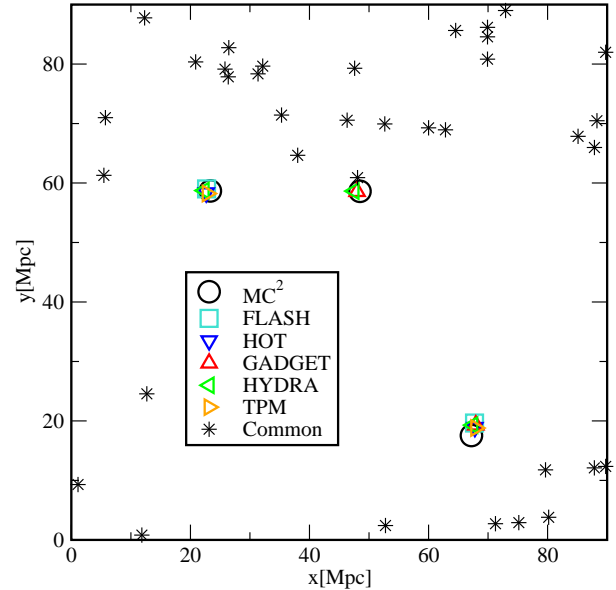}
\caption{Comparison of the positions of the heaviest halos 
  (M$_{halo}>5\cdot 10^{13}$ M$_\odot$) found in the six simulations
  at $z=0$ in the 90~Mpc box. Black stars mark halos which were found
  in all simulations at the same position with less than 0.5\%
  deviation. In order to visually separate the different code symbols,
  the halos from the MC$^2$ simulation were moved by
  0.5~Mpc in the $x$ direction, the HYDRA halos by -0.5~Mpc in the $x$
  direction, the FLASH halos by 0.5~Mpc in the $y$ direction, and the
  TPM halos by -0.5~Mpc in the $y$ direction.}
\label{plotthirtyfive}
\end{figure}

In this section we analyze and compare results for positions, masses,
and velocities for specific halos directly. In contrast to the
individual particles -- for which such an exercise would make little
sense in higher density regions -- halos may be considered, in an
informal sense, to be the fundamental stable mass units of
cosmological N-body simulations. Thus, to the extent that this is
true, we not only expect codes to agree on halo statistics but also on
individual halo properties such as the individual halo masses,
positions, and velocities, as well as individual halo statistics
describing the internal structure of the halos. This last part has
already been considered in our study of the Santa Barbara cluster
where the box size and mass resolution are roughly similar to the
small box $\Lambda$CDM simulation.

We have chosen three different mass bins selecting in turn the very
heaviest halos from the simulations, medium sized halos, and small
mass halos in such a way that every mass bin contains roughly 40
halos. For the lighter mass bins we also employ a spatial cut, i.e.,
restrict ourselves to only a fraction of the simulation volume in
order to keep the numbers of halos manageable. Details of the chosen
cuts are given below.

First we investigate the heaviest halos in the small box.  We identify
all halos above a mass of $5\cdot 10^{13}$ M$_\odot$ which corresponds
to 25000 particles per halo.  By imposing this mass bound we find
between 38 and 39 halos in the six different simulations.  Next we
compare the position for every single halo. The center of each halo as
defined as the particle with the largest number of friends (and thus
the highest density smoothed over the scale of the linking length).
We allow for a discrepancy of 0.5\% in the position (which translates
to 0.5 Mpc) in $x$, $y$, and $z$ directions for identifying the same
halo in all six simulations.  In Fig.~\ref{plotthirtyfive} we show the halos
found in this way with a black star. (All halos are projected onto the
$xy$-plane.)  Among the roughly 39 halos found in every simulation, 36
halos (92\%) were effectively at the same position in all six
simulations. In addition, we also display the halos which are {\em
not} found in every simulation. 

\begin{figure}[t]
\includegraphics[width=80mm]{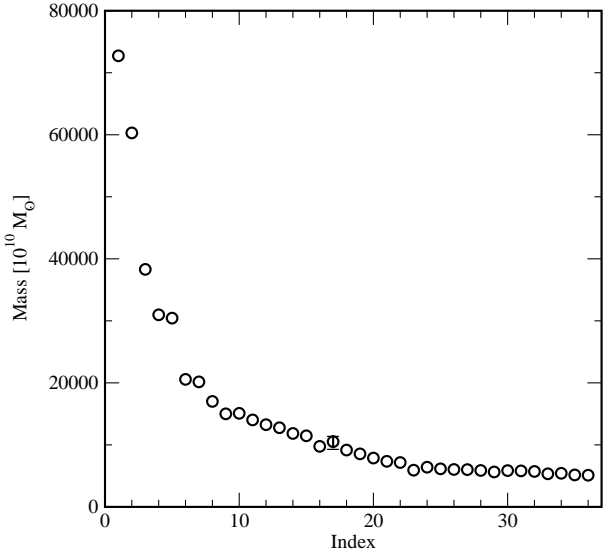}
\caption{Comparison of the masses of the heaviest halos identified 
  in all six codes at $z=0$, 90~Mpc box. The $x$-axis in the figure
  represents an arbitrary index label for each halo. The error bar
  describes the scatter of all codes from the average (see text).} 
\label{plotthirtysix}
\end{figure}

\begin{figure}[t]
\includegraphics[width=80mm]{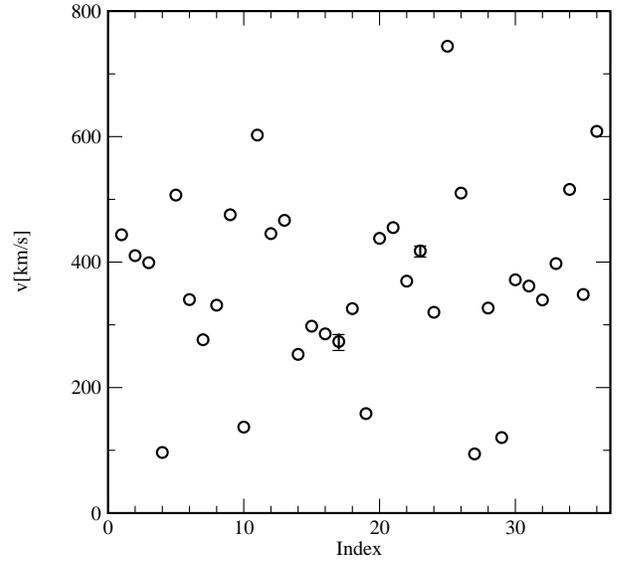}
\caption{Comparison of the velocities of the heaviest halos identified
in all six codes at $z=0$, 90~Mpc box. The $x$-axis shows the same index
as in Fig.~\ref{plotthirtysix}. The error bar describes the scatter
of all codes from the average.}
\label{plotthirtyseven}
\end{figure}

Next, for the halos which are identified to be at the same position in
all simulations, we compute the average mass and velocity and label
every halo with a labeling index.  Fig.~\ref{plotthirtysix} shows the
average mass at each index and Fig.~\ref{plotthirtyseven} shows the average
velocity at each index.  In addition, we have calculated the deviation
of every halo mass and velocity from the overall average:
$\delta_{m_i}=|m_i-\langle m \rangle|$ and $\delta_{v_i}=|v_i-\langle
v \rangle|$, $i=1,..$.  The average of these deviations:
$\frac{1}{6}\sum_{j=1,6}\delta_{m_j}$ and
$\frac{1}{6}\sum_{j=1,6}\delta_{v_j}$ describing the scatter from the
average are displayed as error bars; if no error bar is present the
error bar is smaller than the symbol itself; we find that the scatter
for most halos is smaller than 2\% for masses and velocities. The
agreement in masses and in velocities is excellent; the only error
bars distinguishable from the symbols are still much smaller than the
naive expected statistical scatter.

\begin{figure}
\includegraphics[width=80mm]{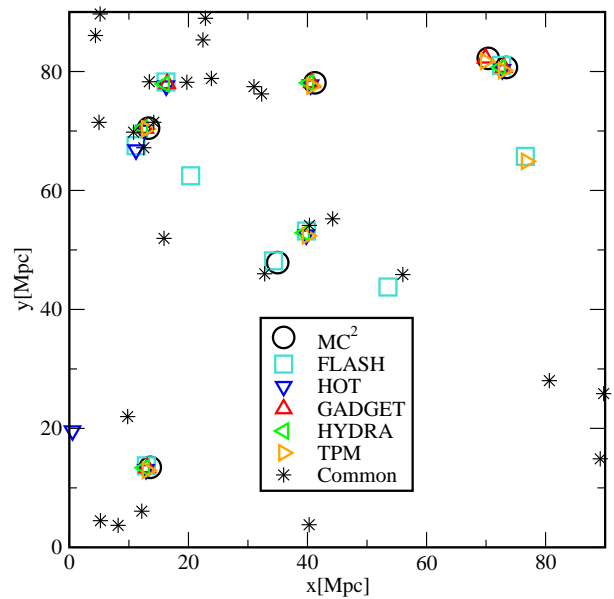}
\caption{Halo positions from all six codes following the conventions of
Fig.~\ref{plotthirtyfive}, but for halos containing 1500 to 10000 particles.
Again, to improve visibility the halos from the different codes were
shifted by $\pm$0.5~Mpc.}
\label{plotthirtyeight}
\end{figure}

\begin{figure}[t]
  \includegraphics[width=80mm]{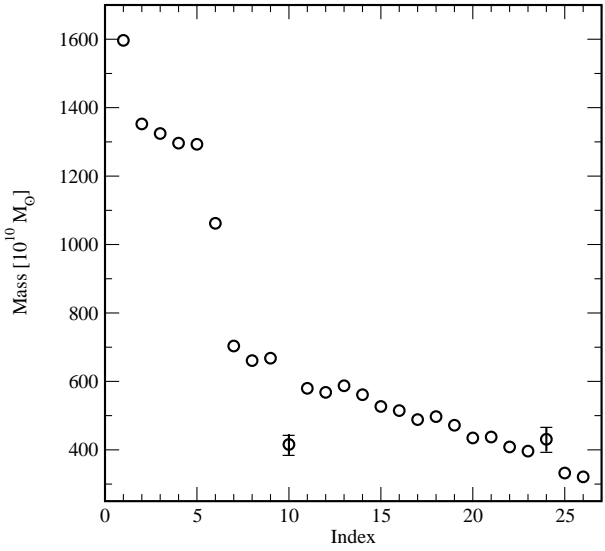}
\caption{Comparison of halo masses from all six codes for halos containing 
  1500 to 10000 particles, 90~Mpc box.}
\label{plotthirtynine}
\end{figure}

\begin{figure}[t]
\includegraphics[width=80mm]{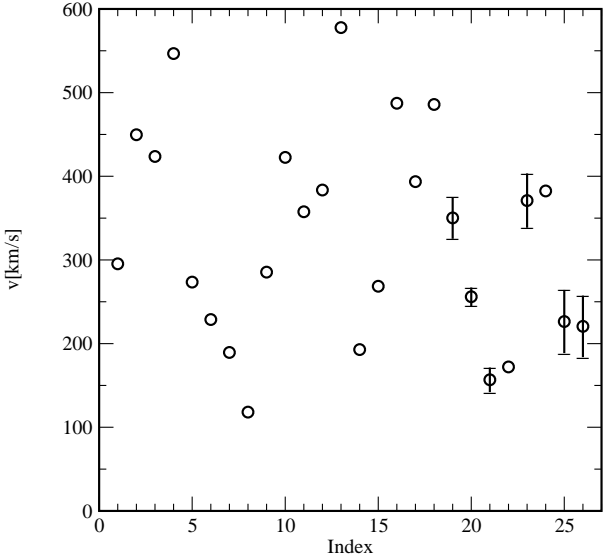}
\caption{Comparison of halo velocities for halos containing 1500 to 
  10000 particles, 90~Mpc box.}
\label{plotfourty}
\end{figure}

The medium mass halos are analyzed in the same way. The mass slice in
this case is chosen to be between $3\cdot 10^{12}$~M$_\odot$ and
$2\cdot 10^{13}$~M$_\odot$ which corresponds to roughly 1500 to 10000
particles.  In order to obtain roughly 40 halos, a spatial cut is
necessary.  The region analyzed in this case is a 10~Mpc slice in $z$
from $z=40$~Mpc to $z=50$~Mpc. In Fig.~\ref{plotthirtyeight} the identical
halos from all six simulations are marked by black stars and all
additionally found halos from all codes are shown.  In this mass bin,
26 out of 39 halos (66\%) are at the same positions in all six codes.
As carried out for the heaviest halos, we calculate the average mass
and velocity for each of the 26 identified halos and measure the
scatter.  The results are shown in Figs.~\ref{plotthirtynine} and
\ref{plotfourty}. If no errorbar is shown, the scatter in mass and
velocity is for most halos again below 2\%.  The masses agree very
well, and only two halos show a small amount of scatter around the
average. The result for the velocities is not as impressive;
nevertheless, only five of the halos show significant scatter.

\begin{figure}[b]
\includegraphics[width=80mm]{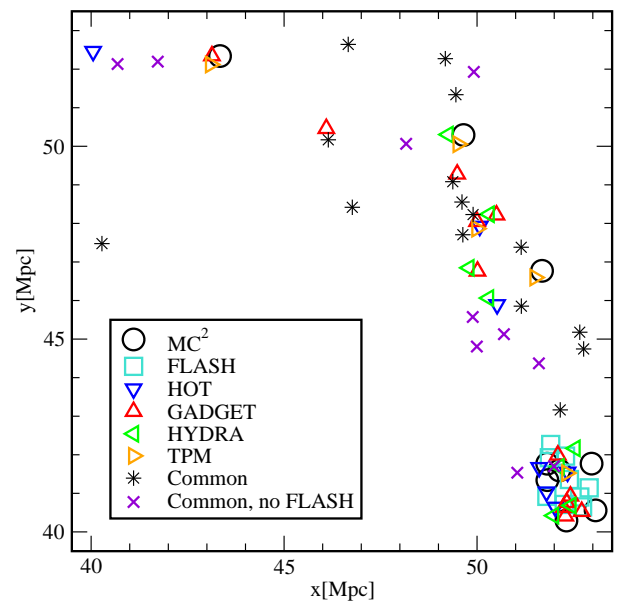}
\caption{Positions of halos, following Fig.~\ref{plotthirtyfive}, for
  halos containing 30 to 300 particles, 90~Mpc box. The additional
  points marked with crosses are halos found by all the codes,
  excluding FLASH. In this plot, the halos from MC$^2$, FLASH, HYDRA,
  and TPM were shifted by only 0.2~Mpc in the four different
  directions.} 
\label{plotfourtyone}
\end{figure}

\begin{figure}[t]
\includegraphics[width=80mm]{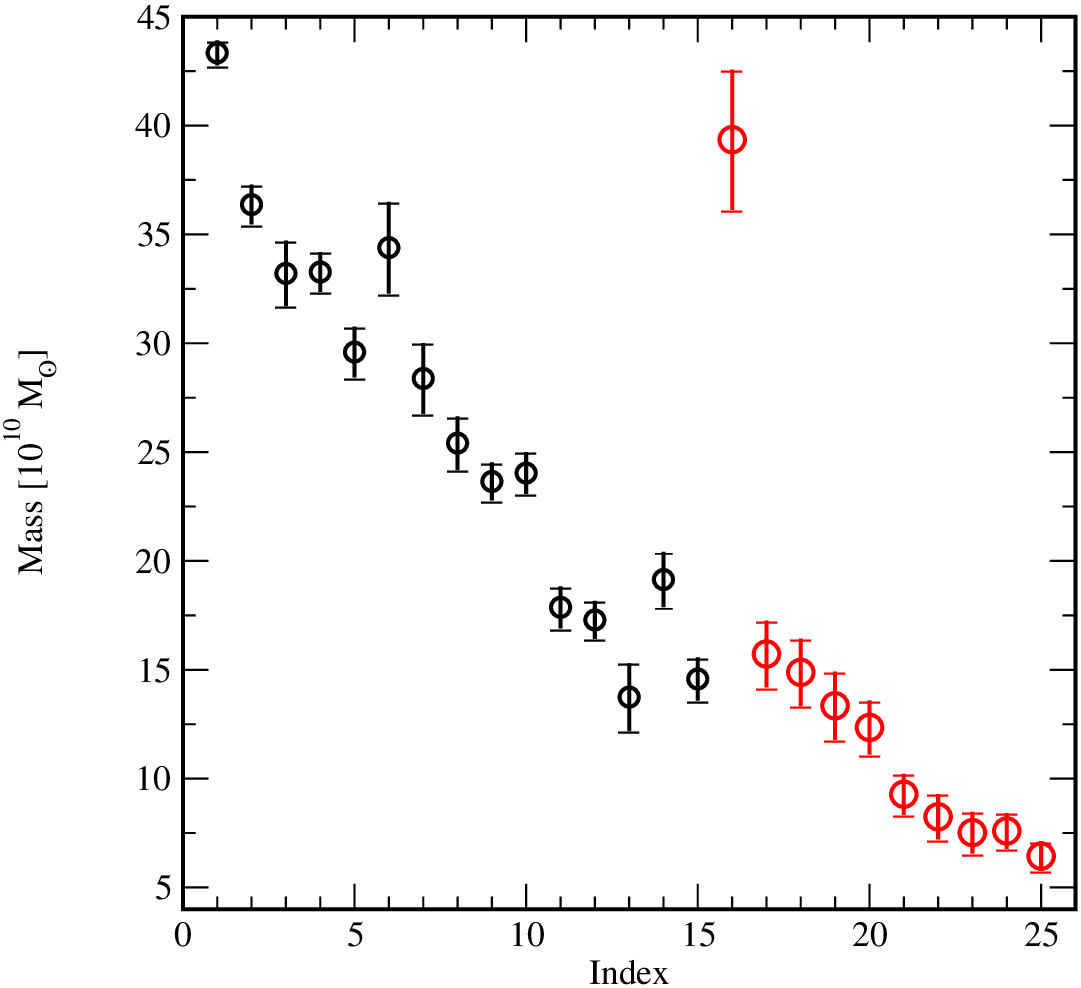}
\caption{Comparison of halo masses for halos containing 30 to 300 
  particles, 90~Mpc box. From halo 16 onwards, the FLASH results were
  not taken into account.}
\label{plotfourtytwo}
\end{figure}

Finally, we analyze a subset of the lightest halos in the simulations.
While our halo catalog contains halos identified at the level of
10~particles per halo, for this test we only consider halos with more
than 30~particles as this is often the lowest number used in
cosmological simulations.  The mass range in this case is $6\cdot
10^{10}$~M$_\odot$ to $6\cdot 10^{11}$~M$_\odot$ which translates to
$30 - 300$~particles.  In addition we restrict ourselves to the inner
cube between 40~Mpc and 53~Mpc in the $x$, $y$, and $z$ directions.
Again, we identify all halos which have the same positions in the six
simulations allowing for 0.5\% deviation. We find roughly 40 halos in
this volume and mass bin in the different codes. All codes agree on 15
halos; since the FLASH results are known to be deficient in low-mass
halos, a check without FLASH reveals agreement on 25 halos
(Fig.~\ref{plotfourtyone}). Nine of the missing halos are all at the
low-mass end (Fig.~\ref{plotfourtytwo}).

\begin{figure}[t]
\includegraphics[width=80mm]{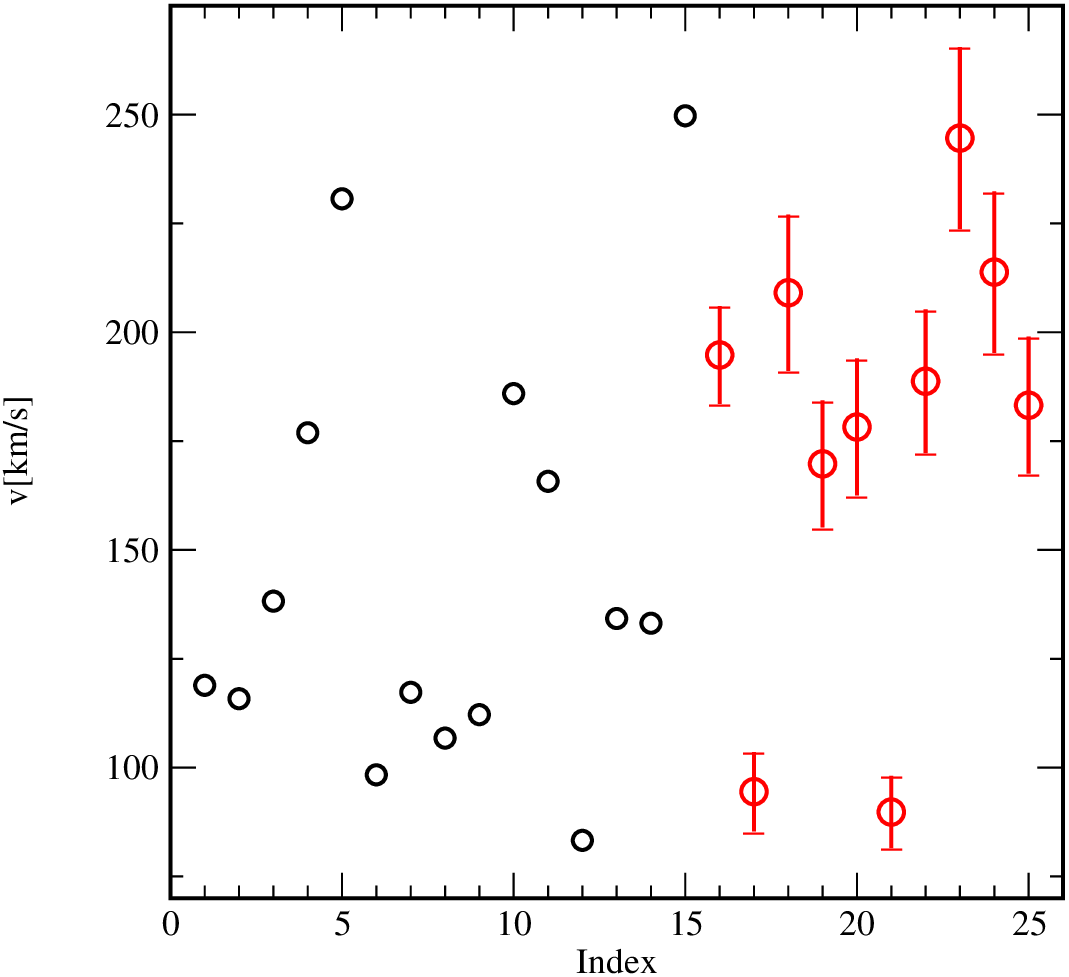}
\caption{Comparison of halo velocities for halos containing 30 to 300 
  particles, 90~Mpc box. From halo 16 onwards, the FLASH results were
  not taken into account.}
\label{plotfortythree}
\end{figure}

The agreement in the low-mass bin is -- excluding the FLASH results --
much the same as in the medium mass case. The average masses and their
scatter from the 25 identified halos are shown in
Fig.~\ref{plotfourtytwo}. Here the scatter, though still small, is
noticeably bigger than in the higher-mass bins. The halo velocity
distribution shows excellent consistency for 15 halos (out of the
heaviest 16 halos) with significantly more scatter for the remaining
halos (Fig.~\ref{plotfortythree}). If the velocity scatter is not
shown in Fig.~\ref{plotfortythree}, it is smaller than 2\%.

Overall the results of the direct halo comparison for the small box
are very good. Most halos are found at identical places and with very
close masses and velocities in all the simulations. (Only in the
smallest mass bin were some halos lost in the lower resolution FLASH
simulation.)

\begin{figure}[b]
\includegraphics[width=80mm]{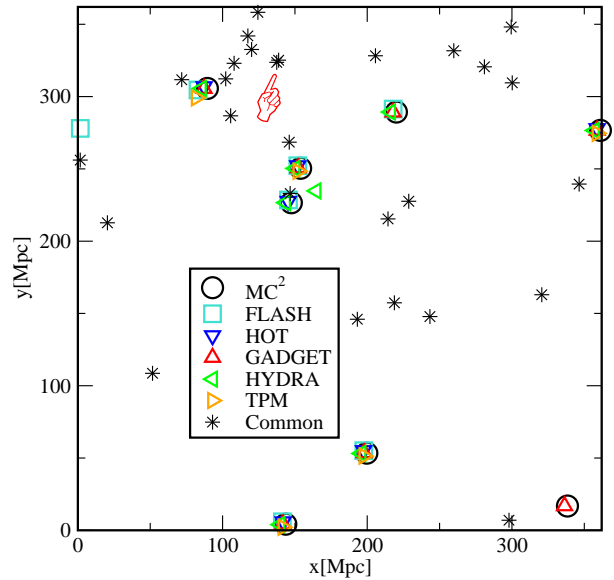}
\caption{Comparison of the positions of the heaviest halos 
  (M$_{halo}>7.32\cdot 10^{14}$~M$_\odot$) found in the six simulations
  at $z=0$ in the 360~Mpc box.  Black stars mark halos which were
  found in all simulations at the same position with less than 0.5\%
  deviation. To improve visibility of the code symbols, the remaining
  halos from MC$^2$ had their positions shifted by 2~Mpc in $x$, the
  HYDRA halos by -2~Mpc in $x$, the FLASH halos by 2~Mpc in $y$, and
  the TPM halos by -2~Mpc in the $y$ direction; the positions of the
  halos from GADGET and HOT were not changed.} 
\label{plotfourtyfour}
\end{figure}

\begin{figure}
\includegraphics[width=80mm]{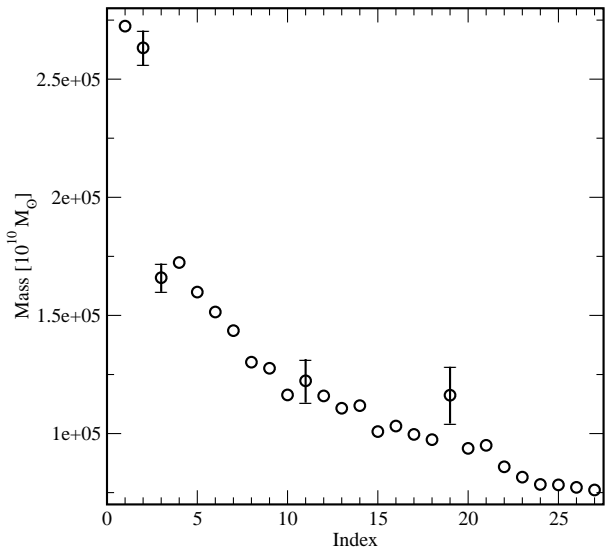}
\caption{Comparison of the masses of the heaviest halos
  identified in all six codes at $z=0$, 360~Mpc box. The $x$-axis
  represents an arbitrary index label for each halo. The error bar
  describes the scatter of all codes from the average.} 
\label{plotfourtysix}
\end{figure}

\begin{figure}
\includegraphics[width=80mm]{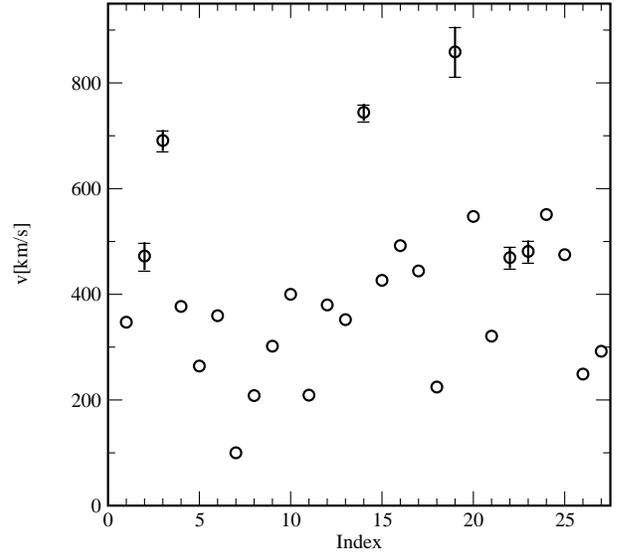}
\caption{Comparison of the velocities of the heaviest halos identified
  in all six codes at $z=0$, 360 Mpc box. The $x$-axis shows the same 
  index as in Fig.~\ref{plotfourtysix}.}
\label{plotfourtyseven}
\end{figure}

Next, the large box results are analyzed in a similar way.  As before,
we begin by considering the heaviest halos from the six simulations.
All halos above a mass of $7.32\cdot 10^{14}$~M$_\odot$ corresponding
to roughly 6000~particles are considered.  The heaviest halo found has
a mass of almost $2.8\cdot 10^{15}$~M$_\odot (\approx
22000$~particles).  Overall 38 different halos were identified in this
mass bin.  From these 38 halos, 27 (71\%) are at the same position in
all six simulations within 0.5\% corresponding to roughly 1.5~Mpc. The
positions of the halos are shown in Fig.~\ref{plotfourtyfour}, black stars
again showing the 27 halos found at the same position in all six
codes.  The halos which are not identified in all the codes are still
found in most of the codes or in all the codes with a slightly larger
difference in the position with only two exceptions. Remarkably, all
the identified 27 halos are the 27 heaviest halos, i.e., if we would
have chosen the mass cut slightly higher the agreement would have been
100\%.  Next we analyze the masses and velocities of the 27 common
halos, the results being shown in Figs.~\ref{plotfourtysix} and
\ref{plotfourtyseven}.  While the scatter in mass is negligible for most
halos (below 3\%), halo~2, halo~3, and halo~11 show a scatter between
roughly 5 to 10\%, still within expected statistical errors. The mass
of halo~19 varies significantly in the different simulations, with the
scatter lying at 20\%.

\begin{figure}
\includegraphics[width=80mm]{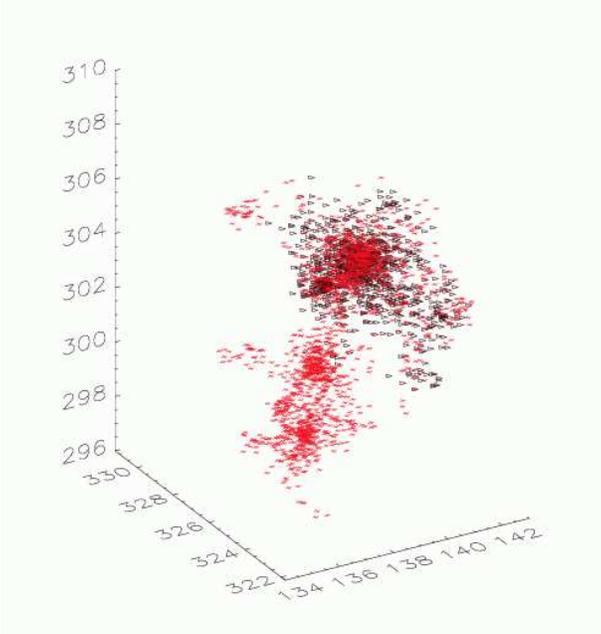}
\caption{Halo~19 from the MC$^2$ (red crosses) and GADGET simulations 
  (black triangles). The FOF halo finder has bridged two particle
  ``lumps'' into one halo for the MC$^2$ simulation, but not for the
  GADGET simulation.}
\label{plotfourtyfive}
\end{figure}

In order to understand this scatter, we have analyzed halo~19 in more
detail: in Fig.~\ref{plotfourtyfour} the red marker indicates the position of
this specific halo.  Halo~2 is located very close to halo~19.  We show
halo~19 from the MC$^2$ simulation (which has a relatively large mass
of $1.36\cdot 10^{15}$ M$_\odot$) and from the GADGET simulation (with
a lower mass of $9.42\cdot 10^{14}$ M$_\odot$) in
Fig.~\ref{plotfourtyfive}. The discrepancy is immediately clear: While in the
GADGET simulation, a more or less spherical halo was found by the FOF
halo finder, in the MC$^2$ simulation some particles close to the host
halo built a bridge to a smaller satellite halo below the big
halo. This is a well-known problem with FOF halo finders (see
e.g. Gelb \& Bertschinger 1994; Summers et al 1995). Due to the choice
of a specific linking length, two halos which would be identified by
eye as separate can be identified as being one halo if ``bridging''
particles between the halos are very close to each other. Spherical
density algorithms are an alternative to FOF halo finders (see
e.g. Lacey \& Cole 1994), but this method also does not identify
satellite halos separately. In the last several years, improved halo
finding algorithms have appeared, especially targeted to finding
subhalos in high resolution simulations. These methods are based on
hierarchical FOF and bound density maxima algorithms (see e.g. Klypin
et al. 1999). Since in this paper we are mainly interested in medium
resolution results, a basic FOF halo finder suffices.  Nevertheless,
the reader should keep in mind that results as found for halo~19 can
occur with a halo finding method which is insensitive to small scale
structure. Indeed, the minor code discrepancies found so far for halos
could easily result from the nondefinitiveness of FOF halo-finding.

The result for the velocities (Fig.~\ref{plotfourtyseven}) is very
similar to the results for the masses: six halos show significant
scatter in the velocities, three of them also displaying scatter in
the mass. For all six halos the scatter is around 10\%, therefore
below the naive statistical error. For all other halos the scatter is
smaller than 3\%. Overall the agreement of the heaviest halos in
position, mass, and velocity is rather good.

\begin{figure}
\includegraphics[width=80mm]{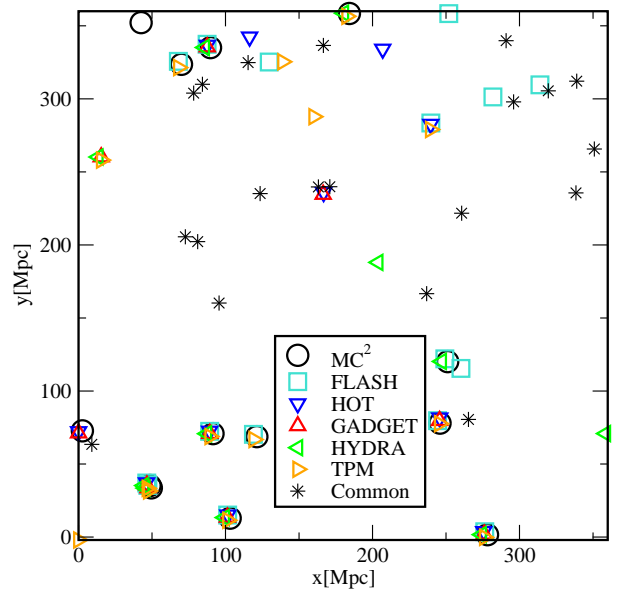}
\caption{Comparison of halo positions, following 
  Fig.~\ref{plotfourtyfour}, for halos containing 3000 to 5000
  particles, 360~Mpc box.  As for the largest halos, the positions of
  the colored halos were shifted by $\pm$2~Mpc to improve visibility.}
\label{plotfourtyeight}
\end{figure}

\begin{figure}
\includegraphics[width=80mm]{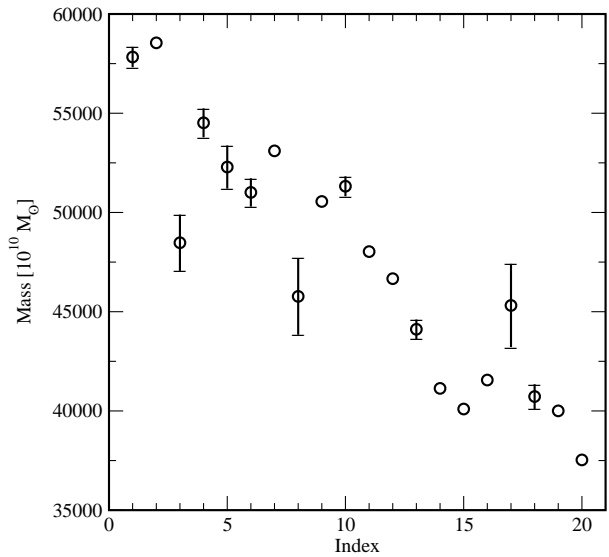}
\caption{Comparison of halo masses for halos containing 3000 to 5000 
  particles, 360~Mpc box.}
\label{plotfourtynine}
\end{figure}

The investigation of the medium mass halos begins by choosing a mass
slice between $3.682\cdot 10 ^{14}$ M$_\odot$ and $6.137\cdot 10
^{14}$ M$_\odot$ corresponding to $3000 - 5000$ particles per halo and
a spatial cut in $z$ between 135~Mpc and 325~Mpc. Altogether 46
different halos are identified in the different simulations. MC$^2$
has 32 halos in this mass and spatial bin, FLASH has 38, HOT has 32,
GADGET has the lowest number of halos in this bin with 30, HYDRA has
31, and in the TPM simulation 34 halos are found. We again ask which
of these halos are at the same position if we allow for an error of
0.5\% and this time find only 20 halos. In Fig.~\ref{plotfourtyeight} these 20
halos are shown with black stars and the rest of the halos colored
according to the simulation they are found in. In this mass bin, more
halos are identified only in single simulations than is the case for
the very heavy halos: This could be due to the fact that on the edges
of the mass and spatial cuts, halos from some simulations are getting
lost. Next we compare the masses and velocities of the 20 halos common
to all six simulations.  The results are shown in Figs.~\ref{plotfourtynine}
and \ref{plotfifty}.  The largest scatter in the mass is around 10\%,
still of the expected statistical fluctuation. If no errorbar is
displayed, the scatter is well below 2\%. More medium mass halos show
a measurable scatter (half of them) than for the heaviest halos
(15\%).  The scatter of the velocities is very small (below 2\% if no
errorbar is shown), the agreement of all codes being excellent.
Overall, the agreement for the medium mass halos is not as good as for
the heaviest halos but still satisfactory.

\begin{figure}
\includegraphics[width=80mm]{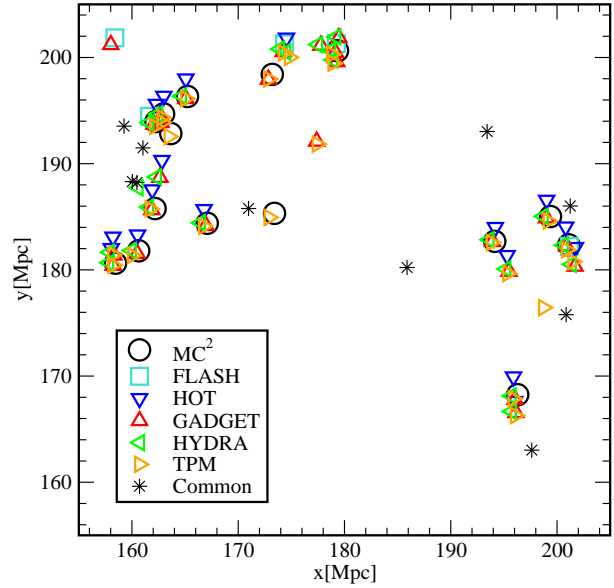}
\caption{Comparison of halo positions, following Fig.~\ref{plotfourtyfour}, 
  for halos containing 30 to 300 particles, 360~Mpc box. The code
  symbols were shifted by 0.3~Mpc for MC$^2$, FLASH, HYDRA, and TPM in
  the four directions to improve visibility.}
\label{plotfiftyone}
\end{figure}

\begin{figure}
\includegraphics[width=80mm]{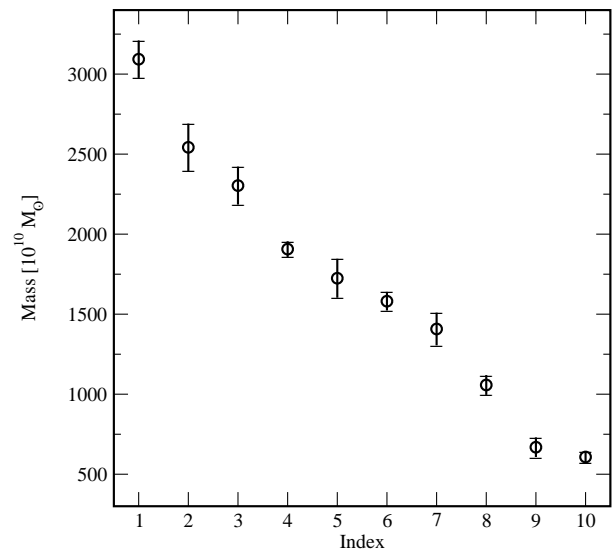}
\caption{Halo masses for halos containing 30 to 300 particles.}
\label{plotfivtytwo}
\end{figure}

The last mass range we consider consists of halos in the mass range
between $3.682\cdot 10 ^{12}$ M$_\odot$ and $3.68\cdot 10 ^{13}$
M$_\odot$, corresponding to $30 - 300$ particles. We cut out an inner
cube of the simulation with $x$, $y$, and $z$ between 158~Mpc and
202~Mpc.  Overall, 43 different halos were found in all six
simulations. The MC$^2$ simulation has 25 halos in this bin, FLASH
only 15, HOT 27, GADGET 35, HYDRA 33, and TPM 35. From these 43 halos,
10 are found in all simulations, again allowing for 0.5\% of
spatial deviation.  Fig.~\ref{plotfiftyone} shows these 10 halos with a
black star as well as all other halos. This number appears to be very
small but is mainly due to the fact that the resolution of FLASH is
not sufficient to resolve very small halos (as observed and discussed
already for the mass function). The scatter in the masses and
velocities for the 10 halos is shown in Figs.~\ref{plotfivtytwo} and
\ref{plotfiftythree}. The result here is very good, and all halos agree
within expected statistical fluctuations.

In summary, the results for the direct halo comparison for the large
box are very consistent with our expectations. The very heavy halos
show excellent agreement across all the codes, while for the very
small halos the effects of force resolutions for the different codes
become apparent and the results are not as good. Nevertheless, given
the fact that the resolution of the two mesh codes MC$^2$ and FLASH is
an order of magnitude worse than of the other codes in these tests,
the agreement for even the smallest halos is surprisingly good.

\begin{figure}
\includegraphics[width=80mm]{vel2ind.eps}
\caption{Comparison of halo velocities for halos containing 3000 to 5000 
  particles, 360~Mpc box.}
\label{plotfifty}
\end{figure}

\begin{figure}[t]
\includegraphics[width=80mm]{vel4ind.eps}
\caption{Comparison of halo velocities for halos containing 30 to 300 
  particles, 360~Mpc box.}
\label{plotfiftythree}
\end{figure}

\section{Discussion and Conclusions}
\label{sec:disc}

In this paper we have reported the results from a detailed code
comparison study with 6 codes for different test problems.  For all
tests the codes were given exactly the same initial conditions and the
analysis of the simulation results was carried out in an identical
fashion.  This ensured that we only analyzed deviations arising from
differences in the algorithms and their implementations in simulating
the evolution of dark matter tracer particles.  We restricted
ourselves to medium resolution simulations, resolutions between
$\sim$~10-100~kpc being sufficient for many applications.  The codes
investigated herein represented a variety of algorithms, including a
plain PM-code, an AMR-code, tree-codes, a tree-PM code, and an AP$^3$M
code.  Four of the codes are publicly available (FLASH, GADGET, HYDRA,
TPM), and three were largely developed by us (MC$^2$, HOT, FLASH).  We
analyzed particle and halo statistics in considerable detail.  All
results (particle and halo catalogs) and the initial conditions are
publicly available: We encourage other researchers to test their codes
against our results and to devise new comparisons.

The tests employed force and mass resolutions typical of modern
state-of-the-art codes used for analyzing gravitational clustering in
cosmology.  Overall, the results obtained are the following: the codes
agreed to better than 5\% in some tests and at the approximately 10\%
level for others. Larger systematic disagreements were usually due to
understandable causes such as insufficient force resolution. While the
purpose of this paper is to present a broad suite of results, certain
specific observations can and should be investigated in more detail,
e.g., the results for the power spectrum, where the ``only'' 5\%
agreement may already provide cause for some concern.

We began our series of investigations with the Zel'dovich pancake
collapse test. The PM code MC$^2$ passed this test very easily even
with relatively high force resolution while the other codes (including
the AMR-code at the same force resolution) could not track the
caustics forming in the nonlinear regime. We concluded that the reason
for this failure is the inability of the high-resolution codes to
maintain the planar symmetry of the problem and not because of
unphysical collisions. Tests with cosmological initial conditions
showed that the failure of the pancake test from the non-PM codes does
not lead to significant errors in these more realistic situations. The
encouraging agreement of Richardson extrapolation-improved MC$^2$ and
GADGET results for the particle velocity distribution is also strong
evidence against the presence of unphysical collisionality.

Next we investigated the dark matter component of the Santa Barbara
cluster simulation running all codes with identical particle initial
conditions, obtaining much better agreement between the different
codes than in the original paper. Two other reasons that contributed
to this were that we analyzed all simulations in the same way, and the
resolution range of the simulations did not vary as much as in the
original paper.

Finally, we simulated a $\Lambda$CDM cosmology using two boxes, one
being 90~Mpc on a side, the other, 360~Mpc. Particle statistics
(including velocity statistics, power spectra, and correlation
functions) agreed well, the lower resolution of the mesh codes
becoming apparent, as expected, in the power spectrum and the
correlation functions. The agreement for halo statistics was further
improved, with even the plain PM-code MC$^2$ obtaining results very
close to those from the high-resolution codes. Only FLASH and TPM
deviated from the other codes, for the most part with known underlying
causes. The results for the halo statistics are particularly
significant since for many problems (generating mock catalogs,
evolution of the halo mass function, etc.), cosmologically relevant
information resides more in the spatial and mass distribution of halos
and their internal structure, and less in the statistics of the
underlying tracer particles.
 
While our findings are in general satisfying -- the codes showed
almost no unexpected or unexplainable behavior and, despite the use of
differing algorithms, agreed relatively closely -- in the coming age of
``precision cosmology'' one has to ask whether the level of agreement
found here is sufficient for application to certain next-generation
observations and, if not, where improvements are needed. As one
example, weak gravitational lensing observations promise to deliver
measurements of the mass power spectrum to an accuracy of 1\%~(Huterer
\& Takada 2004). At this level one may worry about feedback effects
from baryons (White 2004; Zhan \& Knox 2004) and neutrinos (Abazajian
et al. 2004) as well as intrinsic code errors in modeling the
distribution of dark matter (including the implementation of initial
conditions, not addressed here).

As a starting point, the present agreement over a broad range of tests
is gratifying, nevertheless, the lack of a rigorous quantification of
error for gravitational N-body solvers is a serious barrier to future
progress. As error control requirements become more severe, the need
for such a theory becomes further manifest. In addition, as more
(uncontrolled) physics is added, and subgrid modeling incorporated as
an essential part of the simulations, it becomes ever harder to
extract error-controlled results. Finally, in order to be useful,
numerical results must be connected to observations. Often this can be
done only indirectly and rather imprecisely, independent of the
quality of the observations themselves.
   
Thus, to constrain the cumulative error from all links in the
simulation/observation chain to less than 1\%, the development of a
multi-step error-control methodology is necessary. It is possible that
for some applications this will remain a hopeless task, but for others
it may be viable. Although the present paper in no way pretends to
address the global problem, we hope that it contains useful hints for
taking some of the initial steps.

\acknowledgements

S.H., K.H., and M.S.W. acknowledge support from the Department of
Energy via the LDRD program of Los Alamos National Laboratory. P.M.R.
acknowledges support from the University of Illinois at
Urbana-Champaign and the National Center for Supercomputing
Applications (NCSA).  P.M.R., K.H., and S.H. acknowledge the
hospitality of the Aspen Center for Physics where part of this work
was carried out.  The calculations described herein were performed
using the computational resources of NCSA, NERSC, and Los Alamos
National Laboratory.  A special acknowledgement is due to
supercomputing time awarded to us under the LANL Institutional
Computing Initiative.  Development of FLASH was supported by DOE under
grant number B341495 to the Center for Astrophysical Thermonuclear
Flashes at the University of Chicago.  We wish to thank our many
colleagues who have developed cosmological simulation and diagnostic
tools and have made the tools and the results available for public
use.  We are indebted to Kev Abazajian, Gus Evrard, Nick Gnedin,
Stefan Gottl\"ober, Daniel Holz, Gerard Jungman, Anatoly Klypin,
Andrey Kravtsov, Adam Lidz, Zarija Lukic, Adrian Melott, Ue-li Pen,
James Quirk, Robert Ryne, Sergei Shandarin, Ravi Sheth, Volker
Springel, Martin White, and Yongzhong Xu for discussions and helpful
advice.


\end{document}